\documentclass{article}

\usepackage{arxiv}

\usepackage[utf8]{inputenc} 
\usepackage[T1]{fontenc}    
\usepackage{hyperref}       
\usepackage{url}            
\usepackage{booktabs}       
\usepackage{amsfonts}       
\usepackage{nicefrac}       
\usepackage{microtype}      
\usepackage{graphicx}
\usepackage[square,numbers]{natbib}
\usepackage{doi}
\usepackage{amsmath}
\usepackage{xcolor}
\usepackage[flushleft]{threeparttable}
\usepackage{algorithm}
\usepackage{algpseudocode}
\usepackage{ulem}

\title{One-sample survival tests in the presence of non-proportional hazards in oncology clinical trials}

\author{ \href{https://orcid.org/0009-0000-4788-9386}{\includegraphics[scale=0.06]{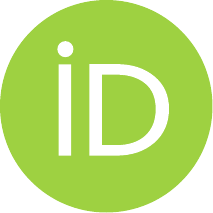}\hspace{1mm}Chlo\'e Szurewsky} \\
	Oncostat, CESP, INSERM U1018\\
	University Paris-Saclay\\
	Villejuif, France \\
	\texttt{chloe.szurewsky.pro@gmail.com} \\
	\And
	\href{https://orcid.org/0000-0003-3276-1392}{\includegraphics[scale=0.06]{orcid.pdf}\hspace{1mm}Guosheng Yin} \\
	Department of Statistics and Acturial Science\\
	University of Hong Kong\\
	Hong Kong, China \\
	\texttt{gyin@hku.hk} \\
	\AND
    \href{https://orcid.org/0000-0003-3292-0939}{\includegraphics[scale=0.06]{orcid.pdf}\hspace{1mm}Gw\'ena\"el Le Teuff} \\
	Oncostat, CESP, INSERM U1018\\
	University Paris-Saclay\\
	Villejuif, France \\
	\texttt{gwenael.leteuff@gustaveroussy.fr} \\
}

\date{}

\renewcommand{\headeright}{}
\renewcommand{\undertitle}{}

\hypersetup{
pdftitle={One-sample survival tests in the presence of non-proportional hazards in oncology clinical trial},
pdfsubject={q-bio.NC, q-bio.QM},
pdfauthor={Chlo\'e Szurewsky, Guosheng Yin, Gw\'ena\"el Le Teuff},
pdfkeywords={Combination tests, Non-proportional hazards, One-Sample Log-Rank Test, l, Score test, Single-arm trials},
}

\begin{document}
\maketitle

\begin{abstract}
    In oncology, conducting well-powered time-to-event randomised clinical trials can be challenging due to the limited number of patients. Several single-arm trial (SAT) designs have recently been proposed to overcome this issue. They rely on the (modified) one-sample log-rank test (OSLRT) under the proportional hazards assumption to compare the survival of an experimental group with that of an external control. We extend Finkelstein's formulation of the OSLRT as a score test by incorporating a piecewise exponential model to capture early, middle, and delayed treatment effects and an accelerated hazards model to accommodate crossing hazards. We further adapt the restricted mean survival time based test and propose a combination test procedure (max-Combo) for SATs. The performance of the developed tests is evaluated through a simulation study. The score tests control the type I error similarly to the OSLRT and achieve the highest power when the data-generating mechanism matches the assumed model. The max-Combo test is more powerful than the OSLRT across all considered scenarios and provides a robust alternative. Uncertainty in the estimation of the survival curve for the external control group, as well as model misspecification, may have a significant impact on performance. The proposed tests are further illustrated using three real data examples.
\end{abstract}


\renewcommand\thefootnote{\fnsymbol{footnote}}
\setcounter{footnote}{1}

\section{Introduction} \label{intro}

In oncology, conducting well-powered phase II randomised clinical trials (RCTs) with time-to-event (TTE) outcomes may be challenging in certain situations, such as rare cancers (e.g. paediatric cancers), personalised medicine where targeted therapies are evaluated for specific tumour biomarkers, or for ethical reasons \citep{FDA2019_rare, FDA2019_demonstrating}. As these situations arise more often when evaluating new therapies \citep{FDA2019_rare}, there is a growing need for innovative clinical trial designs to accelerate clinical research. One proposal is single-arm trials (SATs) that compare the survival of an experimental group to that of an external or historical control group \citep{Davi2020}. This external or historical control group can be constructed using information from the literature of a previous trial or reconstructed from published Kaplan-Meier curves. In this context, it is not appropriate to use the classical two-sample log-rank test as the variance would be incorrectly estimated, leading to invalid p-values \citep{Wu2014}. \\

\noindent Over the last decade, several one- or two-stage designs have been proposed for SATs with a TTE endpoint \citep{Kwak2014, Wu2016, Schmidt2018, Wu2020, Abbas2022}, analogous to Simon's two-stage design \citep{Simon1989} for a binary endpoint. These designs rely on the one-sample log-rank test (OSLRT) \citep{Breslow1975, Woolson1981, Finkelstein2003} and its modified version (mOSLRT) \citep{Wu2014}, both under the proportional hazards (PH) assumption. For example, Kwak and Jung \citep{Kwak2014} evaluate an experimental treatment compared to an external control treatment with a two-stage design based on the OSLRT. During the first stage, $n_1$ patients are recruited and treated up to the interim analysis at time $\tau$, before the end of the planned accrual period. At this interim analysis, the OSLRT is computed, and the treatment is rejected for futility if the statistic is greater than an early stopping value $c_1$. Otherwise, the trial continues to accrue and treat patients. Based on this two-stage approach, they propose two optimal designs: one minimising the expected sample size or the expected accrual period and one minimising the maximum sample size or the maximum accrual period. Abbas et al \citep{Abbas2022} propose a three-arm two-stage design with no formal test for futility or efficacy at interim analysis. At the first stage, the three randomised arms are compared to the historical control arm with a drop-the-loser approach. The mOSLRT is computed for each treatment arm, and the one with the largest statistic is selected for the second stage. \\

\noindent However, the PH assumption may be violated when the relative treatment effect between the experimental and external control groups is time-dependent. For example, this situation may arise in immuno-oncology clinical trials, where a delayed treatment effect is observed. In such cases, the use of the OSLRT and its modified version can lead to erroneous conclusions. So far, no survival SAT design currently exists that accommodates a time-dependent treatment effect between an experimental and an external control group, except for Chu et al. \citep{Chu2020}, who propose a design for long-term survivors (cure model) with a random delayed effect based on a piecewise exponential proportional hazards model. \\

\noindent The objectives of this work are (i) to develop statistical tests based on the OSLRT for SATs when the PH assumption does not hold and (ii) to evaluate alternative approaches, including an RMST (Restricted Mean Survival Time) based test and a combination test for SATs. These two alternative approaches are commonly used in RCTs in the presence of non-PH. The former quantifies the area under the survival curve and tests for differences in this quantity between the two groups \citep{Royston2013, Uno2014}. The latter combines different statistical tests when no prior knowledge about the form of the treatment effect over time exists, which allows different treatment effect patterns to be tested \citep{Karrison2016}. \\

\noindent The paper is organised as follows. Section 2 presents the definition of the OSLRT statistic. Sections 3 and 4 describe the OSLRT-based score tests and the two alternative testing approaches extended to SATs. In Section 5, the simulation study of SATs mimicking different treatment effects over time is described with its simulation parameters. Section 6 investigates the impact of variability in the survival curve estimate for the external control group and its model misspecification. In Section 7, the proposed methods are applied to three real data examples: a phase II SAT in adults with a high-grade astrocytoma treated with an inhibitor \citep{Kelly2023}, a phase II SAT in children with neuroblastoma \citep{Fox2014} and a subgroup of patients from a phase III RCT in patients with small-cell lung cancer \citep{Liu2021}. The endpoint is overall survival (OS) for these three oncology clinical trials. The main results and discussion are presented in Section 8.

\section{One-sample log-rank test} \label{sec1}

\subsection{Notations}

Let $T_i$ and $C_i$ be the individual failure and censoring times of the $i^{th}$ patient ($i=1, ..., n$) in the experimental group of size $n$, assumed to be independent. The observed event time is defined as $X_i = \min(T_i, C_i)$ and the associated failure indicator as $\delta_i = \mathrm{I}(T_i \leq C_i)$. The hazard, cumulative hazard and survival functions of the external control group are $\lambda_0(t)$, $\Lambda_0(t)$, and $\mathrm{S}_0(t)$, respectively and those of the experimental group are $\lambda_1(t)$, $\Lambda_1(t)$, and $\mathrm{S}_1(t)$. Assume that the external control group is known without sampling variability for the purpose of design and analysis of SATs \citep{Wu2021} and that there are no differences in patient populations. In a survival SAT, the hypotheses are expressed as:
\begin{equation}
    \mathrm{H}_0 : \mathrm{S}_1(t) \leq \mathrm{S}_0(t) \text{ vs } \mathrm{H}_a : \mathrm{S}_1(t) > \mathrm{S}_0(t) \text{ for all $t > 0$}
\end{equation}

\subsection{Formulation}

Considering the proportional hazards model $\mathrm{S}_1(t) = \mathrm{S}_0(t)^{\mathrm{HR}}$, with HR the hazard ratio of the experimental group versus the external control group, the OSLRT statistic \citep{Breslow1975, Woolson1981, Finkelstein2003} is defined as
\begin{equation}
    \mathrm{OSLRT = \frac{O-E}{\sqrt{E}}}
    \label{OSLRT}
\end{equation}
where $\mathrm{O} = \sum\limits_{i=1}^n \delta_i$ is the observed number of events and $\mathrm{E} = \sum\limits_{i=1}^{n} \Lambda_0(X_i)$ is the expected number of events calculated from a parametric estimate of the cumulative hazard function of the historical control group. Different survival distributions (e.g., exponential, Weibull, log-logistic, log-normal) can be used to model the cumulative hazard function of the external control group. In practice, the exponential distribution is typically used, regardless of its adequacy to fit the data. Choosing other distributions, such as Weibull or log-logistic, can substantially impact sample size calculations \citep{Wu2017, Wu2021}. A parametric distribution is used to compute $\mathrm{E}$ because a non-parametric based test does not preserve the empirical type I error and power, in particular when the sample size is small \citep{Wu2021}.\\

\noindent Finkelstein et al. reformulate the OSLRT as a score statistic under the PH assumption \citep{Finkelstein2003}. A score statistic is defined as follows:
\begin{equation*}
    \mathrm{Score} = \frac{\left. \frac{\partial \log(L)}{\partial \beta} \right| _{\beta = 0}}{\sqrt{-\left. \frac{\partial^2 \log(L)}{\partial \beta^2}\right| _{\beta = 0}}},
\end{equation*}
Denoting $\beta = \log(\mathrm{HR})$, the PH model introduced above is equivalent to $\Lambda_1(t) = \exp(\beta)\Lambda_0(t)$ and $\lambda_1(t) = \exp(\beta)\lambda_0(t)$. The log-likelihood of the survival data from the experimental group can be written as:
\begin{equation}
    {\rm \log(L)} = \sum\limits_{i = 1}^n \left[\delta_i \log({\rm f_1}(X_i)) + (1-\delta_i)\log({\rm S_1}(X_i))\right] = \sum\limits_{i = 1}^n \left[\delta_i \log(\lambda_1(X_i)) - \Lambda_1(X_i)\right]
    \label{log-like}
\end{equation}
where $\mathrm{f_1}$ is the probability density function of the time-to-event data of the experimental group. \\
The score statistic is obtained by evaluating the first and second derivatives of the log-likelihood with respect to $\beta$ under the null hypothesis ($\beta = 0$). The first derivative evaluated at $\beta = 0$, gives
\begin{equation*}
    \left. \frac{\partial \log(L)}{\partial \beta} \right| _{\beta = 0} = \sum_{i = 1}^n \left(\delta_i - \Lambda_0(X_i)\right) = \mathrm{O - E}
\end{equation*}
while the negative second derivative gives
\begin{equation*}
    \left. - \frac{\partial^2 \log(L)}{\partial \beta^2}\right| _{\beta = 0} = \sum_{i = 1}^n \Lambda_0(X_i) = \mathrm{E}
\end{equation*}
Therefore, the score statistic under the PH model is
\begin{equation*}
    \mathrm{Z_{PH} = \frac{O - E}{\sqrt{E}} = OSLRT}
\end{equation*}
Additional algebraic details of the derivation are provided in Appendix \ref{OSLRT details}. \\

\noindent As the OSLRT is a conservative test, even for large sample sizes \citep{Kwak2014, Wu2014, Sun2011}, Wu \citep{Wu2014} proposes a modified version of the OSLRT statistic (mOSLRT), defined as:
\begin{equation}
    \mathrm{mOSLRT = \frac{O-E}{\sqrt{\frac{O+E}{2}}}}.
    \label{mOSLRT}
\end{equation}

\noindent Under the null hypothesis, the OSLRT and mOSLRT statistics follow a standard normal distribution $\mathcal{N}(0, 1)$ asymptotically. Hence, the null hypothesis $\mathrm{H}_0$ is rejected when $\mathrm{OSLRT} < -z_{1-\alpha}$ where $z_{1-\alpha}$ is the $100(1-\alpha)$ percentile of the standard normal distribution.

\section{Score tests for non-proportional hazards}\label{sec2}

As described in Section \ref{sec1}, Finkelstein et al. \citep{Finkelstein2003} expressed the OSLRT as a score statistic under the PH assumption. To address the issue of non-proportional hazards between the experimental and external control groups in a SAT, we consider a piecewise exponential proportional hazards (PEPH) model, which allows the HR to vary across pre-specified time intervals \citep{Chu2020, He2013}: 
\begin{equation*}
    \lambda_1(t) =
    \begin{cases}
        r_1 \lambda_0(t)  &\text{if $t \leq k_1$}\\
        r_2 \lambda_0(t)  &\text{if $k_1 < t \leq k_2$}\\
        \vdots \\
        r_{l+1} \lambda_0(t)  &\text{if $t \geq k_l$}
    \end{cases}
\end{equation*}
where $k_i$ denote the change-points (CPs) of the model with $0 = k_0 < k_1 < k_2 < \cdots < k_l < k_{l+1} = \infty$ and $r_i$ the HR, assumed constant within each interval. The number and locations of CPs must be specified a priori. The PEPH survival model also allows for directly expressing the log-likelihood function of the survival data of the experimental group (Equation \ref{log-like}) in terms of the cumulative hazard function of the external control group $\Lambda_0{(t)}$. To construct a one-dimensional score test, we limit the number of CPs to one for the early and delayed effects, and two for the middle effect. \\

\noindent In the following subsections, we derive score tests corresponding to an early effect (Section \ref{EE}), a middle effect (Section \ref{ME}), a delayed effect (Section \ref{DE}) and a crossing hazards effect (Section \ref{CH}).

\subsection{Early effect} \label{EE}

Let us assume an early treatment effect, representing an initial benefit of the experimental arm up to a specific time $k$, which then diminishes. The hazard and survival functions of an early effect can be modelled as
\begin{equation*}
    \lambda_1(t) =
    \begin{cases}
        \exp(\beta)\lambda_0(t)  &\text{if $t \leq k$}\\
        \lambda_0(t)  &\text{if $t > k$}
    \end{cases}
    \Longleftrightarrow
    \mathrm{S}_1(t) =
    \begin{cases}
        \mathrm{S}_0(t)^{\exp(\beta)}  &\text{if $t \leq k$}\\
        \mathrm{S}_0(k)^{\exp(\beta)-1}\mathrm{S}_0(t)  &\text{if $t \geq k$}
    \end{cases}
\end{equation*}
where $\exp(\beta)$ is the HR on the interval [0, $k$] and $k$ is the CP. The early effect (EE) score statistic is then defined as follows (see Appendix \ref{EE details}):
\begin{equation}
    \mathrm{Z_{EE}} = \frac{\sum\limits_{i: X_i \leq k}(\delta_i - \Lambda_0({X_i})) - \sum\limits_{i: X_i \geq k} \Lambda_0{(k)}}{\sqrt{\sum\limits_{i: X_i \leq k} \Lambda_0({X_i}) + \sum\limits_{i: X_i \geq k}\Lambda_0{(k)}}}
    \label{Score EE}
\end{equation}
The numerator is the difference between two components: the first term includes patients with time-to-event $X_i \leq k$, capturing the contrast between each patient's status and the expected number of events, while the second term includes patients with time-to-event $X_i \geq k$. When $k = \infty$, the score statistic $\mathrm{Z_{EE}}$ reduces to the OSLRT statistic (Equation (\ref{OSLRT})). 

\subsection{Middle effect} \label{ME}

When considering a middle treatment effect with a benefit of the experimental arm compared to the external control group over a specific time interval [$k_1$, $k_2$], the corresponding hazard and survival functions can be modelled as:
\begin{equation*}
    \lambda_1(t) =
    \begin{cases}
        \lambda_0(t)  &\text{if $t \leq k_1$}\\
        \exp(\beta)\lambda_0(t)  &\text{if $k_1 < t \leq k_2$}\\
        \lambda_0(t)  &\text{if $t > k_2$}
    \end{cases}
\end{equation*}
is equivalent to
\begin{equation*}
    \mathrm{S}_1(t) =
    \begin{cases}
        \mathrm{S}_0(t)  &\text{if $t \leq k_1$}\\
        \mathrm{S}_0(k_1)^{1-\exp(\beta)}\mathrm{S}_0(t)^{\exp(\beta)}  &\text{if $k_1 \leq t \leq k_2$}\\
        \mathrm{S}_0(k_1)^{1-\exp(\beta)}\mathrm{S}_0(k_2)^{\exp(\beta)-1}\mathrm{S}_0(t)  &\text{if $t \geq k_2$}
    \end{cases}
\end{equation*}
where $\exp(\beta)$ is the HR on the interval [$k_1$, $k_2$] and $k_1$, $k_2$ are the CPs. The middle effect (ME) score statistic is then defined as follows (see Appendix \ref{ME details}):
 \begin{equation}
        \mathrm{Z_{ME}} = \frac{\sum\limits_{i: X_i \in (k_1; k_2]}(\delta_i - \Lambda_0({X_i})) + \sum\limits_{i: X_i \geq k_1} \Lambda_0({k_1}) - \sum\limits_{i: X_i \geq k_2}\Lambda_0({k_2})}{\sqrt{\sum\limits_{i: X_i \in [k_1; k_2]}\Lambda_0({X_i}) - \sum\limits_{i: X_i \geq k_1}\Lambda_0({k_1}) + \sum\limits_{i: X_i \geq k_2}\Lambda_0({k_2})}}
        \label{Score ME}
    \end{equation}
The numerator consists of three components. The first term consists of the difference between the patient's status and the expected number of events for patients with $X_i$ in $(k_1; k_2]$. The second term represents the expected number of events for patients with $X_i \geq k_1$, and the third term, subtracted from the previous two, represents the expected number of events for patients with $X_i \geq k_2$. When $k_1 = 0$ and $k_2 = k$, the score statistic reduces to that of the score statistic of the early effect (Equation (\ref{Score EE})) and, conversely, when $k_1 = k$ and $k_2 = \infty$, it reduces to the delayed effect score statistic (Equation (\ref{Score DE})). Furthermore, when $k_1 = 0$ and $k_2 = \infty$, the score statistic $\mathrm{Z_{ME}}$ is equivalent to the OSLRT statistic (Equation (\ref{OSLRT})). 

\subsection{Delayed effect} \label{DE}

For a delayed treatment effect representing a benefit of the experimental arm compared to the external control group after a certain time $k$, the corresponding hazard and survival functions can be modelled as:
\begin{equation*}
        \lambda_1(t) =
        \begin{cases}
            \lambda_0(t)  &\text{if $t \leq k$}\\
            \exp(\beta)\lambda_0(t)   &\text{if $t > k$}
        \end{cases}
        \Longleftrightarrow
        \mathrm{S}_1(t) =
        \begin{cases}
            \mathrm{S}_0(t)  &\text{if $t \leq k$}\\
            \mathrm{S}_0(k)^{1-\exp(\beta)}\mathrm{S}_0(t)^{\exp(\beta)}   &\text{if $t \geq k$}
        \end{cases}
    \end{equation*}
where $\exp(\beta)$ is the HR on the interval $[k, \infty)$ and $k$ is the CP. The derived score statistic is then defined as follows (see Appendix \ref{DE details}):
 \begin{equation}
        \mathrm{Z_{DE}} = \frac{\sum\limits_{i: X_i > k}(\delta_i - \Lambda_0({X_i}) + \Lambda_0({k}))}{\sqrt{\sum\limits_{i: X_i > k}(\Lambda_0({X_i}) - \Lambda_0({k}))}}
        \label{Score DE}
\end{equation}
The numerator is computed only for patients with $X_i > k$. It sums the difference between each patient's status and the expected number of events, plus the cumulative hazard evaluated at time $k$. When $k = 0$, the score statistic $\mathrm{Z_{DE}}$ reduces to the OSLRT statistic (Equation (\ref{OSLRT})). 

\subsection{Crossing effect} \label{CH}

When the sign of the treatment effect changes over time, a situation referred to as a crossing effect, we construct a one-dimensional score test using an accelerated hazards model \citep{Moreau1992}:
\begin{equation*}
    \lambda_1(t) = \exp(\beta)\lambda_0(t)\Lambda_0({t})^{\exp(\beta)-1}
    \Leftrightarrow
    \mathrm{S}_1(t) = \exp\left(-\Lambda_0({t})^{\exp(\beta)}\right)
\end{equation*}
The crossing effect considered here corresponds to a scenario where the experimental treatment initially provides lower survival probabilities, followed by a survival benefit at later time points for $\beta <0$. \\
This model also allows the determination of the crossing time of the hazard curves \citep{Zhang2009}:
\begin{equation}
    \mathrm{T_{crossing}} = \Lambda_0^{-1}\left(\exp\left(\frac{-\beta}{\exp(\beta)-1}\right)\right)
    \label{Tcross}
\end{equation}
where $\Lambda_0{(t)^{-1}}$ is the inverse function of the cumulative hazard function of the external control group. The derived score statistic is defined as follows (see Appendix \ref{CH details}):
\begin{equation}
    \mathrm{Z_{CH}} = \frac{\sum\limits_{i = 1}^n \left(\delta_i - \left(\Lambda_0(X_i) - \delta_i\right) \log(\Lambda_0(X_i))\right)}{\sqrt{-\sum\limits_{i = 1}^n \left[\delta_i - \Lambda_0(X_i)\{1+\log(\Lambda_0(X_i))\}\right]\log(\Lambda_0(X_i))}}
    \label{Score CH}
\end{equation}
Unlike the three previous score statistics, the $\mathrm{Z_{CH}}$ statistic uses all available information without any restriction on follow-up time.\\ \\

\noindent All of the developed score statistics share the same asymptotic distribution as the OSLRT statistic, i.e. a standard normal distribution. Therefore, we reject the null hypothesis if $\mathrm{Z_{**}}<-z_{1-\alpha}$ where $z_{1-\alpha}$ is the $100(1-\alpha)$ percentile of the standard normal distribution and $\mathrm{Z_{**}}$ refers to $\mathrm{Z_{EE}, Z_{ME}, Z_{DE}}$ and $\mathrm{Z_{CH}}$.

\section{Alternative tests for non-PH for single-arm trials}

We also consider two alternative approaches commonly used in RCTs when the PH assumption does not hold. The first is a test based on the restricted mean survival time (RMST) \citep{Royston2013, Uno2014}, and the second is a combination test procedure known as max-Combo \citep{Lee1996, Roychoudhury2021}. The max-Combo test belongs to a broader class of versatile testing procedures that combine multiple test statistics and can accommodate all types of non-proportionality without prior knowledge of the PH or non-PH patterns of the treatment effect in RCTs.

\subsection{RMST-based test}

The RMST \citep{Royston2013, Uno2014, Royston2011, Liao2020, Deboissieu2024} is a clinically meaningful alternative to the HR for quantifying the treatment effect in RCTs with time-to-event outcomes. One advantage of the RMST is that it does not rely on the PH assumption. However, RMST requires the definition of a time window $[0,\tau]$. Correct estimation of the RMST can be performed up to time $\tau$, defined as the last follow-up time, under a mild condition on the censoring distribution \citep{Tian2020}. In the case of an SAT assuming no sampling variability in the survival curve estimate of the external control group, the RMST-based test statistic is expressed as:
\begin{equation}
    \mathrm{dRMST_{SA}(\tau) = \frac{\widehat{RMST_{1}}- RMST_{0}}{\sqrt{Var(\widehat{RMST_{1}})}}}
    \label{RMST SA}
\end{equation}
where $\widehat{\mathrm{RMST}_{1}} = \int\limits_{0}^{\tau}\widehat{\mathrm{S}}_1(t)dt$ is the RMST for the experimental arm, with $\widehat{\mathrm{S}}_1(t)$ the Kaplan-Meier (KM) estimator of the survival function. Numerical integration of $\widehat{\mathrm{S}_1}(t)$ is performed using the trapezoidal method. The variance of $\widehat{\mathrm{RMST}_{1}}$ is estimated using the Greenwood plug-in estimator:
\begin{equation*}
    \mathrm{Var(\widehat{RMST_{1}})} = \sum\limits_{X_i} \left(\int_{X_i}^{\tau}\widehat{\mathrm{S}}_{1}(t)dt\right)^2 \frac{d_i}{n_i(n_i - d_i)}
\end{equation*}
where $d_i$ is the number of deaths at $X_i$ and $n_i$ the number of patients still at risk at $X_i$. The RMST of the external control group is defined as $\mathrm{RMST_{0}} = \int\limits_{0}^{\tau}{\mathrm{S}_0(t)}dt$, considered as the true value under the strong assumption that $\mathrm{S}_0(t)$ is the true survival function of the external control group. We also need to select a parametric survival distribution such as exponential, Weibull, log-normal, log-logistic or generalised gamma (tractable form) for $\mathrm{S}_0(t)$ to allow analytical integration to estimate $\mathrm{RMST}_{0}$. We could also estimate $\mathrm{RMST}_{0}$ using the Kaplan-Meier estimate of the historical control group survival curve; however, this introduces variability that must be accounted for in the variance of the test. \\

\noindent Although several approaches exist to define the time-window $[0,\tau$], we follow the method proposed by Huang and Kuan \citep{Huang2018} for small sample sizes:
\begin{equation*}
    \tau = \min(\max(X_{i, 1}), \max(X_{i, 0}))
\end{equation*}
where $X_{i, 1}$ and $X_{i, 0}$ are the observed survival times in the experimental and external control groups, respectively. Note that $\max(X_{i, 0})$ is known prior to the start of the SAT. 

\subsection{Max-Combo test} \label{max-combo}

In RCTs, the combination test procedure known as max-Combo \citep{Karrison2016} is defined as the maximum of different tests, for example, multiple Fleming-Harrington $\mathrm{FH}(\rho,\gamma)$ \citep{Harrington1982} weighted log-rank tests, allowing for consideration of different PH and non-PH treatment effects. We develop a similar test to accommodate all types of non-proportionality and/or different change-point values. Since all proposed statistics are one-sided and reject the null hypothesis for sufficiently negative values, more negative values of the test statistic indicate stronger evidence against the null hypothesis. Therefore, the combination statistic is defined as the maximum of the negated statistics, which is equivalent to taking the minimum of the original statistics. For instance, the following max-Combo statistic combines the mOSLRT statistic (Equation (\ref{mOSLRT})) with score statistics for early (Equation (\ref{Score EE})) and delayed effects (Equation (\ref{Score DE})) at two distinct CPs, which may differ for the early and delayed score statistics:
\begin{equation}
    \mathrm{max\text{-}Combo} = \mathrm{\min(mOSLRT, Z_{EE_{k_1}}, Z_{EE_{k_2}}, Z_{DE_{k_1'}}, Z_{DE_{k_2'}})} 
    \label{Test_max_combo}
\end{equation}

\noindent To address multiplicity, we consider two approaches: the first applies the Hochberg correction \citep{Westfall1999} to individual p-values, while the second computes the exact p-value through multiple integrations, as the combined test statistic asymptotically follows a multivariate normal distribution. As the covariance matrix is challenging to calculate, we use the following relationship:
\begin{equation*}
    \mathrm{Cov(Z_i, Z_j)} = \rho_{ij}\sqrt{\mathrm{Var(Z_i)Var(Z_j)}}
\end{equation*}
where $\rho_{ij}$ denotes the correlation between two test statistics noted $\mathrm{Z}_i$ and $\mathrm{Z}_j$. Since the mOSLRT and the developed score statistics follow the standard normal distribution asymptotically, their variances are equal to 1, so the covariance matrix coincides with the correlation matrix. This correlation matrix can be defined using the ratio of the expected number of events, as considered by Abbas et al \citep{Abbas2022}. Consequently, we obtain the following variance-covariance matrix for the max-Combo test statistic:
\begin{center}
    $\boldsymbol{\Sigma} =  
    \begin{pmatrix}
        1 \\
        \sqrt{\frac{\mathrm{E_{EE, k = k_1}}}{\mathrm{E_{mOSLRT}}}} & 1\\
        \sqrt{\frac{\mathrm{E_{EE, k = k_2}}}{\mathrm{E_{mOSLRT}}}} & \sqrt{\frac{\mathrm{E_{EE, k = k_1}}}{\mathrm{E_{EE, k = k_2}}}} & 1 \\
        \sqrt{\frac{\mathrm{E_{DE, k = k_1'}}}{\mathrm{E_{mOSLRT}}}} & 0 & 0 & 1 \\
        \sqrt{\frac{\mathrm{E_{DE, k = k_2'}}}{\mathrm{E_{mOSLRT}}}} & 0 & 0 & \sqrt{\frac{\mathrm{E_{DE, k = k_2'}}}{\mathrm{E_{DE, k = k_1'}}}} & 1
    \end{pmatrix}$
\end{center}
where $\mathrm{E_{EE, k = t_k}} = \sum\limits_{i = 1}^{n} \Lambda_0{(X_i)I(X_i \leq t_k)} + \Lambda_0{(t_k)I(X_i \geq t_k)}$, $\mathrm{E_{DE, k = t_k}} = \sum\limits_{i = 1}^{n} \left[\Lambda_0(X_i) - \Lambda_0(t_k)\right]\mathrm{I}(X_i > t_k)$ and $\mathrm{E_{mOSLRT}} = \sum\limits_{i = 1}^{n} \Lambda_0(X_i)$. \\

\noindent Hence, the null hypothesis is rejected when max-Combo $< -c$, where $c$ corresponds to the critical value for a nominal level of 5\%. \\

\noindent Here, we choose to construct a max-Combo test that includes statistical tests for early and delayed effects at two different CPs, as these statistics can capture non-PH patterns commonly observed in practice. In fact, early and delayed treatment effects are the most frequently observed forms of non-PH hazards in oncology clinical trials, whereas middle effects and crossing hazards are less common. However, the proposed max-Combo statistic can be redefined by combining different score statistics, including the $\mathrm{Z_{CH}}$ statistic, or by considering alternative CPs.

\section{Simulation study} \label{Simu}

\subsection{Parameters} \label{Param}

\begin{figure*}[htb]
\centerline{\includegraphics[width=\textwidth]{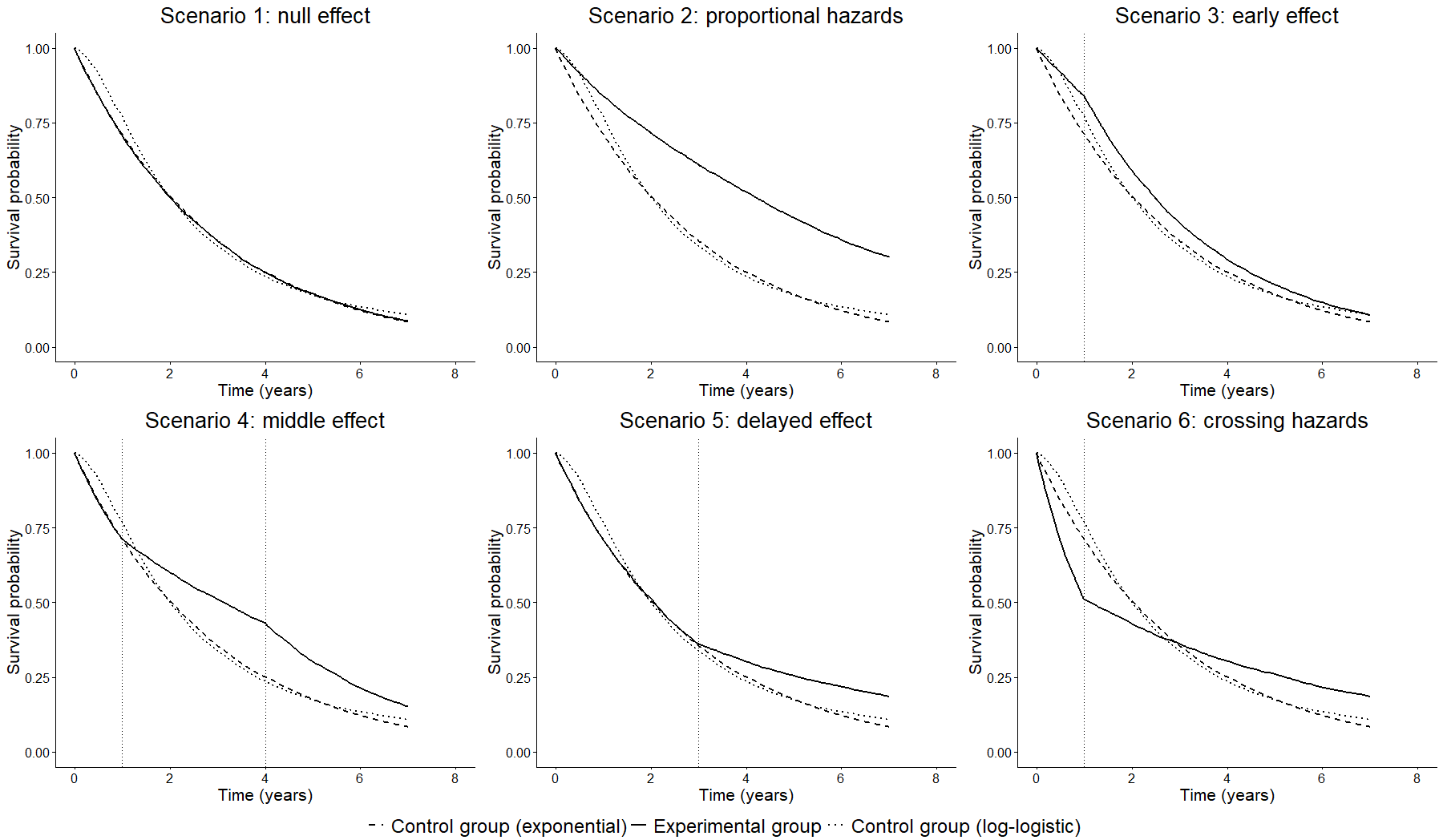}}
\caption{True survival curves under different scenarios of single-arm trials: scenario 1corresponds to the null effect, scenario 2 to a PH treatment effect, scenarios 3-6 to early, middle, delayed and crossing treatment effects, respectively. The dashed and dotted curves represent the survival curves for the external control group, simulated using exponential (dashed) and log-logistic (dotted) models. The solid curve represents the survival curve generated from a piecewise exponential model. The vertical dotted lines represent the change-points used to generate the models for scenarios 3, 4, 5 and 6 ($k = 1$, $k_1 = 1$, $k_2 = 4$, $k = 3$ and $k = 1$, respectively)
\label{Survival}}
\end{figure*}

We conducted a simulation study to evaluate the operating characteristics of an SAT design with a TTE endpoint using one of the developed tests under various PH and non-PH scenarios (Figure \ref{Survival}), including the OSLRT and mOSLRT as benchmark methods. Six scenarios (null effect, PH, early effect, middle effect, delayed effect and crossing hazards) representing a range of practical situations were investigated. Survival times for the experimental group were simulated using an exponential model for PH scenarios (scenarios 1 and 2) and a PEPH model for non-PH scenarios: early effect (scenario 3), middle effect (scenario 4), delayed effect (scenario 5) and crossing hazards (scenario 6). Patients in the experimental group were recruited uniformly over the first 3 years, with a 4-year follow-up period, consistent with typical paediatric oncology SATs. CPs were specified a priori based on the trial duration: $k = 1$ year for scenario 3, $k_1 = 1$ and $k_2 = 4$ for scenario 4, $k = 3$ for scenario 5 and $k = 1$ for scenario 6. For scenario 6, this CP corresponds to the crossing time of the hazard curves, as defined in Equation (\ref{Tcross}). It should be noted that the crossing time of the hazard functions differs from that of the survival curves. The external or historical control group was generated using an exponential distribution with a median survival time of 2 years ($\lambda$ = 0.35). For this simulation study, the max-Combo test statistic (see Equation (\ref{Test_max_combo})) was defined with the following CPs: $k_{1} = 1$ and $k_{2} = 3$ for $\mathrm{Z_{EE}}$ and $k_{1}' = 3$ and $k_{2}' = 5$ for $\mathrm{Z_{DE}}$. \\

\noindent The performance (type I error and power) was estimated using 10 000 Monte Carlo replications. The Monte Carlo standard error of the estimated probabilities (type I error or power) was computed as $\sqrt{\frac{\hat{p}(1-\hat{p})}{N}}$, where $\hat{p}$ denotes the empirical type I error or power and $N$ the number of replications. With $N = 10 000$, this standard error is at most 0.005. The tests were performed at a one-sided significance level of 0.05, as commonly used in phase II oncology trials. The simulation parameters for each scenario were (i) the sample size of the experimental group $n = \{20, 30, 50, 60, 80, 100, 150, 200\}$, (ii) the exponential censoring rate $\{0, 5, 15, 25, 35\%\}$ and (iii) the relative treatment effect $\mathrm{HR} = \{0.5, 0.7, 0.8, 1\}$. \\

\noindent Since the developed score tests require specifying the number and values of CPs a priori, we performed a sensitivity analysis to evaluate the impact on performance when the value of the CP deviates from its true value. The deviation corresponds to adding or subtracting 3 or 6 months to the true value. We also investigated the performance of the tests when accounting for variability in the survival curve estimate of the external control group. \\

\noindent All simulations were performed using software \texttt{R~4.0.3}, and the scripts are available on GitHub Oncostat \url{https://github.com/Oncostat/oslrt_non_PH}. 

\subsection{Results} \label{Results}

\begin{figure*}[htb]
\centerline{\includegraphics[width=\textwidth]{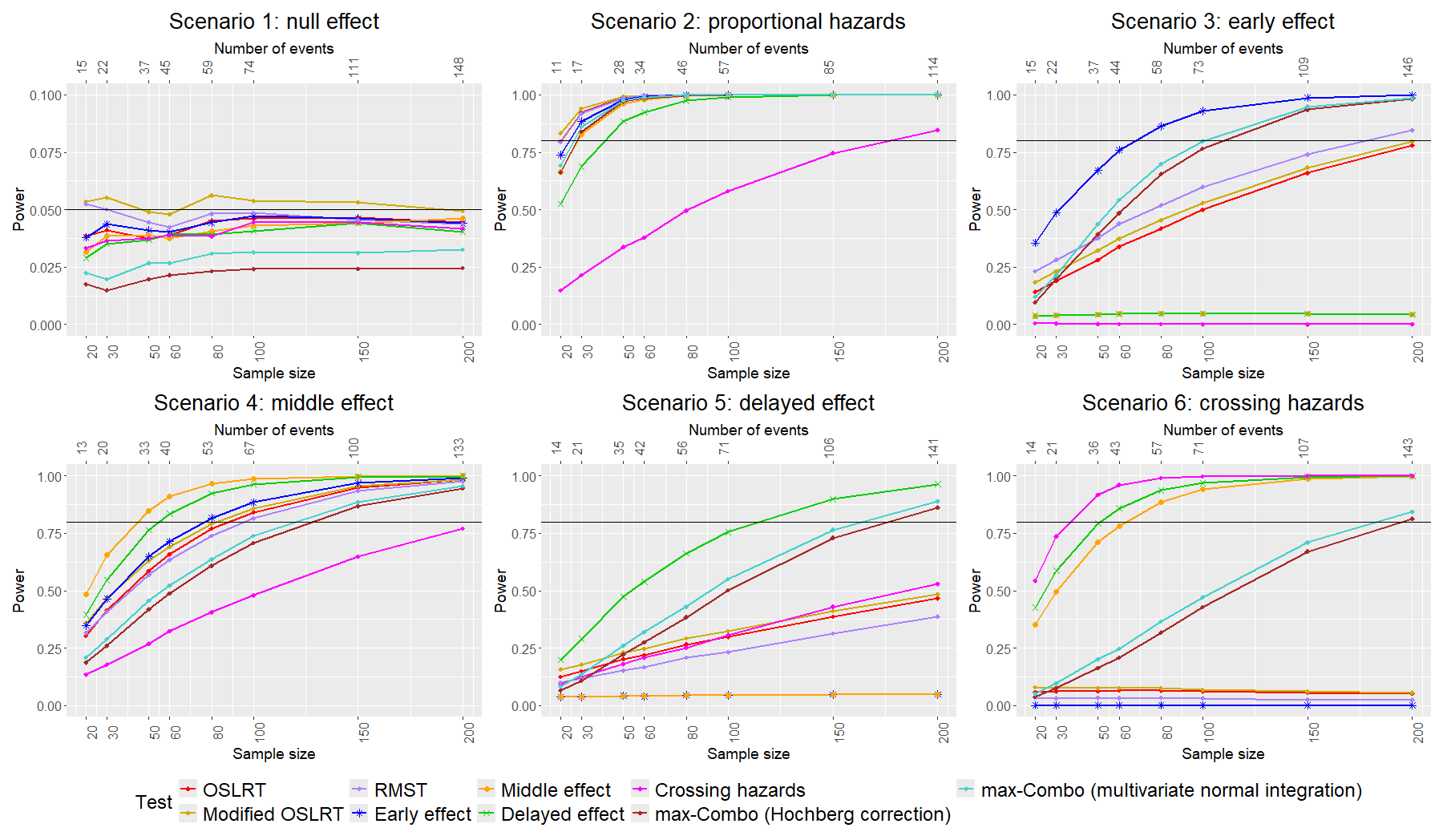}}
\caption{Type I error (scenario 1) and power (scenarios 2-6) of the OSLRT, mOSLRT, developed score tests for early ($\mathrm{Z_{EE}}$ with $k = 4$ for scenarios 1-2, $k = 1$ for scenarios 3 and 6, $k = 4$ for scenario 4 and $k = 3$ for scenario 5), middle ($\mathrm{Z_{ME}}$ with $k_1 = 1$ and $k_2 = 6$ for scenarios 1-2, $k_1 = 1$ and $k_2 = 7$ for scenario 3, $k_1 = 1$ and $k_2 = 4$ for scenarios 4 and 6, $k_1 = 0$ and $k_2 = 3$ for scenario 5) and delayed effects ($\mathrm{Z_{DE}}$ with $k = 2$ for scenarios 1-2, $k = 1$ for scenarios 3-4 and 6 and $k = 3$ for scenario 5), crossing hazard ($\mathrm{Z_{CH}}$) together with the RMST-based test ($\tau = 7$) and max-Combo test (Hochberg and multivariate normal integration), under 15\% censoring and a true HR of 0.5. Black horizontal lines indicate the nominal 5\% type I error (scenario 1) and the80\% power level (scenarios 2-6).
\label{Power}}
\end{figure*}

The main results of the simulation study are presented in Figure \ref{Power} for a true HR of 0.5. Within a scenario, the type I error (scenario 1) and power (scenarios 2-6) (y-axis) are presented as a function of the sample size of the experimental group (bottom x-axis), assuming a censoring rate of 15\%. Results for the other censoring rates are provided in the Appendix (Figure \ref{Power_compl}). The corresponding number of events is reported on the top x-axis. \\
\\
\noindent Under the null effect (scenario 1), the OSLRT exhibits an empirical type I error rate below the nominal 5\% level, whereas the mOSLRT performs as expected, with a rate close to 5\%. The developed score tests show similar behaviour to the OSLRT, with type I error rates converging to approximately 4.5\%. The RMST-based test shows a type I error close to 5\% for small sample sizes but becomes more conservative as the sample size increases, approaching the OSLRT. The max-Combo test is conservative, particularly when applying the Hochberg correction. Similar patterns are observed regardless of the censoring rate. \\

\noindent In scenarios with a treatment effect, the test matching the data-generating mechanism achieves the highest power. Other tests show more variable performance, whereas the max-Combo test remains relatively robust across scenarios. \\

\noindent Under the PH scenario (scenario 2), the OSLRT and mOSLRT achieve the highest power and gradually lose power as the PH assumption is violated (scenarios 3 to 6). \\

\noindent In non-PH scenarios, score tests targeting specific alternatives perform best in their corresponding settings: the $\mathrm{Z_{EE}}$ test for early effect (scenario 3), the $\mathrm{Z_{ME}}$ test for middle effect (scenario 4) and the $\mathrm{Z_{DE}}$ test for delayed effect (scenario 5). These tests remain competitive in neighbouring scenarios (e.g. the $\mathrm{Z_{EE}}$ test in scenario 4) but show reduced power when their underlying assumptions are strongly violated. The $\mathrm{Z_{CH}}$ test achieves the highest power under crossing hazards (scenario 6). Notably, the $\mathrm{Z_{ME}}$ test also performs well under the crossing hazards scenario (scenario 6) due to its focus on a relevant time interval. In addition, in the middle effect scenario (scenario 4), the $\mathrm{Z_{DE}}$ test surprisingly outperforms its performance obtained in the delayed effect setting, which can be explained by a higher number of patients within the relevant time interval. \\

\noindent The RMST-based test, with $\tau=7$ years, shows power comparable to the OSLRT and mOSLRT under the PH (scenario 2) and middle effect scenarios (scenario 4), slightly higher power in the early effect scenario (scenario 3), and lower power in the delayed effect scenario (scenario 5). As expected, its power is close to zero in the crossing hazards scenario (scenario 6), as the area under the survival curves before and after the crossing time cancels out. \\

\noindent The performance of the max-Combo test, regardless of the correction for multiple testing, reflects its construction (Equation \ref{Test_max_combo}). It shows reduced power compared to the optimal test in the PH (scenario 2), early effect (scenario 3) and delayed effect (scenario 5) scenarios due to multiplicity correction. However, it provides a robust alternative and outperforms the OSLRT and RMST-based test in the early effect (scenario 3), delayed effect (scenario 5) and crossing hazards (scenarios 6) scenarios for sample size $n \geq 50$. \\
Sensitivity analyses were performed to investigate the impact of modifying the composition of the max-Combo test defined in Equation (\ref{Test_max_combo}). Specifically, we evaluated reduced max-Combo tests by removing component statistics ($\mathrm{Z_{DE}}$ for scenarios 1, 2 and 3 and $\mathrm{Z_{EE}}$ for scenarios 1, 4 and 5), as well as extended max-Combo tests incorporating additional statistics ($\mathrm{Z_{ME}}$, $\mathrm{Z_{CH}}$, $\mathrm{Z_{EE}}$ with a third CP $k_3 = 2$ and $\mathrm{Z_{DE}}$ with a third CP $k_3' = 1$). The results when some components are removed are presented in Figure \ref{DE_remove} (removing $\mathrm{Z_{DE}}$) and Figure \ref{EE_remove} (removing $\mathrm{Z_{EE}}$). The type I error converges to that of the mOSLRT, and the power is improved up to 10\% better when two statistics are left out. The results of the analyses adding components to the max-Combo test across the six scenarios are presented in Figure \ref{ME_CH_add} ($\mathrm{Z_{CH}}$ and $\mathrm{Z_{ME}}$) and Figure \ref{EE_DE_add} ($\mathrm{Z_{EE}}$ and $\mathrm{Z_{DE}}$). The type I error is similar to that of the original version of the max-Combo (Equation (\ref{Test_max_combo})) and the power decreases for all alternative max-Combo procedures in scenarios 2 (proportional hazards), 3 (early effect) and 5 (delayed effect), with a maximum decrease of 10\% observed when two statistics are added. In scenarios 4 (middle effect) and 6 (crossing hazards), the power is improved for all alternative procedures except when only $\mathrm{Z_{EE}}$ with $k_3 = 2$ is added, for which the power decreases.\\

\noindent The same pattern of results is observed regardless of the censoring rate (see Figure \ref{Power_compl} in the Appendix) and the magnitude of treatment effect (HR = 0.7 or 0.8), with power decreasing as the hazard ratio approaches 1 (see Figure \ref{Power_HR0.7} for HR = 0.7 and Figure \ref{Power_HR0.8} for HR = 0.8 in the Appendix). \\ \\

\begin{figure*}[htb]
\centerline{\includegraphics[width=\textwidth]{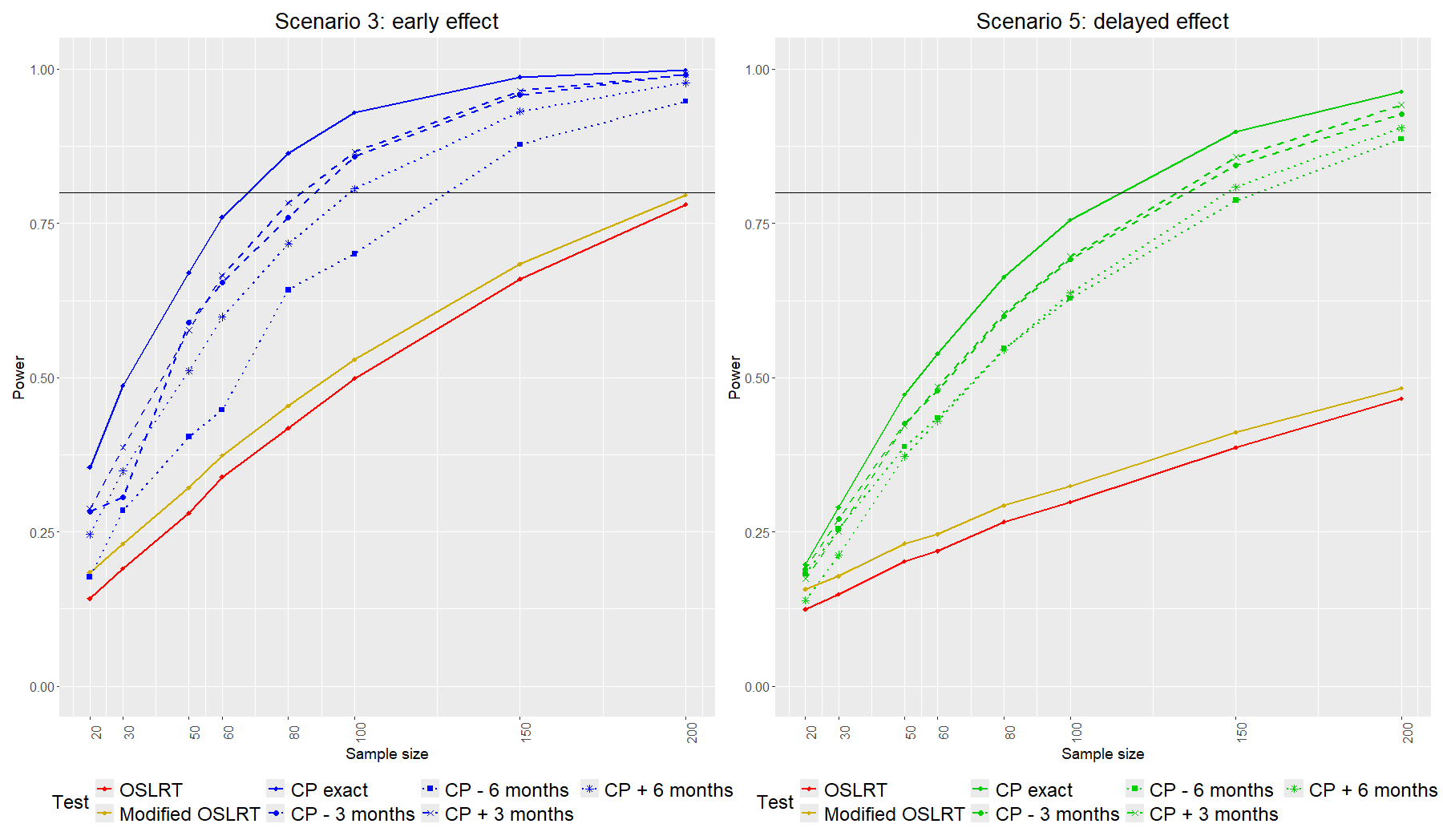}}
\caption{Impact of change-point misspecification on the power of the early and delayed effect score tests under 15\% censoring and a true HR of 0.5. Misspecification is assessed by considering four new CPs: $k_1 = k-3$, $k_2 = k+3$, $k_3 = k-6$ and $k_4 = k+6$ months.
\label{CPs misspe}}
\end{figure*}

\noindent As the $\mathrm{Z_{EE}}$ and $\mathrm{Z_{DE}}$ tests rely on the strong assumption that the pattern of the treatment effect and the number and location of CPs are known a priori, we further evaluate the impact of misspecifying the CP. We compute the power of these score tests when deviations of $\pm$ 3 and $\pm$ 6 months from the true CP value are introduced. The results of this sensitivity analysis are presented in Figure \ref{CPs misspe} for an early (left) and a delayed effect (right). The solid line represents power for the true CP, and the power of the OSLRT and mOSLRT are also reported.
\\
Regardless of the deviation from the true CP, for a censoring rate of 15\%, the power decreases but remains always higher than that of the OSLRT and mOSLRT. This result is consistent across censoring rates (see Figure \ref{CPs misspe comp} in Appendix). So, minor deviations from the true CP have a little impact on the power that remains higher than the OSLRT and mOSLRT.

\section{Variability in the survival curve of the external control group}

We initially assumed that the survival curve of the external control group follows an exponential distribution without sampling variability, but these two assumptions may be questioned. In practice, the survival of the external control group is often estimated from a limited number of patients. We therefore evaluated the impact of misspecification of the survival curve of the external control group on the performance of the different tests as follows. First, we modelled the parameter $\lambda$ of the exponential distribution (used by default in Section \ref{Simu}) as a random variable (Section \ref{Var-expo}). Second, we accounted for sampling variability in the score tests using a correction (Section \ref{Sampling}). Third, we fitted different parametric survival distributions (Weibull, log-normal, etc) to estimate $\Lambda_0(t)$ (Section \ref{Var-distrib}). 

\subsection{Variability in the exponential parameter} \label{Var-expo}

\begin{figure*}[htb]
\centerline{\includegraphics[width=\textwidth]{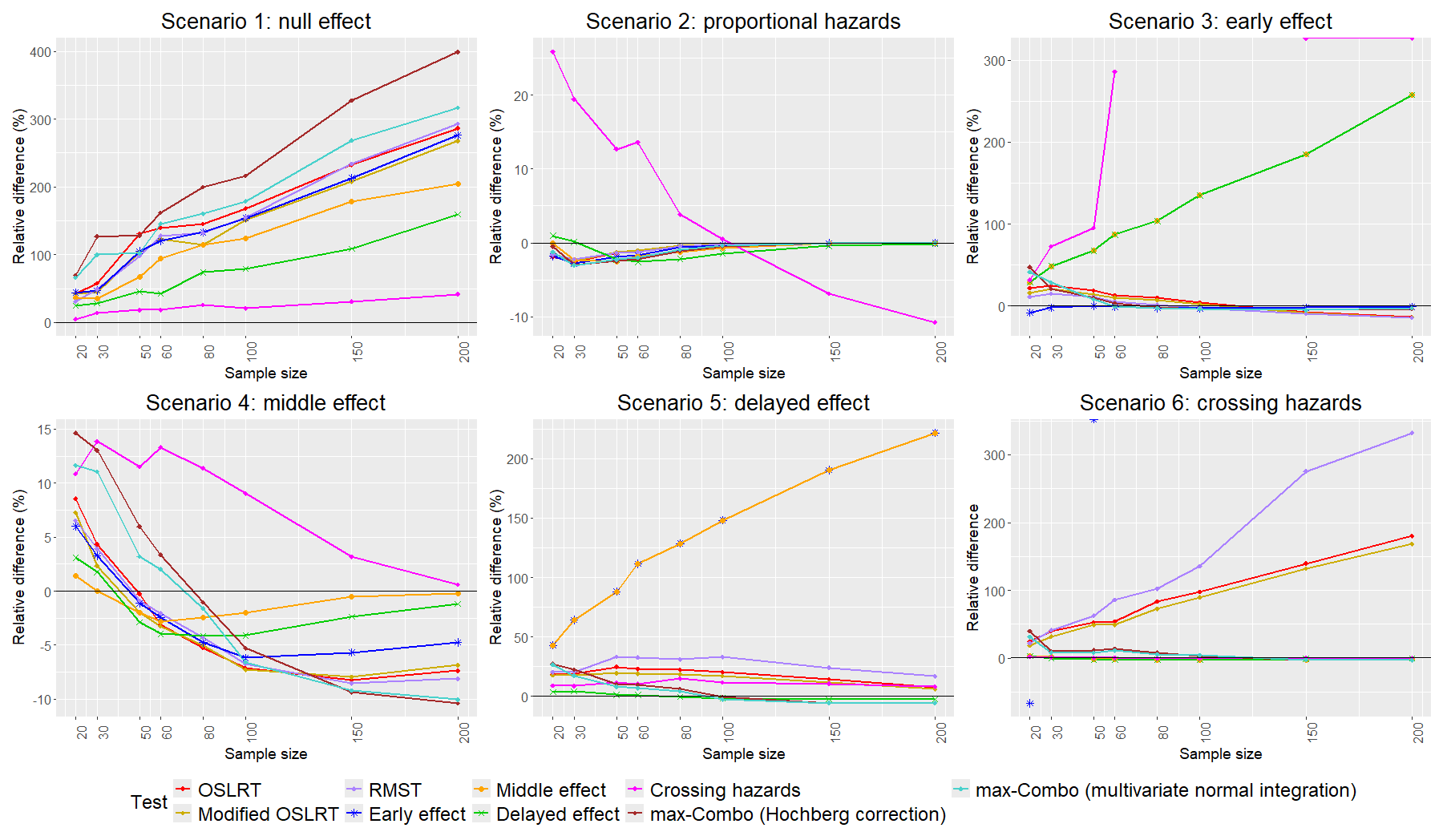}}
\caption{Relative difference in type I error (scenario 1) and power (scenarios 2-6) between analyses incorporating uncertainty in exponential parameter of the external control group and those based on the true parameter. Results are shown for the OSLRT, mOSLRT, developed score tests for an early ($\mathrm{Z_{EE}}$ with $k = 4$ for scenarios 1-2, $k = 1$ for scenarios 3 and 6, $k = 4$ for scenario 4 and $k = 3$ for scenario 5), middle ($\mathrm{Z_{ME}}$ with $k_1 = 1$ and $k_2 = 6$ for scenarios 1-2, $k_1 = 1$ and $k_2 = 7$ for scenario 3, $k_1 = 1$ and $k_2 = 4$ for scenarios 4 and 6, $k_1 = 0$ and $ k_2 = 3$ for scenario 5) and delayed effect ($\mathrm{Z_{DE}}$ with $k = 2$ for scenarios 1-2, $k = 1$ for scenarios 3-4 and 6 and $k = 3$ for scenario 5), crossing hazard ($\mathrm{Z_{CH}}$), RMST-based test ($\tau$ = 7) and max-Combo test (Hochberg and multivariate normal integration) under 15\% of censoring and a true HR of 0.5.
\label{Diff_sensibility}}
\end{figure*}

We reproduced the simulation study with an HR of 0.5, but now introduced variability in the survival curve of the external control group using the algorithm \ref{Variability} (see the details in Figures \ref{Med_miss} and \ref{Lambda_miss}, resulting in scenarios in Figure \ref{Survival_misspe}). 
\begin{algorithm}
\caption{Algorithm for generating variability on the exponential parameter of the external control group} \label{Variability}
\begin{algorithmic}
  \State The external control group theoretically follows an exponential distribution with a median survival of 2 years ($\lambda_{th} = 0.35$) 
  \For{each replication $i$}
  \For{External control group}
  \State Generate median survival: $\mathrm{med}_i \sim Gamma(80,40)$
  \State Calculate exponential parameter: $\lambda_i = \frac{\log(2)}{\mathrm{med}_i}$
  \State Compute survival function: $S_{0, i}(t) = \exp(-\lambda_i t)$ and cumulative hazard function: $\Lambda_{0, i}(t) = \lambda_i t$
  \EndFor
  \EndFor
\end{algorithmic}
\end{algorithm}\\
Figure \ref{Diff_sensibility} presents the results in a similar form to Figure \ref{Power} except that we now report the relative difference (in percentage) in performance when uncertainty is introduced in the parameter $\lambda$ under 15\% of censoring (See Appendix, Figure \ref{Power_sensitivity}, for the crude performance of type I error and power under parameter uncertainty). The scale of the y-axis is intentionally allowed to differ across scenarios to facilitate comparison. Some changes may not be displayed when the denominator is zero. A positive (negative) relative difference indicates an overestimation (underestimation) of the type I error or power. \\

\noindent The main results for each optimal test in a given scenario (see Figure \ref{Diff_sensibility_comp} in Appendix for all censoring rates), as previously identified in Figure \ref{Power}, are: (i) a substantial overestimation of the type I error, increasing with sample size (scenario 1); (ii) a decrease in the relative difference for the OSLRT and mOSLRT from 6\% for $n = 20$ to approximately -10\%  when $n > 50$ (scenario 2); (iii) no meaningful impact on the tests in the early effect scenario regardless of sample size, and no impact on the max-Combo test for $n > 60$  (scenario 3), but a marked increase for $n < 60$ (greater than 15\%); (iv) changes less than 10\% in absolute value for all tests (scenario 4); (v) no impact on the delayed effect test (scenario 5) regardless of sample size and censoring rate; and (vi) no impact on the tests for crossing hazards, middle and delayed effects (scenario 6).

\subsection{Sampling variability of the external control group} \label{Sampling}

As the assumption of no variability in the external control group \citep{Wu2021} leads to inflation of the type I error \citep{Danzer2022_ref, Danzer2022_var}, some authors \citep{Danzer2022_ref, Danzer2022_var} have proposed a correction for the OSLRT to account for this variability (see Appendix \ref{Sampling_details}). This correction requires individual patient data from the external control group; otherwise, an approximation \citep{Danzer2022_ref, Feld2024} can be used based on the ratio $\pi = \frac{n_{\mathrm{exp}}}{n_{\mathrm{control}}}$ where $n_{\mathrm{exp}}$ and $n_{\mathrm{control}}$ denote the number of patients in the experimental and external control groups, respectively. We therefore applied this approximation for the developed score and max-Combo tests. The corresponding corrected test statistics are defined as follows:
\begin{equation*}
    \mathrm{Z_{corrected} = Z_{non\_corrected}} \frac{1}{\sqrt{1+\pi}}
\end{equation*} 

\noindent The correction was not applied to the RMST-based test because its variance is computed differently from that of the OSLRT. A simulation study was conducted with $\pi = \{1, 0.8, 0.6, 0.5\}$, corresponding to $n_{\mathrm{control}} = \{n_{\mathrm{exp}}$, ..., $2 n_{\mathrm{exp}}\}$, while keeping all other parameters as in Section \ref{Simu}. These values of $\pi$ reflect those encountered in practice, particularly in the real data examples presented in Section \ref{Examples}. \\

\begin{figure*}[htb]
\centerline{\includegraphics[width=\textwidth]{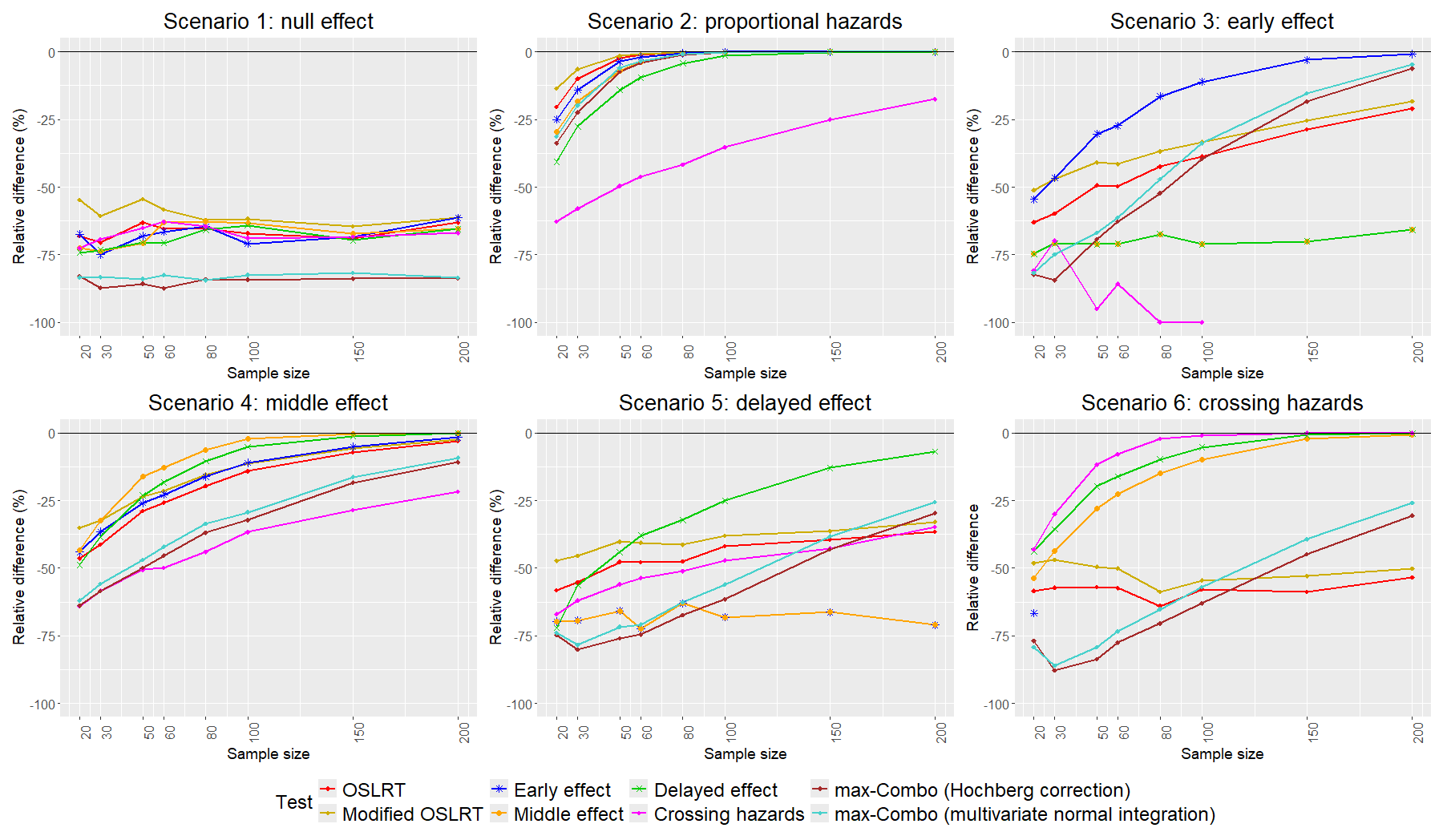}}
\caption{Relative difference (in \%) in type I error (scenario 1) and power (scenarios 2-6) between analyses accounting for sampling variability in the external control group and those assuming no sampling variability. Results are shown for the OSLRT, mOSLRT, developed score tests for an early ($\mathrm{Z_{EE}}$ with $k = 4$ for scenarios 1-2, $k = 1$ for scenarios 3 and 6, $k = 4$ for scenario 4 and $k = 3$ for scenario 5), middle ($\mathrm{Z_{ME}}$ with $k_1 = 1$ and $k_2 = 6$ for scenarios 1-2, $k_1 = 1$ and $k_2 = 7$ for scenario 3, $k_1 = 1$ and $k_2 = 4$ for scenarios 4 and 6, $k_1 = 0$ and $k_2 = 3$ for scenario 5) and delayed effect ($\mathrm{Z_{DE}}$ with $k = 2$ for scenarios 1-2, $k = 1$ for scenarios 3-4 and 6 and $k = 3$ for scenario 5), crossing hazard ($\mathrm{Z_{CH}}$) and max-Combo test (Hochberg and multivariate normal integration) under 15\% of censoring, a true HR of 0.5 and a ratio $\pi = 0.6$.
\label{Diff_variability}}
\end{figure*}

\noindent Figure \ref{Diff_variability} shows the relative difference (in \%) in performance for $\pi = 0.6$, corresponding to $n_{\mathrm{control}} = \{33, 50, 83, 100, 133, 167, 250, 500\}$ patients in the external control group under 15\% of censoring (see Figure \ref{Power_variability} for empirical type I error and power, and Figure \ref{Diff_variability_comp} in Appendix for all censoring rates). \\

\noindent The main results for the optimal tests in each scenario are: (i) a decrease in type I error, consistent with the findings of Danzer et al \citep{Danzer2022_ref, Danzer2022_var}, and (ii) a decrease in power for all tests. For a fixed number of patients in the experimental group (e.g. $n_{\mathrm{exp}} = 100$), as the ratio $\pi$ diminishes from 1 to 0.5, the number of patients in the control group increases (from $n_{\mathrm{control}}  = 100$ to $n_{\mathrm{control}}  = 200$) and the correction factor approaches 1, as the variability of the external control group decreases. Consequently, the corrected test converges to the non-corrected test, reducing the relative difference. Results for $\pi = 1$ are presented in the Appendix (Figures \ref{Power_variability_pi1} and \ref{Diff_variability_pi1}) and results for the other values of $\pi$ are similar (data not shown). 

\subsection{Model misspecification of the survival distribution of the external control curve} \label{Var-distrib}

So far, we have assumed an exponential distribution for the external control group, as is common in practice (see Section \ref{Simu}), and we now evaluate the impact of model misspecification. Using a specific survival distribution for the external control group implies that the expected number of events $\mathrm{E}$ is calculated according to the cumulative hazard function of this distribution. \\

\noindent To examine an "extreme" situation, we chose a log-logistic distribution, whose hazard function (Figure \ref{Hazards} in Appendix) differs substantially from that of the exponential distribution. We re-analysed the simulated data (Section \ref{Results}) using a log-logistic distribution for the external control group instead of an exponential distribution (Figure \ref{Survival}, dashed black line). The log-logistic parameters, shape=1.7 and scale=2, were selected to reasonably approximate the true exponential distribution (Figure \ref{Hazards_cum} in Appendix). The resulting hazard function of the log-logistic distribution (non-monotone) entails that the hazard functions of the experimental and control groups may intersect. \\

\begin{figure*}[htb]
\centerline{\includegraphics[width=\textwidth]{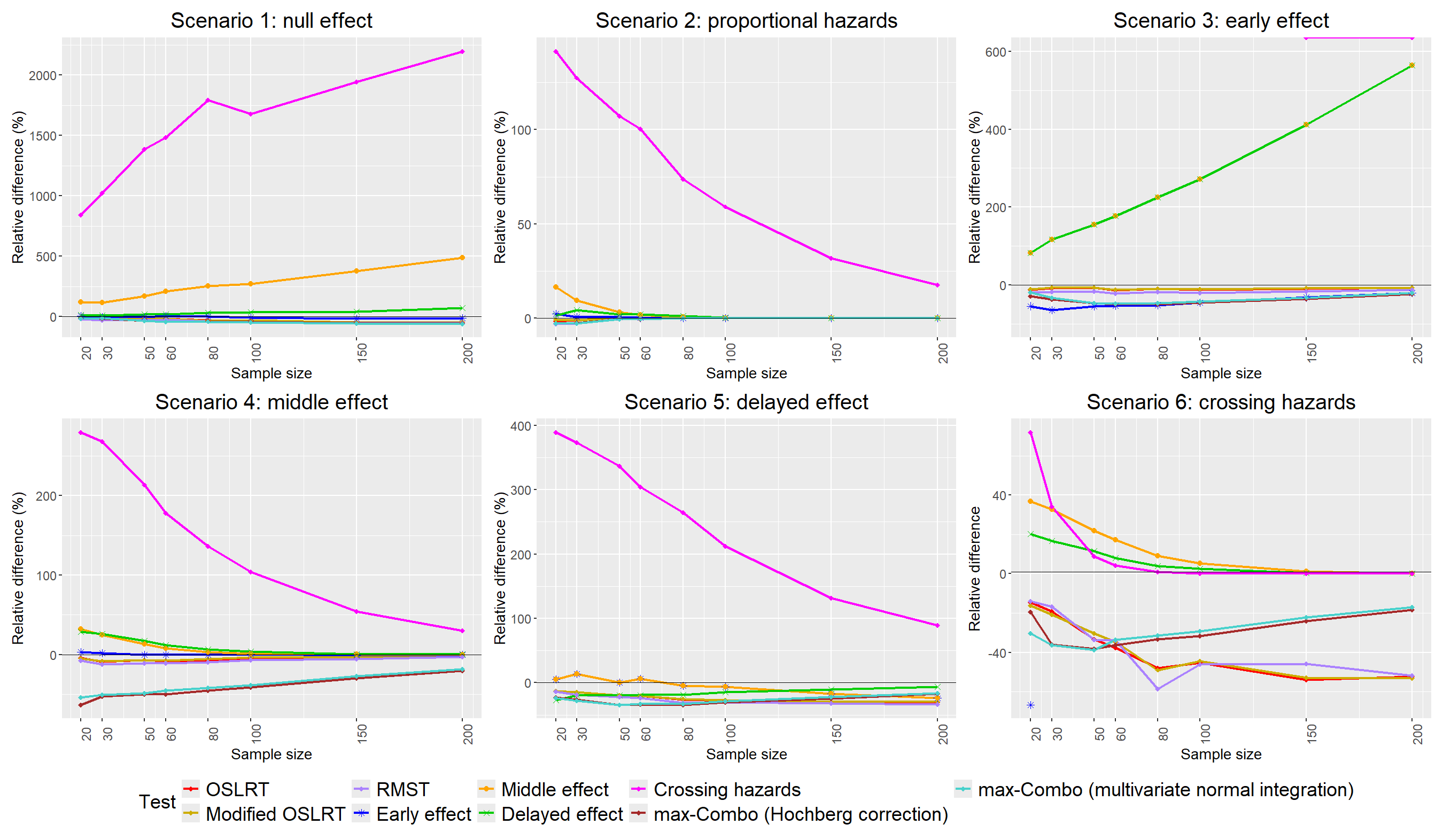}}
\caption{Relative difference (in \%) in power between analyses assuming a misspecified log-logistic distribution and those assuming a correctly specified exponential distribution for the external control group. Results are shown for the OSLRT, mOSLRT, developed score tests for an ($\mathrm{Z_{EE}}$ with $k = 4$ for scenarios 1-2, $k = 1$ for scenarios 3 and 6, $k = 4$ for scenario 4 and $k = 3$ for scenario 5), middle ($\mathrm{Z_{ME}}$ with $k_1 = 1$ and $k_2 = 6$ for scenarios 1-2, $k_1 = 1$ and $k_2 = 7$ for scenario 3, $k_1 = 1$ and $k_2 = 4$ for scenarios 4 and 6, $k_1 = 0$ and $k_2 = 3$ for scenario 5) and delayed effect ($\mathrm{Z_{DE}}$ with $k = 2$ for scenarios 1-2, $k = 1$ for scenarios 3-4 and 6 and $k = 3$ for scenario 5), crossing hazard ($\mathrm{Z_{CH}}$), RMST-based test ($\tau$ = 7) and max-Combo test (Hochberg and multivariate normal integration) under 15\% of censoring and a true HR of 0.5.
\label{Diff_misspe}}
\end{figure*}

\noindent Figure \ref{Diff_misspe} presents the relative difference in performance (in \%) when using a log-logistic distribution compared to an exponential distribution as reference (see Appendix, Figure \ref{Power_misspe} for raw type I error and power). \\

\noindent The performance of the optimal test in each scenario may be affected. This is particularly evident for the early effect test, which shows a substantial decrease in power. Even when not the optimal test, a loss of power is also observed for the max-Combo test in scenarios 3 and 5. Under scenario 2, the power of the $\mathrm{Z_{ME}}$ (for $n < 150$) and $\mathrm{Z_{CH}}$ tests increases, particularly for higher censoring rates for the latter (Figure \ref{Diff_misspe_comp} in Appendix for all results). In scenario 3, the early effect and max-Combo tests never reach 80\% power, decreasing from 90\% without misspecification to 50\% with misspecification when $n = 80$ for the $\mathrm{Z_{EE}}$ test. In scenario 4, the middle effect and delayed effect tests gain power for $n < 100$ (from 85\% to 96\% for the $\mathrm{Z_{ME}}$ test and from 76\% to 90\% for the $\mathrm{Z_{DE}}$ test for $n = 50$), while the crossing hazards test shows a substantial increase (from 40\% to 96\% for $n = 80$), resulting in a power curve similar to the middle and delayed effect tests. In scenario 5, the delayed effect and max-Combo tests lose 20\% and 25\% of power, respectively, while the crossing hazards test gains considerably. For scenario 6, the optimal test increases in power for $n < 80$ (from 92\% to 100\% for $n = 50$), whereas the middle and delayed effects tests show gains for $n < 100$ that decrease towards zero as $n$ increases.

\section{Real data examples} \label{Examples}

We illustrate the developed tests using three clinical trial examples. The first two are SATs with a null and an early treatment effect, respectively, while the third, exhibiting a delayed effect, is derived from a subgroup of patients in a randomised clinical trial. Typically, reporting of SAT results is limited to the KM survival curve for the experimental arm, the median survival time in the external control group, and the OSLRT assuming an exponential distribution for the external control group. Individual patient data (IPD) for the experimental and external control groups were reconstructed from the published survival curves using the \texttt{R} package \texttt{IPDfromKM} \citep{Liu2020}, following the methodology of Guyot et al \citep{Guyot2012}. The suitability of the exponential distribution for modelling the cumulative hazard of the external control group was assessed by comparing different parametric distributions (exponential, Weibull, log-logistic, log-normal, gamma and generalised gamma) using the Akaike Information Criteria (AIC). We used the \texttt{R} package \texttt{flexsurv}  \citep{Jackson2016} to implement these different distributions. In the three examples, the tests are reported for the exponential and Weibull distributions (standard models) as well as for the distribution providing the best fit to the external control group (lowest AIC). For interpretability, estimates of the survival curve $\mathrm{S_0}(t)$ and the cumulative hazard function $\Lambda_0(t)$ are also presented for the fitted parametric distributions.

\subsection{Phase II single-arm trial in adults with high-grade astrocytoma} \label{Ex1}

The first example is a phase II SAT evaluating overall survival (OS) of the addition of TVB-2640 (an inhibitor) to Bevacizumab (a monoclonal antibody) in adults with high-grade astrocytoma \citep{Kelly2023}, a rare cancer. A total of 25 patients (22 deaths, 12\% censoring) were enrolled in the experimental group and compared with an external control group \citep{Taal2014}, comprising 50 patients with recurrent glioblastoma treated with Bevacizumab alone. \\

\begin{figure*}[htb]
    \centerline{\includegraphics[width=\textwidth]{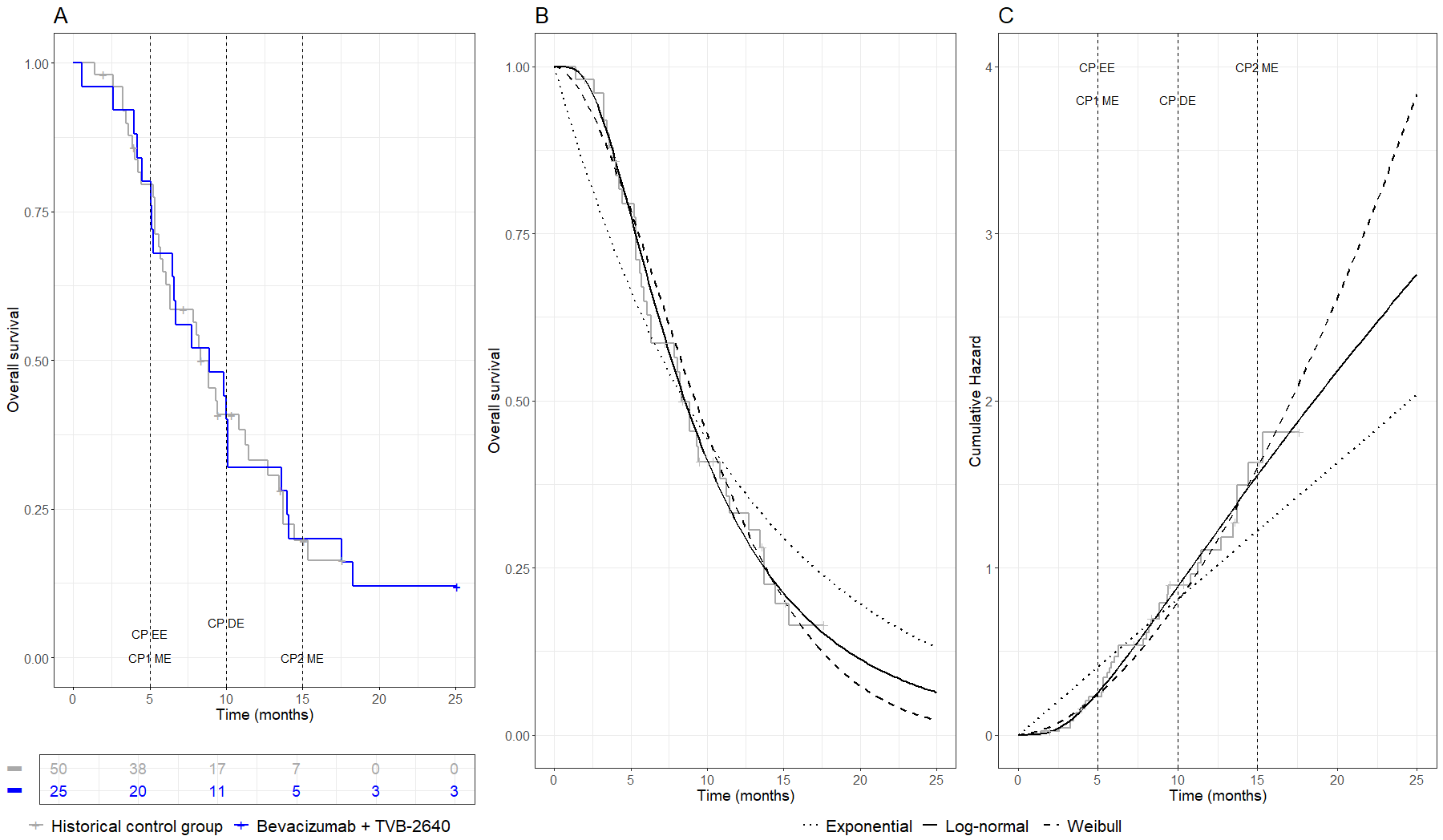}}
    \caption{Phase II single-arm trial in adults with high-grade astrocytoma. (A) Kaplan-Meier estimates of overall survival for the experimental group (Bevacizumab + TVB-2640) and the external control group reconstructed from individual patient data. (B) Kaplan-Meier estimate of overall survival for the external control group with parametric fits based on exponential, Weibull, and log-normal distributions. (C) Non-parametric and parametric (exponential, Weibull, and log-normal) estimates of the cumulative hazard function of the external control group. Vertical dashed lines indicate the change-points used in the statistical tests (5, 10 and 15 years).}
    \label{OS Kelly}
\end{figure*}

\noindent Figure \ref{OS Kelly}A shows the KM estimate of OS for both groups, indicating no statistically significant difference (OSLRT: p = 0.56 \citep{Kelly2023}). The log-normal distribution (solid line) provides the best fit to the external control group (AIC = 243) compared with the exponential (AIC = 258) and Weibull (AIC = 248) distributions (Figure \ref{OS Kelly}B). As the expected number of events $\mathrm{E}$ is a key component of non RMST-based tests, Figure \ref{OS Kelly}C presents the estimated cumulative hazard function $\Lambda_0(t)$ under the three parametric models. Table \ref{tab: pval_RD_Kelly} reports the p-values obtained from the different tests (rows) under each distribution (columns). \\

\noindent The interpretation is consistent with the original publication: (i) the same p-value is obtained for the OSLRT under the exponential distribution, and (ii) most tests indicate no statistically significant difference at the 5\% level, even if some tests are marginally significant. Notably, for the OSLRT and mOSLRT, p-values under the Weibull distribution are smaller than those obtained under the exponential and log-normal distributions, reflecting an overestimation of the expected number of events $\mathrm{E}$ (Figure \ref{OS Kelly}C). This also explains the marginal significance observed for some tests, such as $\mathrm{Z_{DE}}$ (p = 0.0601), as $\Lambda_0(t)$ increases more rapidly after $k = 15$ years compared with the other distributions. These results highlight the importance of selecting an appropriate model for the cumulative hazard function. 

\begin{table*}[htb]
        \caption{P-values for the OSLRT, mOSLRT and the developed tests comparing overall survival of Bevacizumab + TVB-2640 in a phase II single-arm trial in adults with high-grade astrocytoma ($n = 25$) with an external control group ($n = 50$).}
        \label{tab: pval_RD_Kelly}
        \centering
        \begin{tabular}{l c c c}
            \hline
            Tests & Exponential* & Weibull* & Log-normal* \\
            & (AIC = 258) & (AIC = 248) & (AIC = 243) \\
            \hline
            \textbf{Score tests} & & & \\
            \hspace{0.1cm} OSLRT & \textbf{0.5646}** & 0.1644 & 0.3432 \\
            \hspace{0.1cm} mOSLRT & 0.5641 & 0.1525 & 0.3400 \\
            \hspace{0.1cm} $\mathrm{Z_{EE}}$$^{\bf a}$ & 0.0738 & 0.4133 & 0.3941 \\
            \hspace{0.1cm} $\mathrm{Z_{ME}}$$^{\bf a}$ & 0.9806 & 0.6434 & 0.6105 \\
            \hspace{0.1cm} $\mathrm{Z_{DE}}$$^{\bf a}$ & 0.7994 & 0.0601 & 0.3361 \\
            \hspace{0.1cm} $\mathrm{Z_{CH}}$$^{\bf a}$ & 0.9616 & 0.0547 & 0.0815 \\
            \textbf{$\tau$-RMST$^{\bf b}$} & 0.4423 & 0.5581 & 0.5134 \\
            \textbf{max-Combo$^{\bf c}$} & & & \\
            \hspace{0.1cm} Hochberg correction & 0.3692 & 0.0883 & 0.4288 \\
            \hspace{0.1cm} Multivariate normal integration & 0.2210 & 0.0642 & 0.3615 \\
            \hline
        \end{tabular} \\
        \raggedright
        \scriptsize
        * Parametric distributions used to model the cumulative hazard function of the external control group $\Lambda_0(t)$ (the exponential distribution is used by default)  \\
        ** Bold value corresponds to the p-value reported in the original publication \\
        $^{\bf a}$ Score tests for early, middle, delayed and crossing effects, respectively. Change-points: $k = 5$ for early, $k_1 = 5$ and $k_2 = 15$ for middle and $k = 10$ for delayed effects. \\
        $^{\bf b}$ Restricted mean survival time with $\tau = 17.60$ months.\\
        $^{\bf c}$ Max-Combo test combining the mOSLRT, two early effect score tests ($k = 5$ and 10), and two delayed effect score tests ($k = 10$ and 15).
\end{table*}

\subsection{Phase II single-arm trial in children with neuroblastoma} \label{Ex2}

\begin{figure*}[htb]
    \centerline{\includegraphics[width=\textwidth]{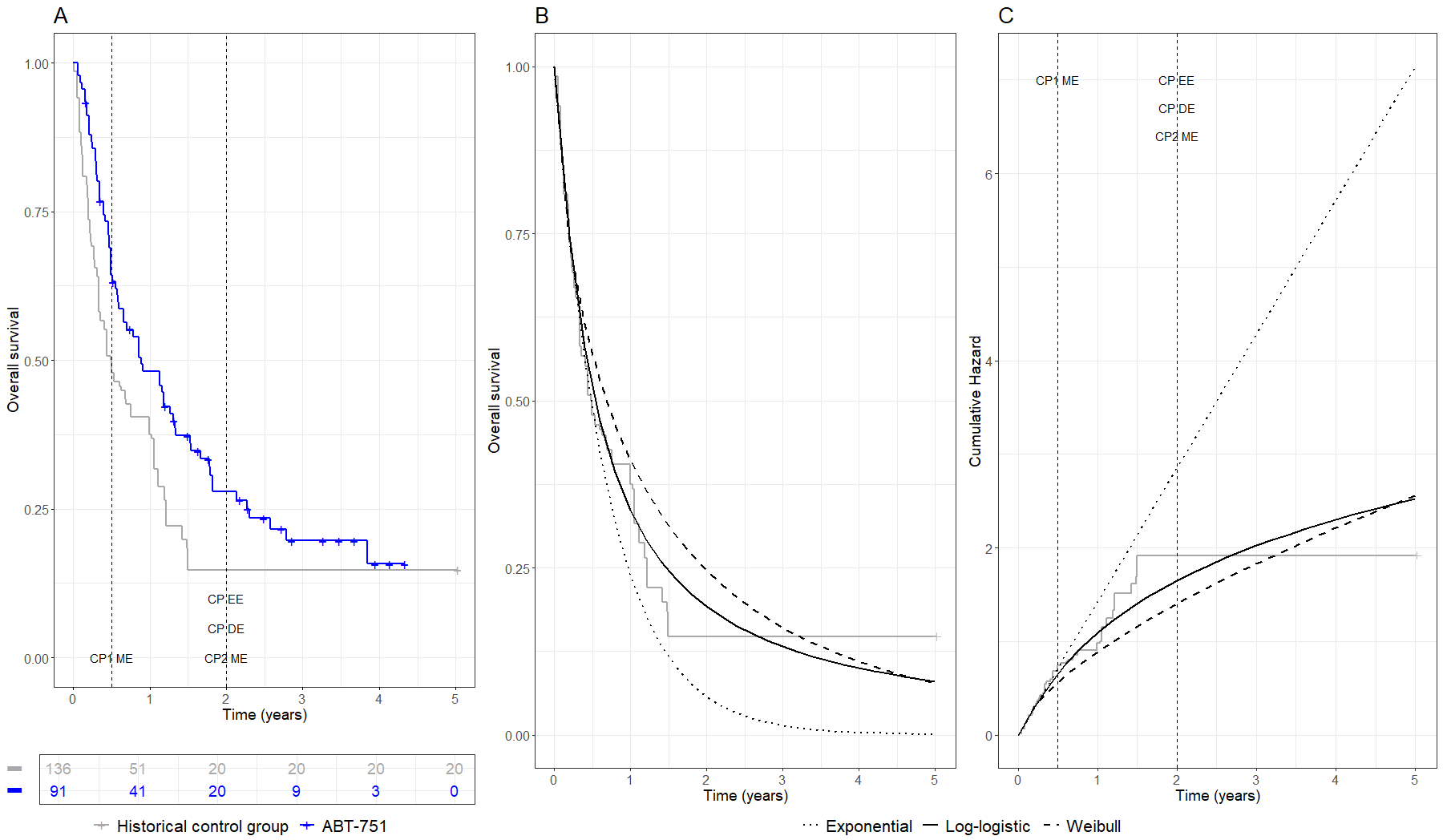}}
    \caption{Phase II single-arm trial in children with neuroblastoma. (A) Kaplan-Meier estimates of overall survival for the experimental group (ABT-751) and the external control group reconstructed from individual patient data. (B) Kaplan-Meier estimate of overall survival for the external control group with parametric fits based on exponential, Weibull and log-logistic distributions. (C) Non-parametric and parametric (exponential, Weibull and log-logistic) estimates of the cumulative hazard function of the external control group. Vertical dashed lines indicate the change-points used in the statistical tests (0.5 and 2 years).}
    \label{OS Fox}
\end{figure*}

In this example, we analyse data from a phase II SAT study conducted by Fox et al. \citep{Fox2014}, which evaluated ABT-751 (a bioavailable sulfonamide inhibitor) in children with relapsed or refractory neuroblastoma. A total of n = 91 patients (68 deaths, 25\% censoring) received ABT-751, and OS in the experimental group was compared with that of an external control group. This external group is composed of 136 patients from 5 previous phase I or II studies. An early treatment effect in favour of the experimental arm is observed within the first two years (Figure \ref{OS Fox}A, blue curve). \\

\noindent Among the different parametric distributions, the log-logistic distribution (AIC = 244) provides the best fit to the external data, compared with the exponential (AIC = 317) and Weibull (AIC = 278) distributions. The OSLRT and mOSLRT are statistically significant regardless of the modelling of $\Lambda_0(t)$ (Table \ref{tab: pval_RD_Fox}). However, the exponential distribution overestimates the expected number of events (Figure \ref{OS Fox}C, dotted line), leading to artificially smaller p-values (first two rows of Table \ref{tab: pval_RD_Fox}) compared with those obtained under the Weibull and log-logistic models. These findings are consistent with those observed in section \ref{Var-distrib} (Figures \ref{Diff_misspe} and \ref{Power_misspe} for scenario 3). \\

\noindent The impact of model misspecification is particularly evident for the $\mathrm{Z_{DE}}$ test, which yields a significant p-value under the exponential distribution but not under the Weibull or log-logistic distributions. Focusing on the best-fitting model (log-logistic, last column of Table \ref{tab: pval_RD_Fox}), the $\mathrm{Z_{EE}}$ test, as expected, detects a significant difference between the experimental and external control groups. This difference is more pronounced than under the Weibull distribution (p = 0.0019 vs 0.0496). This is explained by a smaller expected number of events (underestimation) with the Weibull distribution (dashed line, Figure \ref{OS Fox}C). The max-Combo test, which includes the $\mathrm{Z_{EE}}$ test with two different change-points ($k = 1$ and 2 years), is significant regardless of the multiple testing correction. The RMST-based test also yields a significant p-value, although its magnitude depends on the chosen parametric model. The exponential distribution underestimates OS and therefore inflates the test statistic (Figure \ref{OS Fox}B), whereas the Weibull distribution overestimates OS, resulting in a p-value of approximately 5\%. The log-logistic model provides a more reliable estimate (p = 0.0045).

\begin{table*}[htb]
    \caption{P-value for the OSLRT, mOSLRT and the developed tests comparing overall survival of ABT-751 in a phase II single-arm trial in children with neuroblastoma ($n = 91$) to an external control group ($n = 136$).}
    \centering
    \begin{tabular}{l c c c}
        \hline
        Test & Exponential* & Weibull* & Log-logistic* \\
        & (AIC = 317) & (AIC = 278) & (AIC = 244) \\
        \hline
        \textbf{Score tests} & & & \\
        \hspace{0.1cm} OSLRT & $6.262.10^{-14}$ & 0.0296 & 0.0017 \\
        \hspace{0.1cm} mOSLRT & 0 & 0.0232 & 0.0007 \\
        \hspace{0.1cm} $\mathrm{Z_{EE}}$$^{\bf a}$ & $7.304.10^{-10}$ & 0.0496 & 0.0019 \\
        \hspace{0.1cm} $\mathrm{Z_{ME}}$$^{\bf a}$ & $6.639.10^{-8}$ & 0.4367 & 0.0997 \\
        \hspace{0.1cm} $\mathrm{Z_{DE}}$$^{\bf a}$ & $3.298.10^{-6}$ & 0.1566 & 0.2901 \\
        \hspace{0.1cm} $\mathrm{Z_{CH}}$$^{\bf a}$ & $4.708.10^{-7}$ & 0.9987 & 0.9738 \\
        \textbf{$\tau$-RMST$^{\bf b}$} & $9.164.10^{-8}$ & 0.0521 & 0.0045 \\
        \textbf{max-Combo$^{\bf c}$} & & & \\
        \hspace{0.1cm} Hochberg correction & 0 & 0.1071 & 0.0036 \\
        \hspace{0.1cm} Multivariate normal integration & 0 & 0.0770 & 0.0029 \\
        \hline
    \end{tabular} \\
    \raggedright
    \scriptsize
    * Parametric distributions used to model the cumulative hazard function of the external control group $\Lambda_0(t)$ (exponential is used by default). \\
    $^{\bf a}$ Score tests for early, middle, delayed and crossing effects, respectively. Change-points: $k = 2$ for the early and delay effect. $k_1 = 0.5$ and $k_2 = 2$ for the middle effect. \\
    $^{\bf b}$ Restricted mean survival time with $\tau = 4.33$ years. \\
    $^{\bf c}$ Max-Combo test combining the mOSLRT, two early effect score tests ($k = 1$ and 2) and two delayed effect score tests ($k = 2$ and 3).
    \label{tab: pval_RD_Fox}
\end{table*}

\subsection{Subgroup of patients from a phase III randomized trial in patients with small-cell lung cancer} \label{Ex3}

Using data from a randomised controlled trial, Liu et al \citep{Liu2021} evaluated, in an exploratory biomarker analysis, the effect on OS of Atezolizumab (an immunotherapy) in combination with carboplatin (platinum chemotherapy) and etoposide (CP/ET) ($n = 47$, 19\% censoring) versus placebo plus CP/ET ($n = 59$) in a subgroup of patients (programmed death-ligand 1, PD\-L1 $<$ 5\%) with extensive-stage small-cell lung cancer. \\

\begin{figure*}[htb]
    \centerline{\includegraphics[width=\textwidth]{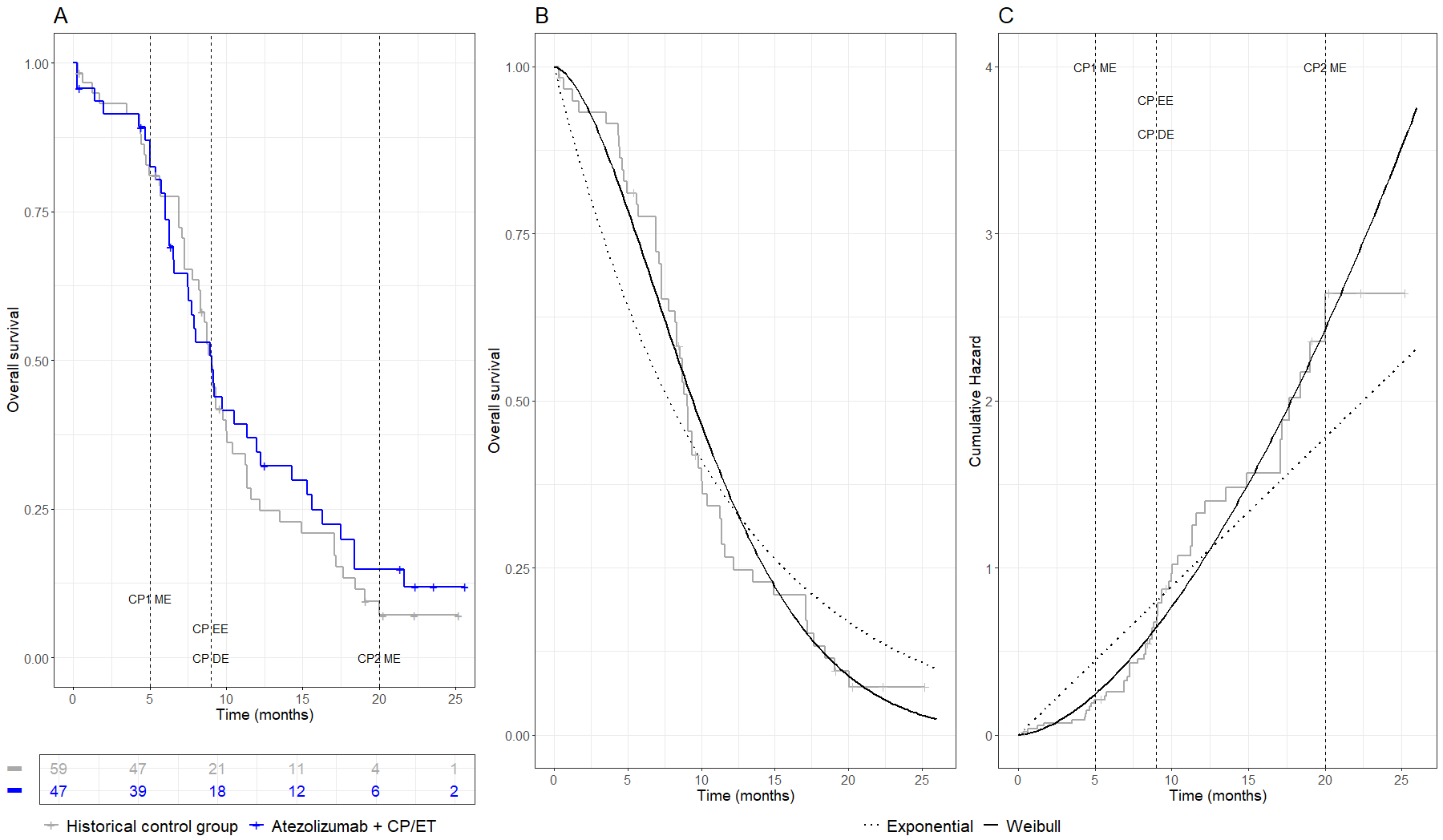}}
    \caption{Subgroup of patients with PD-L1 $<$ 5\% from a phase III randomised controlled trial in patients with extensive-stage small cell lung cancer. (A) Kaplan-Meier estimates of overall survival for the experimental group (Atezolizumab + CP/ET) and the control group reconstructed from individual patient data. (B) Kaplan-Meier estimate of overall survival for the control group with parametric fits based on exponential and Weibull distributions. (C) Non-parametric and parametric (exponential and Weibull) estimates of the cumulative hazard function of the control group. For illustration, the control group is treated as an external control group. Vertical dashed lines indicate the change-points used in the statistical tests (5, 9 and 20 years).}
    \label{OS Liu PD-L1}
\end{figure*}

\noindent As an illustration of a delayed treatment effect (Figure \ref{OS Liu PD-L1}A), we reanalysed this subgroup, treating the control arm as an external control group. A modest benefit of the experimental treatment emerges after approximately 10 months. The external control group are better fitted by a Weibull distribution (AIC = 337) than by an exponential distribution (AIC = 351) (Figure \ref{OS Liu PD-L1}B), as well as compared with other parametric models (data not shown). \\

\noindent As expected, the OSLRT and mOSLRT do not detect a significant difference between the experimental and external control groups (first two rows of Table \ref{tab: pval_RD_Liu_PD-L1}),  reflecting their lack of power from the PH assumption. In contrast, the $\mathrm{Z_{DE}}$ test is significant under the best fitting Weibull model (p = 0.0131) but not under the exponential model (p = 0.5033). The Weibull-based result is probably too liberal due to the influence of a censored patient with the longest follow-up, whose contribution to the numerator of the $\mathrm{Z_{DE}}$ test statistic is large because of a high value of $\Lambda_0(t)$ at the censoring time (Figure \ref{OS Liu PD-L1}C). Removing this patient increases the p-value, although it remains statistically significant (p = 0.0421). By contrast, the exponential model fails to capture the delayed effect, owing to an underestimation of the expected number of deaths after 10 months.\\ 

\noindent Although a delayed effect is visible in Figure \ref{OS Liu PD-L1}A, the survival curve of the external control group slightly exceeds that of the experimental group between 6 and 9 months. This feature likely explains why the conclusion of the $\mathrm{Z_{CH}}$ test is similar to that of the $\mathrm{Z_{DE}}$ test. The RMST-based test is not statistically significant regardless of the distribution. The max-Combo test is significant or marginally significant under both models, regardless of the multiple testing correction. 

\begin{table*}[htb]
    \caption{P-value for the OSLRT, mOSLRT and the developed tests comparing overall survival of Atezolizumab + CP/ET in a subgroup of patients (PDL-1 $<$ 5\%) ($n = 47$) in patients with extensive-stage small cell lung cancer with an external control group ($n = 59$).}
    \centering
    \begin{tabular}{l c c c}
        \hline
        Test & Exponential* & Weibull* \\
        & (AIC = 351) & (AIC = 337) \\
        \hline
        \textbf{Score tests} & & & \\
        \hspace{0.1cm} OSLRT & 0.2258 & 0.1062 \\
        \hspace{0.1cm} mOSLRT & 0.2191 & 0.0954 \\
        \hspace{0.1cm} $\mathrm{Z_{EE}}$$^{\bf a}$ & 0.1863 & 0.6621 \\
        \hspace{0.1cm} $\mathrm{Z_{ME}}$$^{\bf a}$ & 0.9690 & 0.4299 \\
        \hspace{0.1cm} $\mathrm{Z_{DE}}$$^{\bf a}$ & 0.5033 & 0.0131 \\
        \hspace{0.1cm} $\mathrm{Z_{CH}}$$^{\bf a}$ & 0.9821 & 0.0370 \\
        \textbf{$\tau$-RMST$^{\bf b}$} & 0.1539 & 0.2363 \\
        \textbf{max-Combo$^{\bf c}$} & & & \\
        \hspace{0.1cm} Hochberg correction & 0.0074 & 0.0657 \\
        \hspace{0.1cm} Multivariate normal integration & 0.0063 & 0.0497 \\
        \hline
    \end{tabular} \\
    \raggedright
    \scriptsize
    * Parametric distribution used to model the cumulative hazard function of the external control group $\Lambda_0(t)$ (exponential is used by default). \\
    $^{\bf a}$ Score tests for early, middle, delayed and crossing effects, respectively. Change-points: $k = 10$ for early and delayed effects. $k_1 = 5$ and $k_2 = 20$ for middle effect. \\
    $^{\bf b}$ Restricted mean survival time with $\tau = 25.1871$ months. \\
    $^{\bf c}$ Max-Combo test combining the mOSLRT, two early effect score tests ($k = 5$ and 10) and two delayed effect score tests ($k = 10$ and 15).
    \label{tab: pval_RD_Liu_PD-L1}
\end{table*}

\section{Discussion}

We propose, for the first time, a set of statistical tests and alternatives for analysing single-arm trials (SATs) with time-to-event endpoints in the presence of non-proportional hazards. These tests are constructed by reformulating the OSLRT as a score test within standard survival models adapted to different situations of non-proportionality. Specifically, piecewise exponential proportional hazards models are used for early, middle and delayed effects, while an accelerated hazards model is employed for crossing hazards. In addition, the RMST-based test and max-Combo test are extended to the SAT setting. The former has the advantage of being distribution-free, whereas the latter offers flexibility to capture different patterns of non-proportionality. \\

\noindent The developed score tests are (i) as conservative as the OSLRT and (ii) the most powerful under the scenarios for which they are designed, although they require prior specification of change-points (CP). A sensitivity analysis shows that moderate misspecification of these CPs reduces the power, but performance remains higher than that of the OSLRT and the mOSLRT. Larger deviations are expected to reduce power further, in which case the max-Combo test may outperform the corresponding optimal score test. The RMST-based test, which assumes no variability in the external control group, shows at best comparable performance to the OSLRT and mOSLRT under non-proportional hazards. The max-Combo test, combining the mOSLRT with score tests for early and delayed effects across multiple change-points, provides a valuable alternative when the treatment effect pattern is unknown at the design stage. Although conservative, particularly when using the Hochberg correction compared with multivariate normal comparison, it outperforms the OSLRT and mOSLRT in terms of power for sample sizes exceeding approximately 30-50 patients. While not optimal, it maintains reasonably good power (typically within 10-15\% of the optimal test) across different scenarios of non-proportional hazards, except for crossing hazards, indicating a degree of robustness. Given that power is a key consideration in early-phase oncology clinical trials, where controlling type II error is often prioritised compared to type I error. An additional advantage of the max-Combo test is its adaptability to prior clinical knowledge, although a balance must be struck between the number of components (e.g., change-points and the multiple testing correction). \\

\noindent Previous work assumes that the survival curve of the external control group is known without sampling variability and follows the same distribution as that of the experimental group. However, in practice, this distribution is typically estimated from historical data based on limited sample sizes and without access to individual patient data. First, we assessed the impact of variability in the survival estimate by allowing the exponential parameter $\lambda$ to vary. This led to a marked inflation of the type I error across all tests, with only limited impact on the power of the optimal test for a given scenario. For the max-Combo test, the impact remained modest for sample sizes exceeding 60 patients. Second, we apply corrected score tests to account for variability in the external control group, following Danzer et al \citep{Danzer2022_ref, Danzer2022_var} and Feld et al \citep{Feld2024}. As expected, this correction reduced the type I error, but also led to a decrease in power. Third, we investigated the impact of model misspecification in the external control group, i.e., an inaccurate estimation of the expected number of events. Using a log-logistic distribution representing a non-monotone hazard function, we observed (i) an inflation of the type I error for most tests and (ii) a substantial impact on the power of the optimal test for each scenario. The max-Combo test was also impacted, with a noticeable loss of power. Conversely, the score test for crossing hazards showed increased power across scenarios, reflecting the suitability of the accelerated hazards model for non-monotone hazard functions \citep{Zhang2009}. These findings are consistent with the results observed in the real data examples. To mitigate the impact of model misspecification, one may reduce the influence of patients with long follow-up times, or, more generally, truncate the analysis when the number of patients at risk in the external control group becomes too small. These approaches help avoid overestimation of the expected number of events and limit extrapolation beyond the observed follow-up. For example, truncating the experimental data at the maximum follow-up time of the external control group (e.g., 17.60 months instead of 25 months in Example \ref{Ex1}) leads to larger p-values for all tests, except for the RMST-based test. \\

\noindent Some limitations should be discussed. First, the developed score tests require the specification of change-points, which may be difficult in practice, particularly in phase II SATs with limited sample sizes. This limitation may be alleviated by the use of the max-Combo test, which accommodates multiple change-points and treatment effect patterns. Although more flexible or complex parametric or semi-parametric models could be considered, deriving one-dimensional test statistics in small sample size settings would be challenging. Second, the simulation study assumes an identical follow-up period between the experimental and external control groups. However, the three real data examples show that differences in follow-up can substantially affect the results, especially when follow-up is longer in the experimental group. Potential solutions include weighting observations or truncating the data. More generally, the OSLRT, mOSLRT, and score tests are sensitive to extreme time-to-event values, as late observations can disproportionately influence the test statistics through the cumulative hazard function. This issue further highlights the importance of selecting an appropriate parametric model for the external control data.\\

\noindent In conclusion, we propose several survival tests for analysing single-arm oncology trials with a time-to-event endpoint under non-proportional hazards, together with a combination test procedure when the treatment effect pattern is unknown. In practice, careful attention should be paid to the specification of the survival model for the external control group and to differences in follow-up between groups, as both can influence the interpretation. We recommend (i) examining the survival curve of the external control group to define an appropriate time horizon based on the number of patients at risk, (ii) performing analyses truncated at this time point, (iii) assessing the robustness of the results to the choice of the parametric survival model for the external control group through sensitivity analyses, and (iv) evaluating the impact of the selected change-points through additional sensitivity analyses.  \\

\noindent Further work would be extending these methods to sample size calculation, developing Bayesian approaches that allow the incorporation of external data from different sources (control arms of randomised clinical trials, real-world data, registries) to augment the amount of external control data, and designing adaptive single-arm trials, such as basket and umbrella trials, to accelerate the evaluation of multiple treatments in oncology, where rare and specific tumour types are frequently encountered.

\section*{Financial disclosure}

This work was funded by PhD grant MESRI from the doctoral School of Public Health, Paris-Saclay University.

\section*{Conflict of interest}

The authors declared no potential conflict of interest.

\bibliographystyle{plainnat}
\bibliography{Biblio}

\appendix

\section{Derivation of the score tests} \label{Z details}

\subsection{Formulation of the OSLRT as a score test}\label{OSLRT details} 

$\mathrm{S}_1(t) = \mathrm{S}_0(t)^{\mathrm{HR}} \Longleftrightarrow \Lambda_1(t) = \exp(\beta) \Lambda_0(t) \Longleftrightarrow \lambda_1(t) = \exp(\beta) \lambda_0(t)$  with $\mathrm{HR} = \exp(\beta)$\\
The log-likelihood function:
\begin{equation*}
    l = \sum_{i = 1}^n \left(\delta_i \log(\lambda_1(X_i)) - \Lambda_1(X_i)\right) = \sum_{i = 1}^n \left(\delta_i \beta + \delta_i \log(\lambda_0(X_i)) - \exp(\beta) \Lambda_0(X_i)\right)
\end{equation*}
The derivative of the log-likelihood:
\begin{equation*}
    \frac{\partial \log(L)}{\partial \beta} = \sum_{i = 1}^n \left(\delta_i - \exp(\beta)\Lambda_0(X_i)\right)
\end{equation*}
Evaluate this derivative at $\beta = 0$:
\begin{equation*}
    \left. \frac{\partial \log(L)}{\partial \beta} \right| _{\beta = 0} = \sum_{i = 1}^n \left(\delta_i - \Lambda_0(X_i)\right) = \mathrm{O - E}
\end{equation*}
The second derivative of the log-likelihood:
\begin{equation*}
    \frac{\partial^2 \log(L)}{\partial \beta^2} = \sum_{i = 1}^n \left(- \exp(\beta) \Lambda_0(X_i)\right) = - \sum_{i = 1}^n \exp(\beta) \Lambda_0(X_i)
\end{equation*}
Evaluate this second derivative at $\beta = 0$:
\begin{equation*}
    \left. \frac{\partial^2 \log(L)}{\partial \beta^2}\right| _{\beta = 0} = - \sum_{i = 1}^n \Lambda_0(X_i) = - \mathrm{E}
\end{equation*}
The score statistic for the case of the proportional hazards, which is also called the One-Sample Log-Rank Test \citep{Breslow1975, Woolson1981, Finkelstein2003}, is:
\begin{equation*}
    \mathrm{Z_{PH}} =  \frac{\sum\limits_{i = 1}^n \left(\delta_i - \Lambda_0(X_i)\right)}{\sqrt{\sum\limits_{i = 1}^n \Lambda_0(X_i)}} = \mathrm{\frac{O - E}{\sqrt{E}} = OSLRT} \\
\end{equation*}

\subsection{Score test for an early effect} \label{EE details}
Hazard function:
\[
\lambda_1(t) =
\begin{cases}
   \exp(\beta)\lambda_0(t)                &\text{if $t \leq k$}\\
   \lambda_0(t)  &\text{if $t > k$}
\end{cases} \\
\]
Cumulative hazard function:
\[
\Lambda_1(t) =
\begin{cases}
   \exp(\beta)\Lambda_0(t)                &\text{if $t \leq k$}\\
   (\exp(\beta)-1)\Lambda_0(k) + \Lambda_0(t)  &\text{if $t \geq k$}
\end{cases} \\
\]
Survival function:
\[
\mathrm{S}_1(t) =
\begin{cases}
   \mathrm{S}_0(t)^{\exp(\beta)}                &\text{if $t \leq k$}\\
   \mathrm{S}_0(k)^{\exp(\beta)-1}\mathrm{S}_0(t)  &\text{if $t \geq k$}
\end{cases} \\
\]
Log-likelihood function: 
\begin{align*}
    l = \sum_{i = 1}^n & \left(\delta_i \log(\lambda_1(X_i)) - \Lambda_1(X_i)\right) \\
    = \sum_{i = 1}^n & (\delta_i \{\log(\exp(\beta)\lambda_0(X_i))\mathrm{I}(X_i \leq k) + \log(\lambda_0(X_i))\mathrm{I}(X_i > k)\} - \\ & \{\exp(\beta)\Lambda_0(X_i) \mathrm{I}(X_i \leq k) + (\exp(\beta)-1)\Lambda_0(k) \mathrm{I}(X_i \geq k) + \Lambda_0(X_i) \mathrm{I}(X_i \geq k)\}) \\
    = \sum_{i = 1}^n & \left(\delta_i \{\beta \mathrm{I}(X_i\leq k) + \log(\lambda_0(X_i))\} \right. \\ 
    & - \left. \{\exp(\beta)\Lambda_0(X_i) \mathrm{I}(X_i \leq k) + (\exp(\beta)-1)\Lambda_0(k) \mathrm{I}(X_i \geq k) + \Lambda_0(X_i) \mathrm{I}(X_i \geq k)\}\right) \\
    = \sum_{i = 1}^n &\left(\delta_i \beta \mathrm{I}(X_i \leq k) + \delta_i \log(\lambda_0(X_i)) - \exp(\beta) \Lambda_0(X_i) \mathrm{I}(X_i \leq k) - (\exp(\beta)-1) \Lambda_0(k) \mathrm{I}(X_i \geq k) \right. \\
    & - \left. \Lambda_0(X_i) \mathrm{I}(X_i \geq k)\right)
\end{align*}
The derivative of the log-likelihood:
\begin{equation*}
   \frac{\partial \log(L)}{\partial \beta} = \sum_{i = 1}^n \left(\delta_i \mathrm{I}(X_i \leq k) - \exp(\beta)\Lambda_0(X_i) \mathrm{I}(X_i \leq k) - \exp(\beta)\Lambda_0(k) \mathrm{I}(X_i \geq k)\right)
\end{equation*}
Evaluate this derivative at $\beta=0$:
\begin{equation*}
   \left. \frac{\partial \log(L)}{\partial \beta} \right| _{\beta = 0} = \sum_{i = 1}^n \left(\delta_i \mathrm{I}(X_i \leq k) - \Lambda_0(X_i) \mathrm{I}(X_i \leq k) - \Lambda_0(k) \mathrm{I}(X_i \geq k)\right) = \sum_{i : X_i \leq k} (\delta_i - \Lambda_0(X_i)) - \sum_{i : X_i \geq k} \Lambda_0(k)
\end{equation*}
The second derivative of the log-likelihood:
\begin{align*}
    \frac{\partial^2 \log(L)}{\partial \beta^2} & = \sum_{i = 1}^n \left(-\exp(\beta) \Lambda_0(X_i) \mathrm{I}(X_i \leq k) - \exp(\beta) \Lambda_0(k) \mathrm{I}(X_i \geq k)\right) \\
    & = - \left(\sum_{i : X_i \leq k} \Lambda_0(X_i) + \sum_{i : X_i \geq k} \exp(\beta) \Lambda_0(k)\right)
\end{align*}
Evaluate this second derivative at $\beta = 0$:
\begin{equation*}
    \left. \frac{\partial^2 \log(L)}{\partial \beta^2}\right| _{\beta = 0} = - \left(\sum_{i : X_i \leq k} \Lambda_0(X_i) + \sum_{i : X_i \geq k} \Lambda_0(k)\right)
\end{equation*}
The score statistic for the case of early effect with one change-point $k$ is:
\begin{equation*}
    \mathrm{Z_{EE}} = \frac{\sum\limits_{i : X_i \leq k} \left(\delta_i - \Lambda_0(X_i)\right) -\sum\limits_{i : X_i \geq k} \Lambda_0(k)}{\sqrt{\sum\limits_{i : X_i \leq k} \Lambda_0(X_i) + \sum\limits_{i : X_i \geq k} \Lambda_0(k)}}
\end{equation*}

\subsection{Score test for a middle effect} \label{ME details}
Hazard function:
\[
\lambda_1(t) =
\begin{cases}
   \lambda_0(t)  &\text{if $t \leq k_1$}\\
   \exp(\beta)\lambda_0(t)  &\text{if $k_1 < t \leq k_2$}\\
   \lambda_0(t)  &\text{if $t \geq k_2$}
\end{cases} \\
\]
Cumulative hazard function:
\[
\Lambda_1(t) =
\begin{cases}
   \Lambda_0(t)  &\text{if $t \leq k_1$}\\
   (1-\exp(\beta))\Lambda_0(k_1) + \exp(\beta)\Lambda_0(t)  &\text{if $k_1 \leq t \leq k_2$}\\
   (1-\exp(\beta))\Lambda_0(k_1) + (\exp(\beta)-1)\Lambda_0(k_2) + \Lambda_0(t)  &\text{if $t \geq k_2$}
\end{cases} \\
\]
Survival function:
\[
\mathrm{S}_1(t) =
\begin{cases}
   \mathrm{S}_0(t)  &\text{if $t \leq k_1$}\\
   \mathrm{S}_0(k_1)^{(1-\exp(\beta))} \mathrm{S}_0(t)^{\exp(\beta)}  &\text{if $k_1 \leq t \leq k_2$}\\
   \mathrm{S}_0(k_1)^{(1-\exp(\beta))} \mathrm{S}_0(k_2)^{(\exp(\beta)-1)} \mathrm{S}_0(t)  &\text{if $t \geq k_2$}
\end{cases} \\
\]
Log-likelihood function:
\begin{align*}
    l = \sum_{i = 1}^n & \left(\delta_i \log(\lambda_1(X_i)) - \Lambda_1(X_i)\right) \\
     = \sum_{i=1}^n &( \delta_i \{\log(\lambda_0(X_i)) \mathrm{I}(X_i \leq k_1) + \log(\exp(\beta)\lambda_0(X_i)) \mathrm{I}(k_1 < X_i \leq k_2) + \log(\lambda_0(X_i)) \mathrm{I}(X_i > k_2)\} \\
     & - \Lambda_0(X_i) \mathrm{I}(X_i \leq k_1) - (1-\exp(\beta))\Lambda_0(k_1) \mathrm{I}(k_1 \leq X_i \leq k_2) - \exp(\beta)\Lambda_0(X_i) \mathrm{I}(k_1 \leq X_i \leq k_2) \\
     & - (1-\exp(\beta))\Lambda_0(k_1) \mathrm{I}(X_i \geq k_2) - (\exp(\beta)-1) \Lambda_0(k_2) \mathrm{I}(X_i \geq k_2) - \Lambda_0(X_i) \mathrm{I}(X_i \geq k_2)) \\
    = \sum_{i = 1}^n &(\delta_i \log(\lambda_0(X_i)) + \delta_i \beta \mathrm{I}(k_1 < X_i \leq k_2) - \Lambda_0(X_i)(\mathrm{I}(X_i \leq k_1) + \mathrm{I}(X_i \geq k_2)) \\
    & + (\exp(\beta)-1)\Lambda_0(k_1) \mathrm{I}(X_i \geq k_1) - \exp(\beta)\Lambda_0(X_i) \mathrm{I}(k_1 \leq X_i \leq k_2) + (1-\exp(\beta)) \Lambda_0(k_2) \mathrm{I}(X_i \geq k_2))
\end{align*}
Derivative of the log-likelihood:
\begin{align*}
    \frac{\partial \log(L)}{\partial \beta} & = \sum_{i = 1}^n \left(\delta_i \mathrm{I}(k_1 < X_i \leq k_2) + \exp(\beta) \Lambda_0(k_1) \mathrm{I}(X_i \geq k_1) - \exp(\beta)\Lambda_0(X_i)\mathrm{I}(k_1 \leq X_i \leq k_2) \right. \\
    & - \left. \exp(\beta)\Lambda_0(k_2)\mathrm{I}(X_i \geq k_2)\right)
\end{align*}
Evaluate this derivative at $\beta = 0$:
\begin{align*}
    \left. \frac{\partial \log(L)}{\partial \beta} \right| _{\beta = 0} & = \sum_{i = 1}^n \left(\delta_i \mathrm{I}(k_1 < X_i \leq k_2) - \Lambda_0(X_i) \mathrm{I}(k_1 \leq X_i \leq k_2) + \Lambda_0(k_1) \mathrm{I}(X_i \geq k_1) - \Lambda_0(k_2) \mathrm{I}(X_i \geq k_2)\right) \\
    & = \sum_{i : X_i \in ]k_1 ; k_2]} \left(\delta_i - \Lambda_0(X_i)\right) + \sum_{i : X_i \geq k_1} \Lambda_0(k_1) - \sum_{i : X_i \geq k_2} \Lambda_0(k_2)
\end{align*}
The second derivative of the log-likelihood:
\begin{equation*}
    \frac{\partial^2 \log(L)}{\partial \beta^2} = - \sum_{i = 1}^n \left(\exp(\beta) \Lambda_0(X_i) \mathrm{I}(k_1 \leq X_i \leq k_2) - \exp(\beta)(\Lambda_0(k_1) \mathrm{I}(X_i \geq k_1) - \Lambda_0(k_2) \mathrm{I}(X_i \geq k_2))\right)
\end{equation*}
Evaluate this second derivative at $\beta = 0$:
\begin{align*}
    \left. \frac{\partial^2 \log(L)}{\partial \beta^2}\right| _{\beta = 0} & = - \sum_{i = 1}^n \left(\Lambda_0(X_i) \mathrm{I}(k_1 \leq X_i \leq k_2) - \Lambda_0(k_1) \mathrm{I}(X_i \geq k_1) + \Lambda_0k_2) \mathrm{I}(X_i \geq k_2)\right) \\
    & = - \left(\sum\limits_{i: X_i \in [k_1; k_2]}\Lambda_0(X_i) - \sum\limits_{i: X_i \geq k_1}\Lambda_0(k_1) + \sum\limits_{i: X_i \geq k_2}\Lambda_0(k_2)\right)
\end{align*}
The score statistic for the case of middle effect with two change-points $k_1$ and $k_2$ is: 
\begin{equation*}
    \mathrm{Z_{ME}} = \frac{\sum\limits_{i : X_i \in ]k_1 ; k_2]} (\delta_i - \Lambda_0(X_i)) + \sum\limits_{i : X_i \geq k_1} \Lambda_0(k_1) - \sum\limits_{i : X_i \geq k_2} \Lambda_0(k_2)}{\sqrt{\sum\limits_{i: X_i \in [k_1; k_2]}\Lambda_0(X_i) - \sum\limits_{i: X_i \geq k_1}\Lambda_0(k_1) + \sum\limits_{i: X_i \geq k_2}\Lambda_0(k_2})}\\
\end{equation*}

\subsection{Score test for a delayed effect} \label{DE details}
Hazard function:
\[
\lambda_1(t) =
\begin{cases}
   \lambda_0(t)  &\text{if $t \leq k$}\\
   \exp(\beta)\lambda_0(t)   &\text{if $t > k$}
\end{cases} \\
\]
Cumulative hazard function:
\[
\Lambda_1(t) =
\begin{cases}
   \Lambda_0(t)                &\text{if $t \leq k$}\\
   (1-\exp(\beta))\Lambda_0(k) + \exp(\beta)\Lambda_0(t)  &\text{if $t \geq k$}
\end{cases} \\
\]
Survival function:
\[
\mathrm{S}_1(t) =
\begin{cases}
   \mathrm{S}_0(t)                &\text{if $t \leq k$}\\
   \mathrm{S}_0(k)^{1-\exp(\beta)}\mathrm{S}_0(t)^{\exp(\beta)}  &\text{if $t \geq k$}
\end{cases} \\
\]
Log-likelihood function:
\begin{align*}
    l = \sum_{i = 1}^n & \left(\delta_i \log(\lambda_1(X_i)) - \Lambda_1(X_i)\right) \\
    = \sum_{i = 1}^n & \left(\delta_i \{\log(\lambda_0(X_i)) \mathrm{I}(X_i \leq k) + \log(\exp(\beta)\lambda_0(X_i)) \mathrm{I}(X_i > k)\}\right. \\
    & - \left. \{\Lambda_0(X_i) \mathrm{I}(X_i \leq k) + (\exp(\beta)\Lambda_0(X_i) + (1-\exp(\beta))\Lambda_0(k)) \mathrm{I}(X_i \geq k)\}\right) \\
    = \sum_{i = 1}^n & \left(\delta_i \beta - \delta_i \beta \mathrm{I}(X_i \leq k) + \delta_i \log(\lambda_0(X_i)) - \Lambda_0(X_i) \mathrm{I}(X_i \leq k) - \exp(\beta)\Lambda_0(X_i) \mathrm{I}(X_i \geq k) \right. \\
    & + \left. (\exp(\beta)-1)\Lambda_0(k) \mathrm{I}(X_i \geq k)\right)
\end{align*}
Derivative of the log-likelihood:
\begin{align*}
    \frac{\partial \log(L)}{\partial \beta} & = \sum_{i = 1}^n \left(\delta_i - \delta_i \mathrm{I}(X_i \leq k) - \exp(\beta)\Lambda_0(X_i) \mathrm{I}(X_i \geq k) + \exp(\beta)\Lambda_0(k) \mathrm{I}(X_i \geq k)\right) \\
    & = \sum_{i = 1}^n \left(\delta_i \mathrm{I}(X_i > k) - \exp(\beta)\Lambda_0(X_i) \mathrm{I}(X_i \geq k) + \exp(\beta)\Lambda_0(k) \mathrm{I}(X_i \geq k)\right)
\end{align*}
Evaluate this derivative at $\beta =0$:
\begin{equation*}
    \left. \frac{\partial \log(L)}{\partial \beta} \right| _{\beta = 0} = \sum_{i = 1}^n \left(\delta_i \mathrm{I}(X_i>k) - \Lambda_0(X_i) \mathrm{I}(X_i \geq k) + \Lambda_0(k) \mathrm{I}(X_i \geq k)\right) = \sum_{i : X_i > k} \left(\delta_i - \Lambda_0(X_i) + \Lambda_0(k)\right)
\end{equation*}
The second derivative of the log-likelihood:
\begin{equation*}
    \frac{\partial^2 \log(L)}{\partial \beta^2} = - \sum_{i : X_i \geq k} \left(\exp(\beta) \Lambda_0(X_i) - \exp(\beta) \Lambda_0(k)\right)
\end{equation*}
Evaluate this second derivative at $\beta = 0$:
\begin{equation*}
    \left. \frac{\partial^2 \log(L)}{\partial \beta^2}\right| _{\beta = 0} = - \sum_{i : X_i \geq k} \left(\Lambda_0(X_i) - \Lambda_0(k)\right) = - \sum_{i : X_i > k} \left(\Lambda_0(X_i) - \Lambda_0(k)\right)
\end{equation*}
The score statistic for the case of delayed effect with one change-point $k$ is:
\begin{equation*}
    \mathrm{Z_{DE}} = \frac{\sum\limits_{i : X_i > k} \left(\delta_i - \Lambda_0(X_i) + \Lambda_0(k)\right)}{\sqrt{\sum\limits_{i : X_i > k} \left(\Lambda_0(X_i) - \Lambda_0(k)\right)}}
\end{equation*}

\subsection{Crossing hazards} \label{CH details}
Hazard function:
\begin{equation*}
    \lambda_1(t) = \exp(\beta)\lambda_0(t)\Lambda_0(t)^{\exp(\beta)-1}
\end{equation*}
Cumulative hazard function: 
\begin{equation*}
    \Lambda_1(t) = \Lambda_0(t)^{\exp(\beta)}
\end{equation*}
Survival function: 
\begin{equation*}
    \mathrm{S}_1(t) = exp(-\Lambda_0(t)^{\exp(\beta)})
\end{equation*}
Log-likelihood function:
\begin{align*}
    l & = \sum_{i = 1}^n \left(\delta_i \log(\lambda_1(X_i)) - \Lambda_1(X_i)\right) \\
    & = \sum_{i = 1}^n \left(\delta_i \log \left(\exp(\beta)\lambda_0(X_i)\Lambda_0(X_i)^{\exp(\beta)-1}\right) - \Lambda_0(X_i)^{\exp(\beta)}\right) \\
    & = \sum_{i = 1}^n \left(\delta_i \beta + \delta_i \log(\lambda_0(X_i)) + \delta_i \exp(\beta)\log(\Lambda_0(X_i)) - \delta_i \log(\Lambda_0(X_i)) - \Lambda_0(X_i)^{\exp(\beta)}\right)
\end{align*}
Derivative of the log-likelihood function:
\begin{equation*}
    \frac{\partial \log(L)}{\partial \beta} = \sum_{i = 1}^n \left(\delta_i + \exp(\beta) \delta_i \log(\Lambda_0(X_i)) - \Lambda_0(X_i)^{\exp(\beta)}\log(\Lambda_0(X_i))\right)
\end{equation*}
Evaluate this derivative at $\beta = 0$:
\begin{equation*}
    \left. \frac{\partial \log(L)}{\partial \beta} \right| _{\beta = 0} = \sum_{i = 1}^n \left(\delta_i + \delta_i \log(\Lambda_0(X_i)) - \Lambda_0(X_i)\log(\Lambda_0(X_i))\right)
\end{equation*}
The second derivative of the log-likelihood function:
\begin{equation*}
   \frac{\partial^2 \log(L)}{\partial \beta^2} = \sum_{i = 1}^n \left(\exp(\beta) \delta_i \log(\Lambda_0(X_i)) - \Lambda_0(X_i)^{\exp(\beta)}\{1+\log(\Lambda_0(X_i))\}\log(\Lambda_0(X_i))\right)
\end{equation*}
Evaluate this second derivative at $\beta = 0$:
\begin{equation*}
    \left. \frac{\partial^2 \log(L)}{\partial \beta^2}\right| _{\beta = 0} = \sum_{i = 1}^n \{\delta_i - \Lambda_0(X_i)(1+\log(\Lambda_0(X_i)))\}\log(\Lambda_0(X_i))
\end{equation*}
The score statistic for the case of crossing hazards with no pre-specified change-point is:
\begin{equation*}
    \mathrm{Z_{CH}} = \frac{\sum\limits_{i=1}^n \left(\delta_i - \left(\Lambda_0(X_i) - \delta_i\right) \log(\Lambda_0(X_i))\right)}{\sqrt{-\sum\limits_{i = 1}^n \left[\delta_i - \Lambda_0(X_i)(1+\log(\Lambda_0(X_i)))\right]\log(\Lambda_0(X_i))}}
\end{equation*}

\newpage

\section{Simulations}

\subsection{Sampling variability} \label{Sampling_details}

Danzer et al \citep{Danzer2022_ref} reformulate the OSLRT with stochastic processes to include the variability of the external control group in the variance of the statistic, based on their previous work about the variance estimation \citep{Danzer2022_var}:
\begin{equation*}
    \mathrm{Z = \frac{\hat{M_0}(s_{max})}{\hat{\Sigma}(s_{max})}} = \frac{n_B^{-1/2}\{\mathrm{N}_B(s_{max}) - \sum_{i \in \mathcal{N}_B} \hat{\Lambda}_A(s_{max} \wedge X_{B,i})\}}{\sqrt{n_B^{-1} \sum_{i \in \mathcal{N}_B} \hat{\Lambda}_A(s_{max} \wedge X_{B,i}) + n_B^{-1} n_A^{-1} \sum_{i,j \in \mathcal{N}_B} \hat{\sigma}_A^2 (s_{max} \wedge X_{B,i} \wedge X_{B,j})}}
\end{equation*}
where $n_B$ and $n_A$ the number of patients in the experimental and in the external control groups; $\mathcal{N}_B$ is the set of patients in the experimental group; $X_{B, i}$ is the observed failure time for patient i in the experimental group; $\mathrm{N}_B(s) = \sum_{i \in \mathcal{N}_B} \mathrm{I}(T_{B,i} \leq s, T_{B,i} \leq C_{B,i})$ is the number of events in the experimental group; and
$\hat{\Lambda}_A(s)$ is the Nelson-Aalen estimator of the cumulative hazard function of the external control group with $\hat{\sigma}_A^2(s)$ its corresponding estimator of the variance. \\
However, this method requires the individual patient data of the external control group, so an approximation \citep{Danzer2022_ref, Feld2024} is proposed based on the ratio of the group sizes: $\pi = \frac{n_{\mathrm{exp}}}{n_{\mathrm{control}}}$ where $n_{\mathrm{exp}}$ and $n_{\mathrm{control}}$ are respectively the number of patients in the experimental and in the external control group. Then they approximated the factor of under-estimation of the variance of the OSLRT and mOSLRT by:
\begin{equation*}
   \mathrm{R}^2 = \frac{1}{1+\pi}
\end{equation*}
with the assumption that the censoring mechanism is the same in both groups. Thus, they defined a new survival test statistic:
\begin{equation*}
    \mathrm{Z}_{\pi} = \frac{\hat{M}_0(s_{max})}{\hat{\Sigma}_{\mathrm{OSLR}}(s_{max})\sqrt{1+\pi}}
\end{equation*}
where $\hat{M}_0(s) = n_{exp}^{-1/2}\{\sum_{i\in \mathcal{N}_{exp}} \mathrm{I}(T_i\leq s, T_i \leq C_i)-\sum_{i \in \mathcal{N}_{exp}} \hat{\Lambda}_0(s\wedge X_{i})\}$ and the estimator of the variance $\mathrm{Var}(\hat{M}_0(s))$ is $\hat{\Sigma}_{\mathrm{OSLR}}^2(s)=\frac{1}{2}n_{exp}^{-1}\{\sum_{i\in \mathcal{N}_{exp}} \mathrm{I}(T_i\leq s, T_i \leq C_i)+\sum_{i\in\mathcal{N}_{exp}} \hat{\Lambda}_0(s\wedge X_{i})\}$ that are respectively equivalent to $n_{\mathrm{exp}}^{-1/2}\mathrm{(O-E)}$ and $\frac{\mathrm{O+E}}{2n_{\mathrm{exp}}}$ as demonstrated by Wu \citep{Wu2021}.

\newpage

\subsection{Parameters}

\begin{figure*}[!h]
    \centerline{\includegraphics[width=0.93\textwidth]{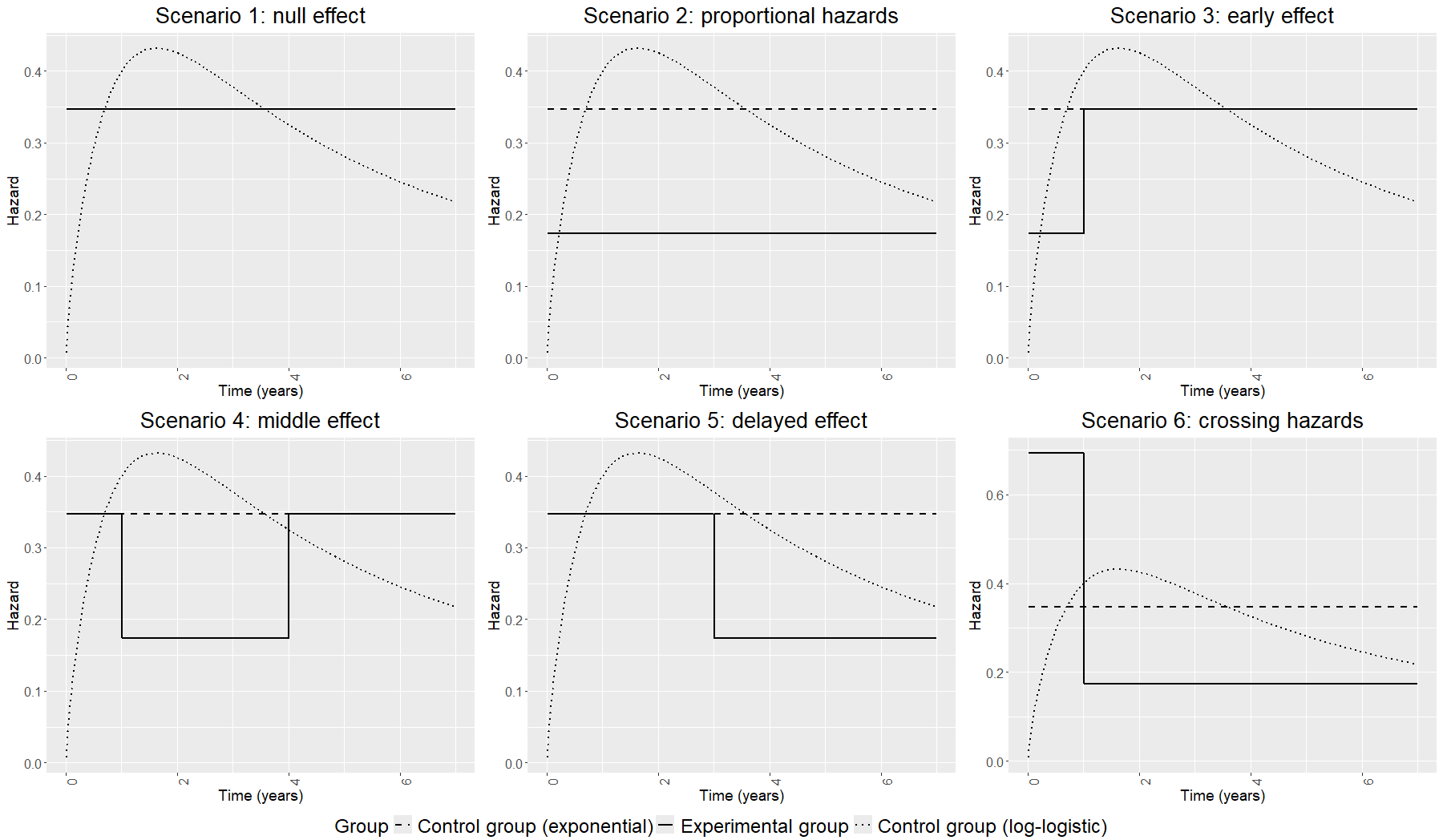}}
    \caption{True hazard functions under different scenarios in single-arm trials: scenario 1 corresponds to null treatment effect, scenario 2 to proportional hazards, scenarios 3-6 to early, middle, delayed and crossing treatment effects, respectively. Dashed and dotted curves represent the hazard function of the external control group simulated using exponential (dashed line) and log-logistic (dotted line) models. The solid curve represents the hazard function generated from a piecewise exponential model.
    \label{Hazards}}
\end{figure*}

\begin{figure*}[!h]
    \centerline{\includegraphics[width=0.7\textwidth]{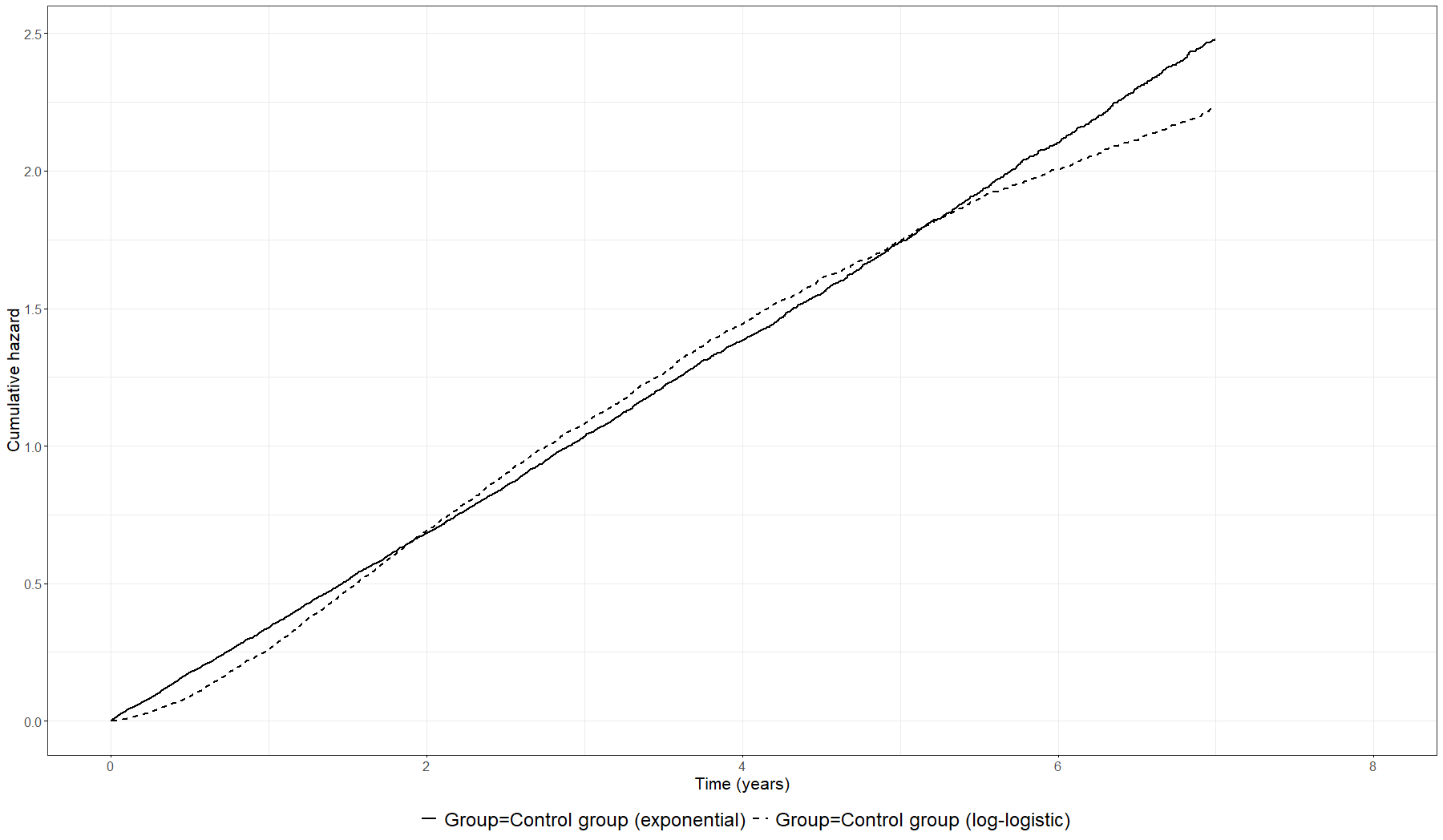}}
    \caption{True cumulative hazard functions of the external control group simulated using an exponential model (solid line) and a log-logistic model (dashed line). 
    \label{Hazards_cum}}
\end{figure*}

\newpage

\subsection{Results}

\begin{figure*}[!h]
    \centerline{\includegraphics[width=\textwidth]{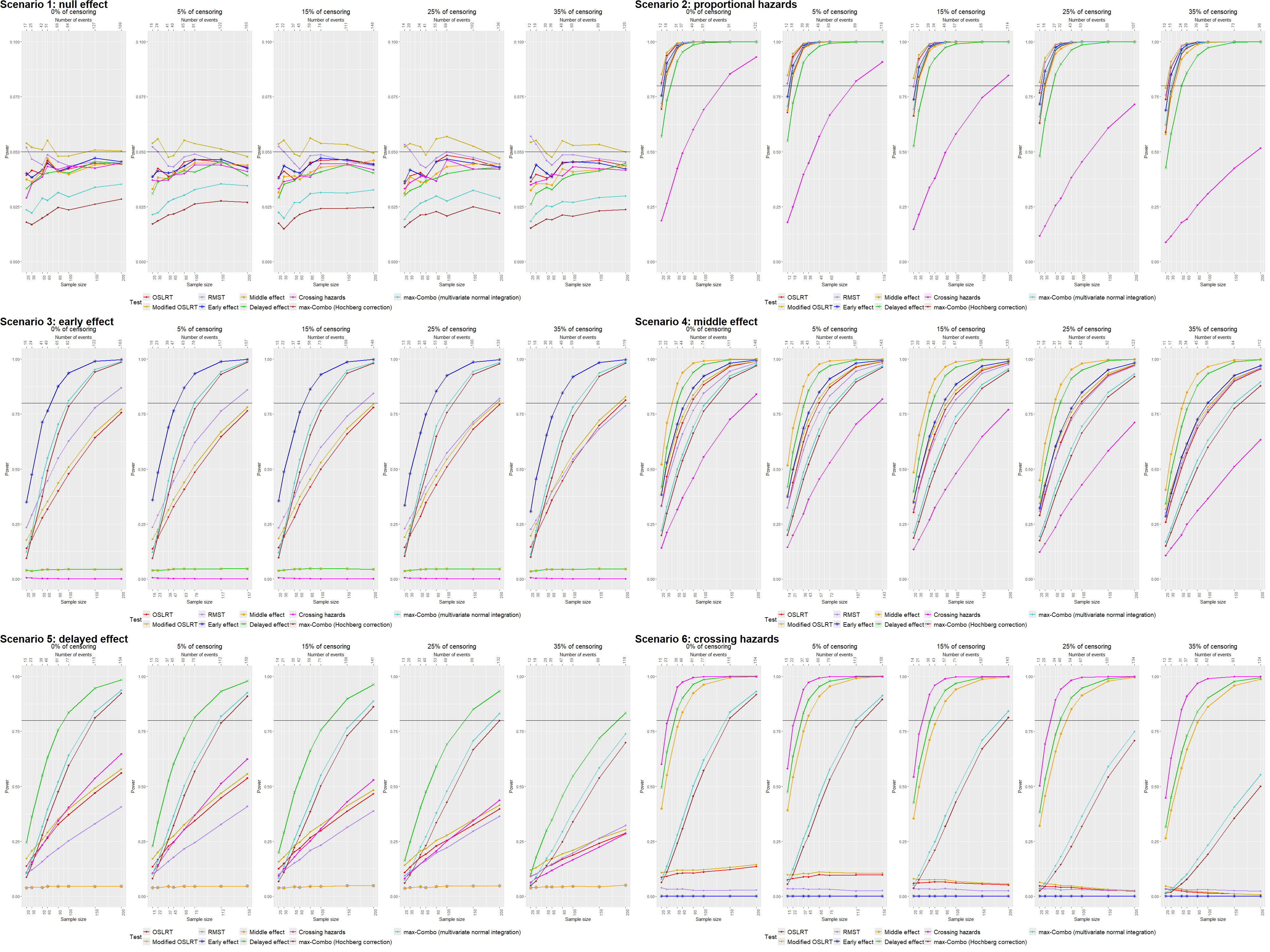}}
    \caption{Type I error (scenario 1) and power (scenarios 2-6) for the OSLRT, mOSLRT and developed score tests for an early ($\mathrm{Z_{EE}}$ with $k = 4$ for scenarios 1-2, $k = 1$ for scenarios 3 and 6, $k = 4$ for scenario 4 and $k = 3$ for scenario 5), middle ($\mathrm{Z_{ME}}$ with $k_1 = 1$ and $k_2 = 6$ for scenarios 1-2, $k_1 = 1$ and $k_2 = 7$ for scenario 3, $k_1 = 1$ and $k_2 = 4$ for scenarios 4 and 6, $k_1 = 0$ and $k_2 = 3$ for scenario 5) and delayed effects ($\mathrm{Z_{DE}}$ with $k = 2$ for scenarios 1-2, $k = 1$ for scenarios 3-4 and 6 and $k = 3$ for scenario 5) and for the crossing hazards ($\mathrm{Z_{CH}}$), RMST-based test ($\tau = 7$) and the max-Combo test (Hochberg and multivariate normal integration) with a true HR of 0.5. Black horizontal lines indicate the nominal 5\% type I error (scenario 1) and 80\% power (scenarios 2-6).
    \label{Power_compl}}
\end{figure*}

\begin{figure*}[!h]
    \centerline{\includegraphics[width=\textwidth]{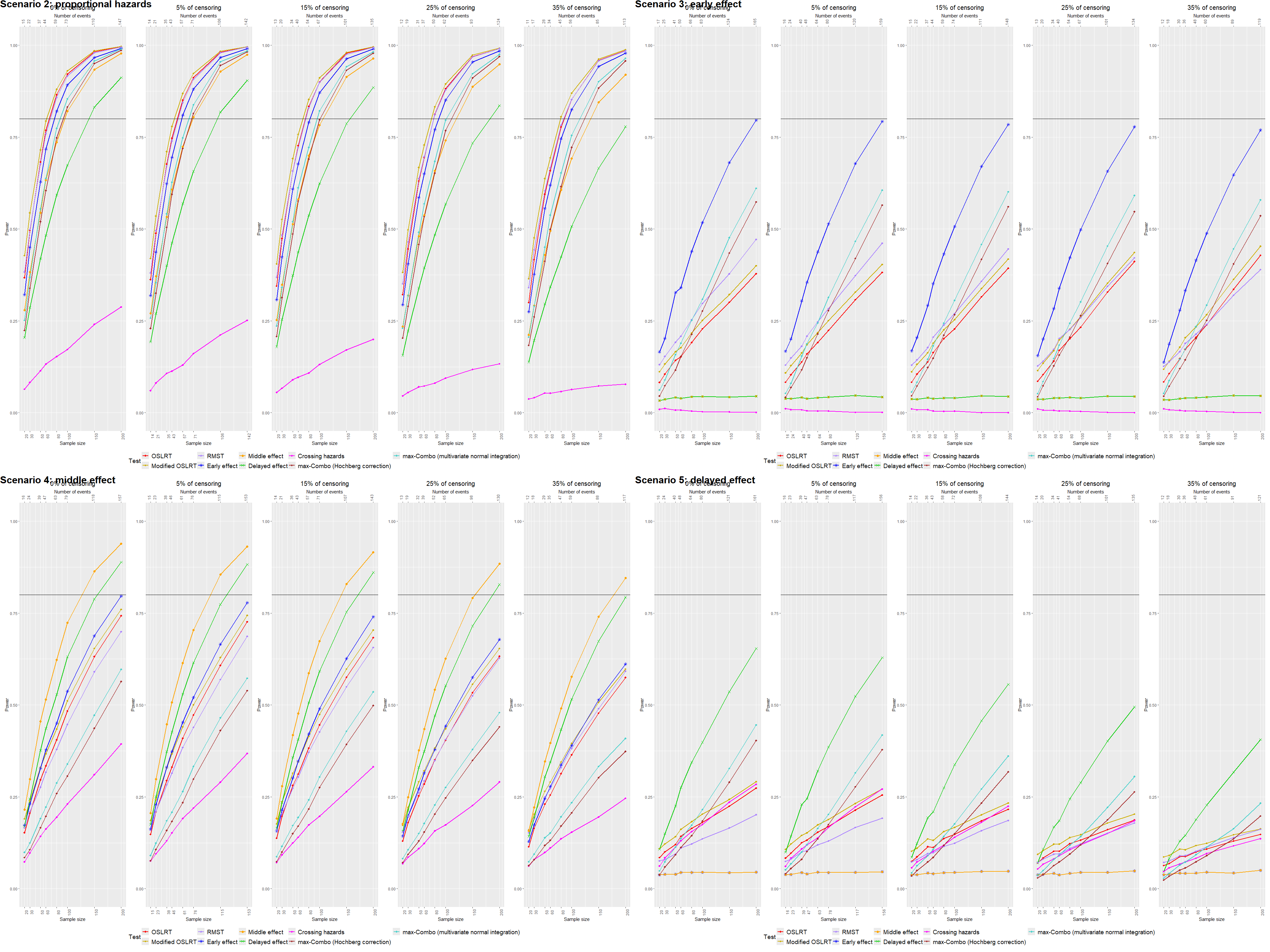}}
    \caption{Power (scenarios 2-5) for the OSLRT, mOSLRT, developed score tests for an early ($\mathrm{Z_{EE}}$ with $k = 4$ for scenarios 2, $k = 1$ for scenarios 3, $k = 4$ for scenario 4 and $k = 3$ for scenario 5), middle ($\mathrm{Z_{ME}}$ with $k_1 = 1$ and $k_2 = 6$ for scenarios 2, $k_1 = 1$ and $k_2 = 7$ for scenario 3, $k_1 = 1$ and $k_2 = 4$ for scenarios 4, $k_1 = 0$ and $k_2 = 3$ for scenario 5) and delayed effects ($\mathrm{Z_{DE}}$ with $k = 2$ for scenarios 2, $k = 1$ for scenarios 3-4 and $k = 3$ for scenario 5), crossing hazards ($\mathrm{Z_{CH}}$), RMST-based test ($\tau = 7$) and max-Combo test (Hochberg and multivariate normal integration) with a true HR of 0.7. Black horizontal lines indicate the 80\% power.
    \label{Power_HR0.7}}
\end{figure*}

\begin{figure*}[!h]
    \centerline{\includegraphics[width=\textwidth]{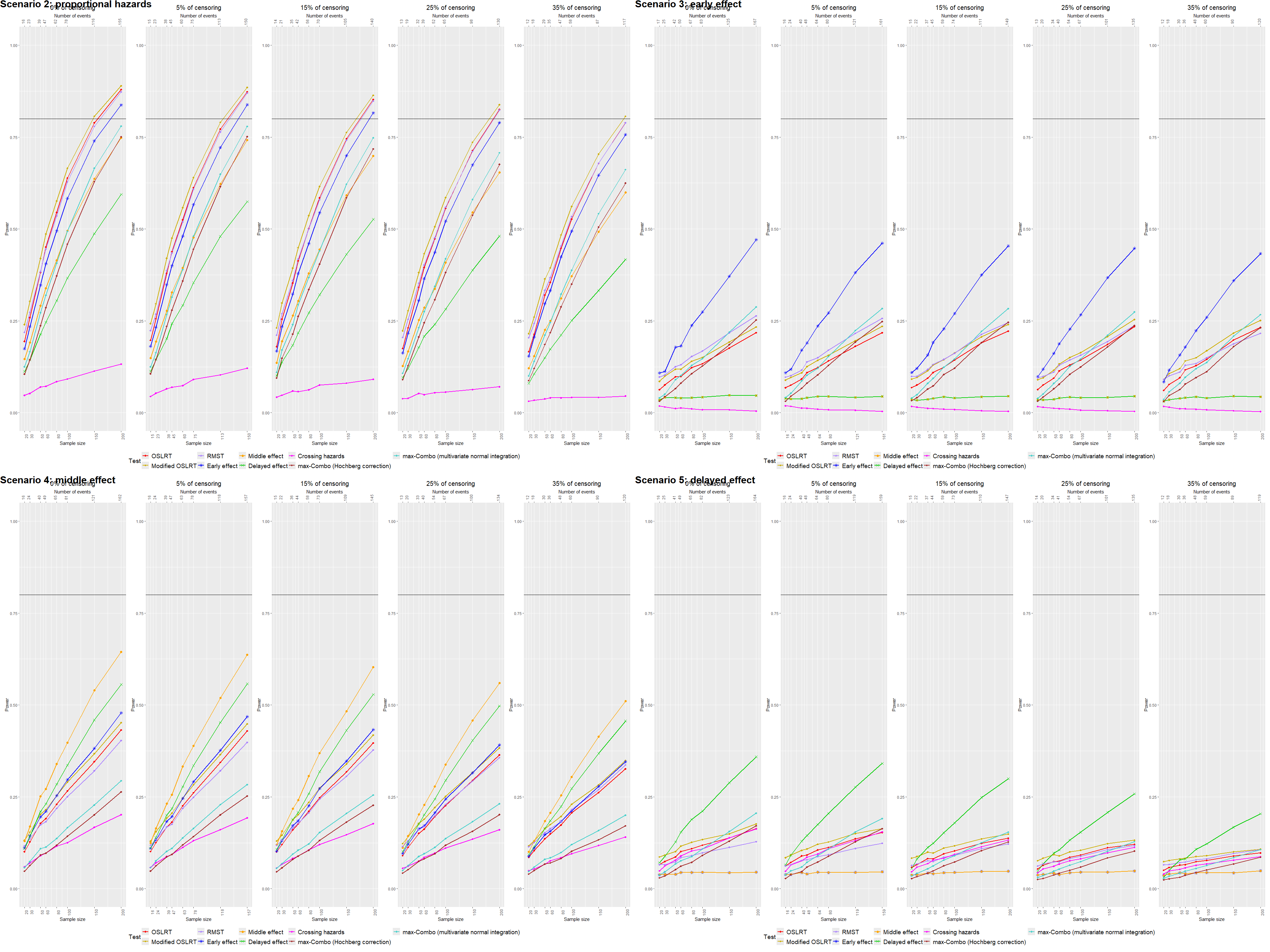}}
    \caption{Power (scenarios 2-5) for the OSLRT, mOSLRT, developed score tests for an early ($\mathrm{Z_{EE}}$ with $k = 4$ for scenarios 2, $k = 1$ for scenarios 3, $k = 4$ for scenario 4 and $k = 3$ for scenario 5), middle ($\mathrm{Z_{ME}}$ with $k_1 = 1$ and $k_2 = 6$ for scenarios 2, $k_1 = 1$ and $k_2 = 7$ for scenario 3, $k_1 = 1$ and $k_2 = 4$ for scenarios 4, $k_1 = 0$ and $k_2 = 3$ for scenario 5) and delayed effects ($\mathrm{Z_{DE}}$ with $k = 2$ for scenarios 2, $k = 1$ for scenarios 3-4 and $k = 3$ for scenario 5), crossing hazards ($\mathrm{Z_{CH}}$), RMST-based test ($\tau = 7$) and max-Combo test (Hochberg and multivariate normal integration) with a true HR of 0.8. Black horizontal lines indicate the nominal 80\% power.
    \label{Power_HR0.8}}
\end{figure*}

\begin{figure*}[!h]
    \centerline{\includegraphics[width=\textwidth]{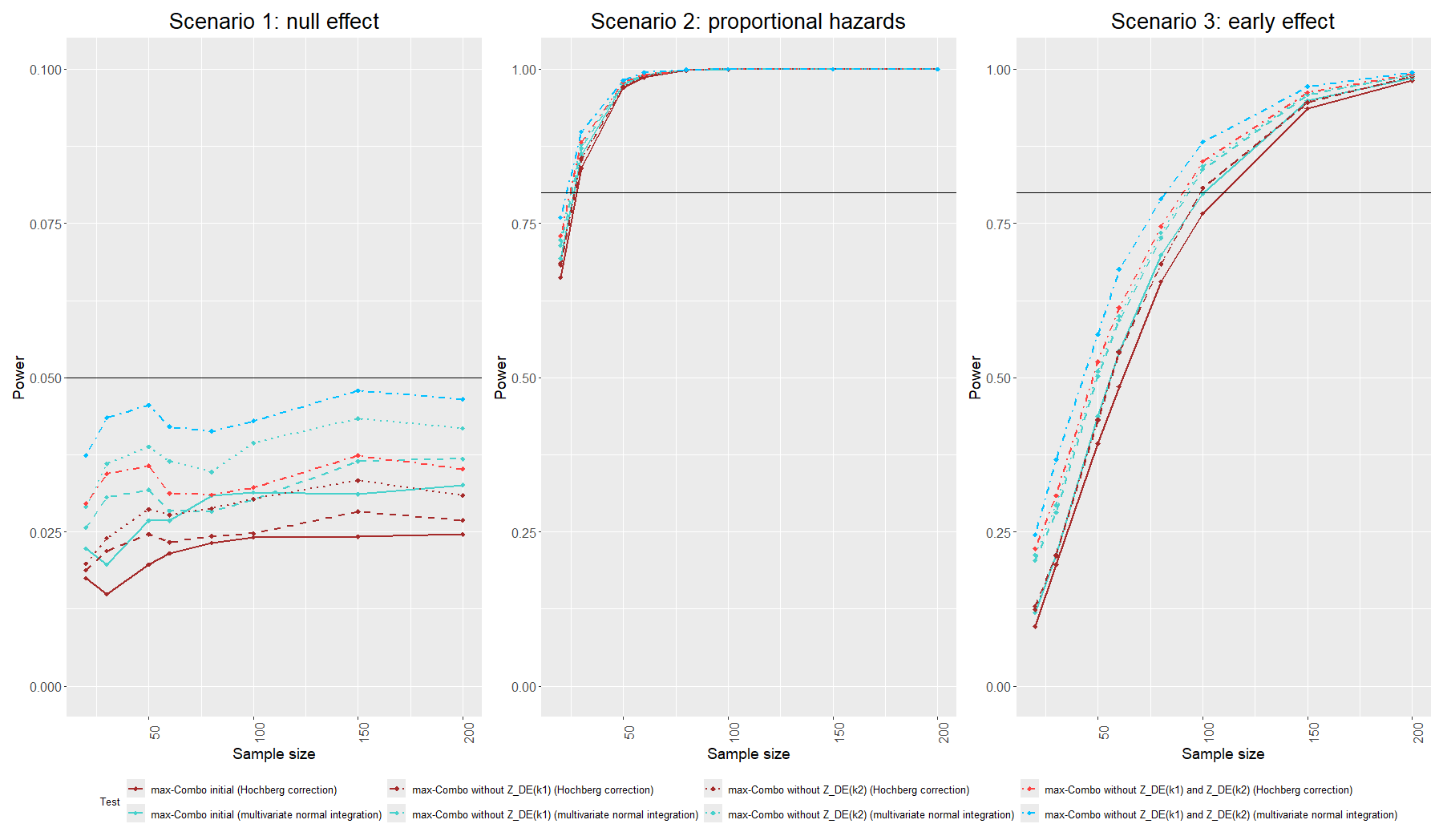}}
    \caption{Impact of removing some worse-performing statistics ($\mathrm{Z_{DE}}(k_1' = 3)$, $\mathrm{Z_{DE}}(k_2' = 5)$ and both) in the max-Combo test for null effect, proportional hazards and early effect scenarios.}
    \label{DE_remove}
\end{figure*}

\begin{figure*}[!h]
    \centerline{\includegraphics[width=\textwidth]{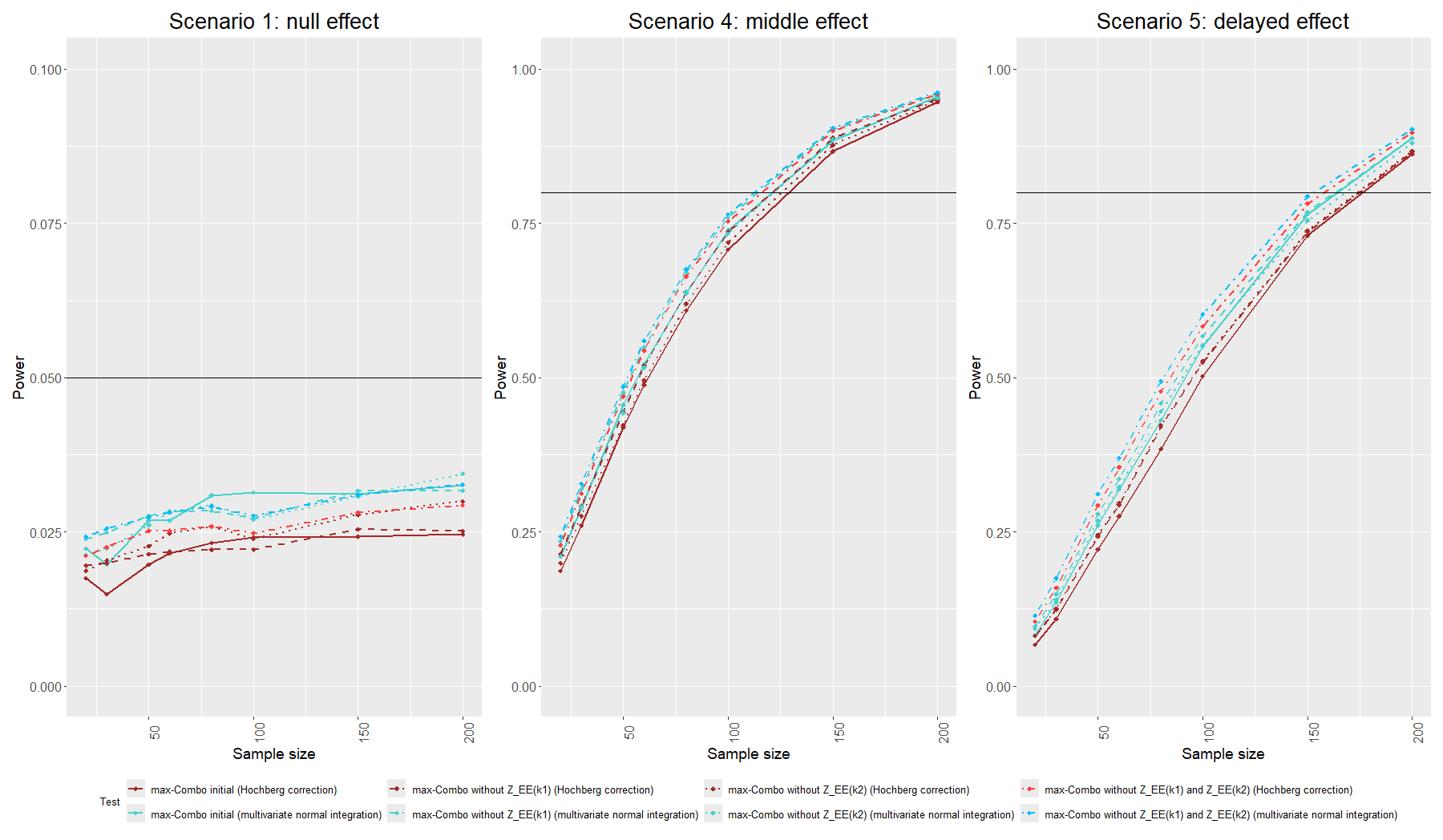}}
    \caption{Impact of removing some worse-performing statistics ($\mathrm{Z_{EE}}(k_1 = 1)$, $\mathrm{Z_{EE}}(k_2 = 3)$ and both) in the max-Combo test for null effect, middle effect and delayed effect.}
    \label{EE_remove}
\end{figure*}

\begin{figure*}[!h]
    \centerline{\includegraphics[width=\textwidth]{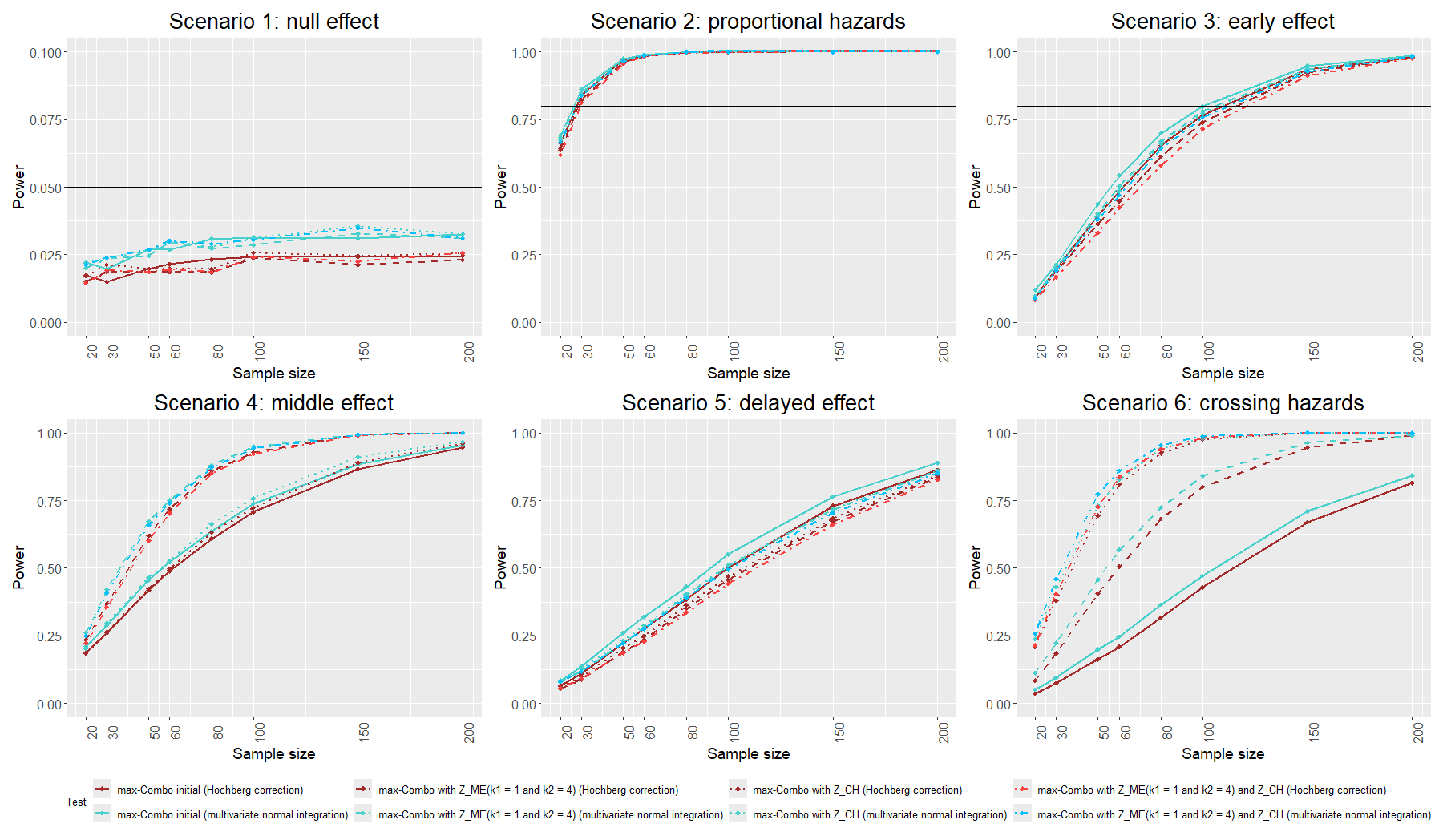}}
    \caption{Impact of adding some statistics ($\mathrm{Z_{ME}}(k_1 = 1, k_2 = 4)$, $\mathrm{Z_{CH}}$ and both) in the max-Combo test.}
    \label{ME_CH_add}
\end{figure*}

\begin{figure*}[!h]
    \centerline{\includegraphics[width=\textwidth]{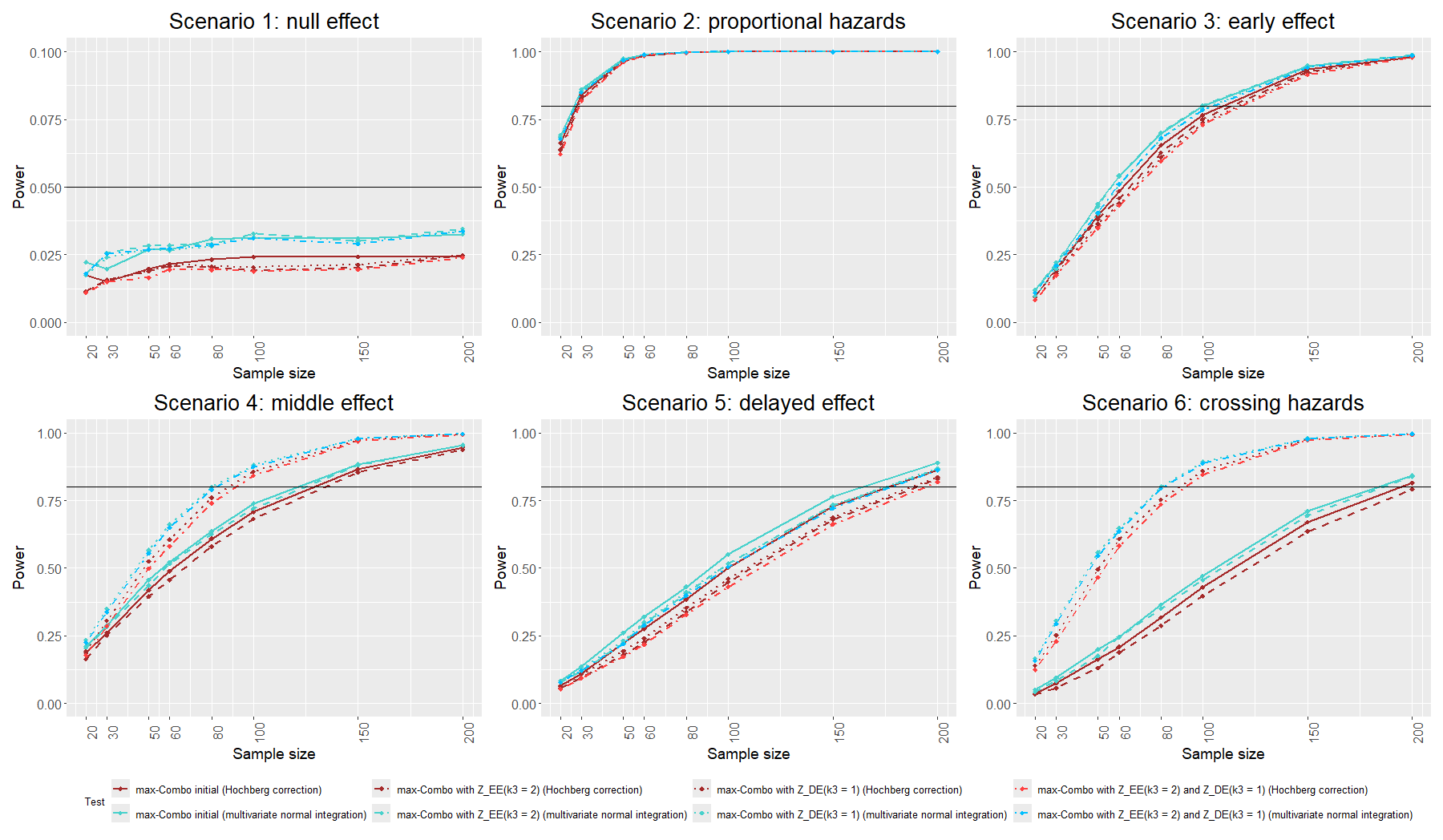}}
    \caption{Impact of adding some statistics ($\mathrm{Z_{EE}}(k_3 = 2)$, $\mathrm{Z_{DE}}(k_3' = 1)$ and both) in the max-Combo test.}
    \label{EE_DE_add}
\end{figure*}

\begin{figure*}[!h]
    \centerline{\includegraphics[width=\textwidth]{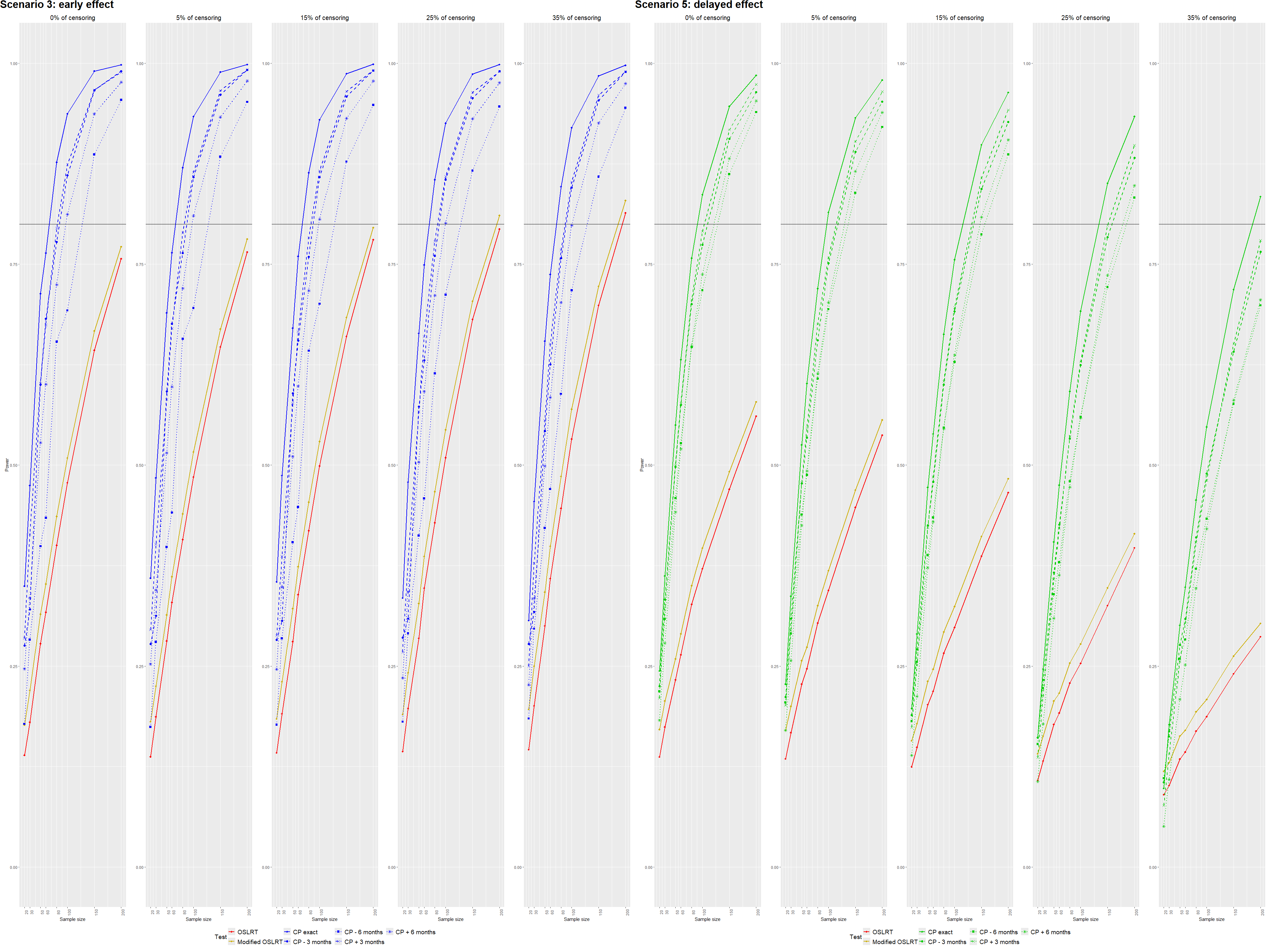}}
    \caption{Impact of change-point misspecification on the power of the early and delayed effect score tests with a true HR of 0.5. This misspecification is evaluated using four new change-points derived from the true value: $k_1 = k-3$, $k_2 = k+3$, $k_3 = k-6$ and $k_4 = k+6$ months.}
    \label{CPs misspe comp}
\end{figure*}

\begin{figure*}[!h]
    \centerline{\includegraphics[width=0.7\textwidth]{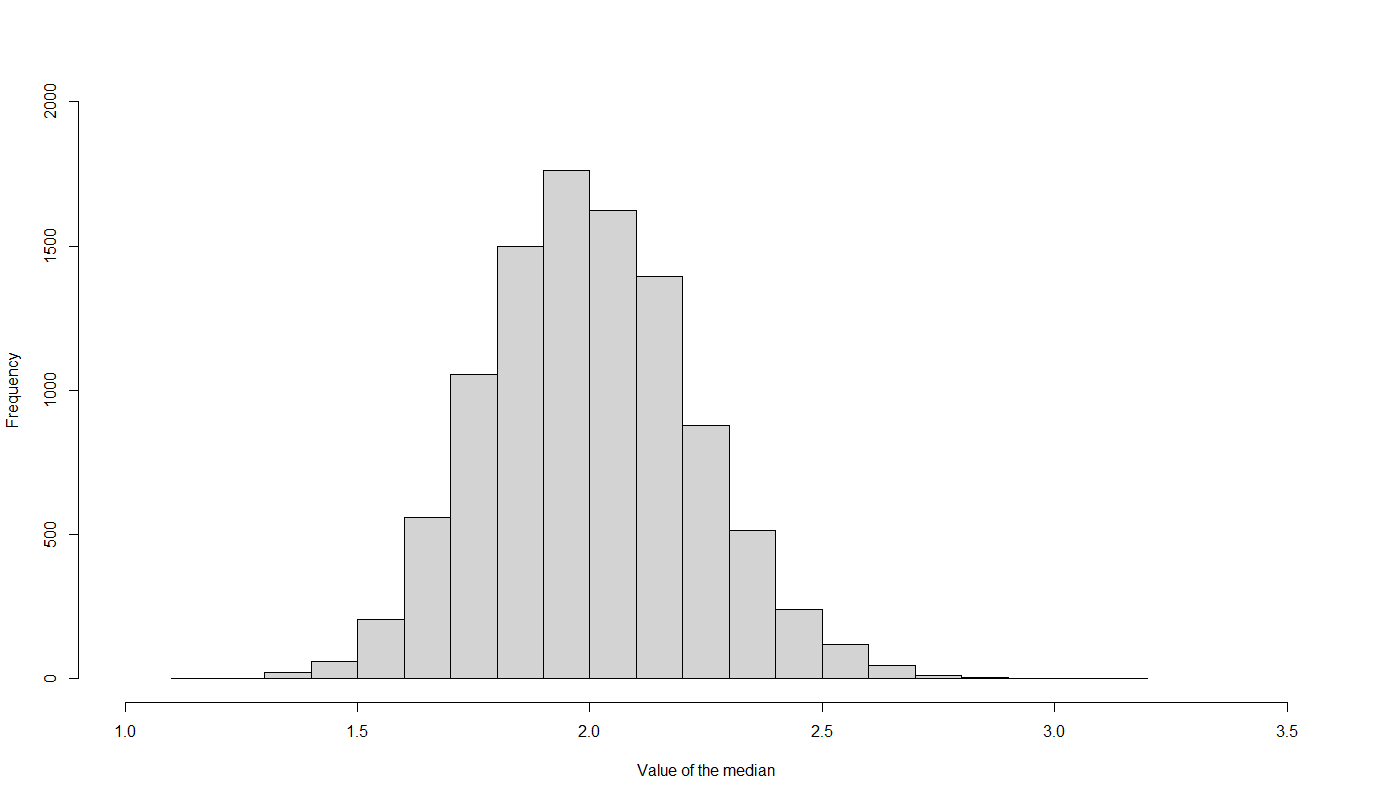}}
    \caption{Distribution of the median survival in the external control group generated from a Gamma distribution $\Gamma(80,40)$.}
    \label{Med_miss}
\end{figure*}

\begin{figure*}[!h]
    \centerline{\includegraphics[width=0.7\textwidth]{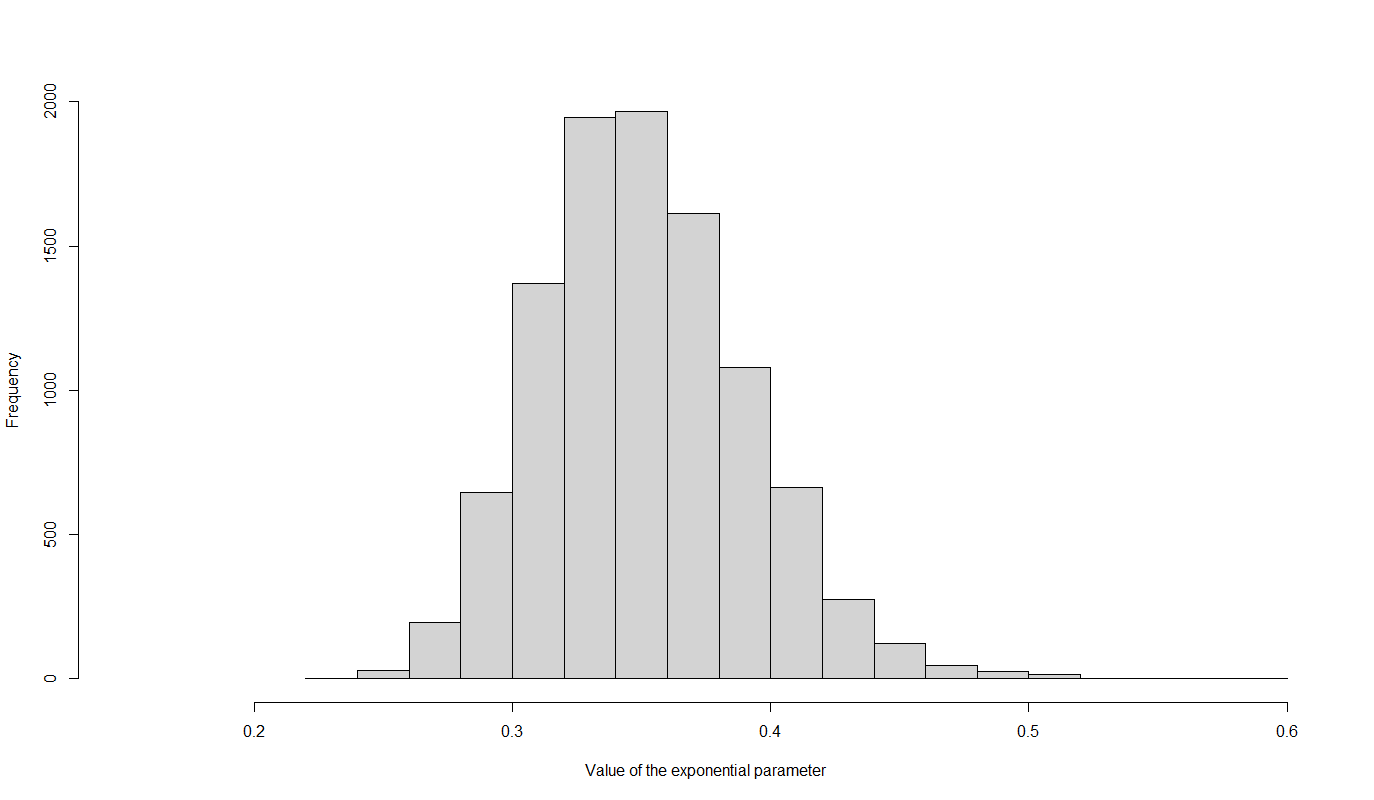}}
    \caption{Distribution of the exponential parameter of the external control group generated from an inverse Gamma distribution $\mathrm{IG}(80, \log(2)x40)$.}
    \label{Lambda_miss}
\end{figure*}

\begin{figure*}[!h]
\centerline{\includegraphics[width=\textwidth]{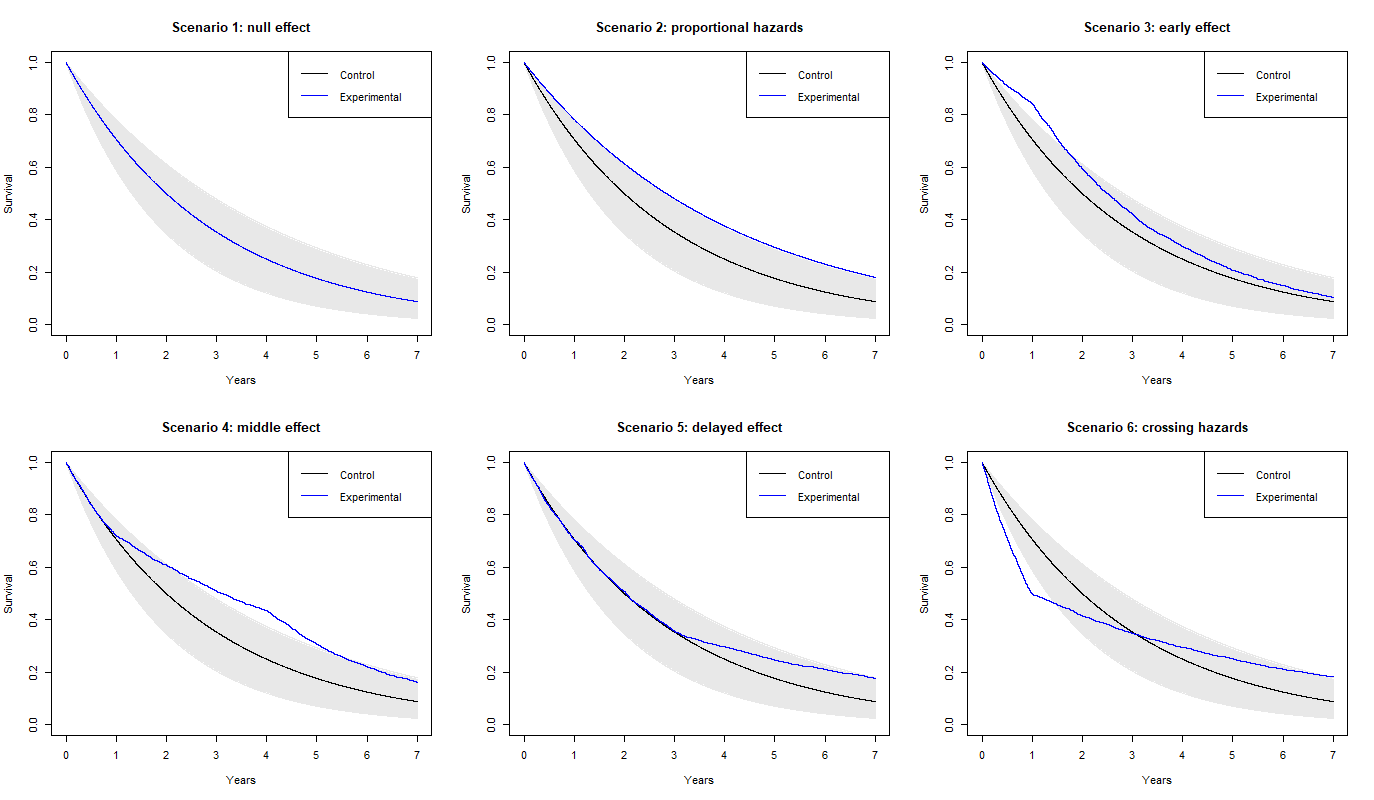}}
\caption{True survival curves under different scenarios: scenario 1 corresponds to null treatment effect, scenario 2 to proportional hazards, scenarios 3-6 to early, middle, delayed and crossing treatment effect. The black curve represents the external control group survival simulated using an exponential model. The blue curve represents the survival curve generated from a piecewise exponential model (scale parameter = $-\log(0.5)/2$). The grey curves correspond each replication of the external control group when the median survival time is generated from a gamma distribution.
\label{Survival_misspe}}
\end{figure*}

\begin{figure*}[!h]
    \centerline{\includegraphics[width=\textwidth]{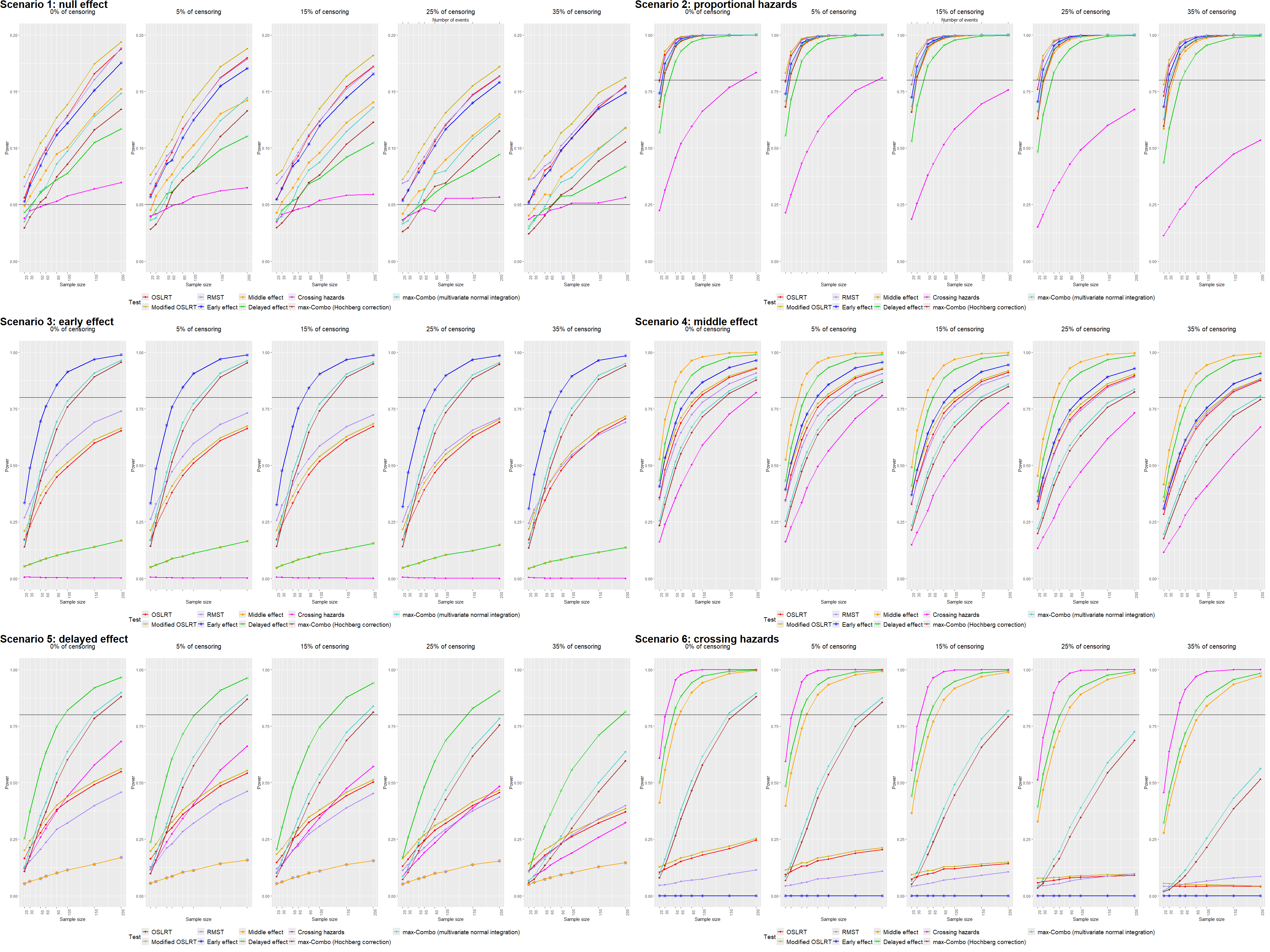}}
    \caption{Type I error (scenario 1) and power (scenario 2-6) for the OSLRT, mOSLRT, developed score tests for an early ($\mathrm{Z_{EE}}$ with $k = 4$ for scenarios 1-2, $k = 1$ for scenarios 3 and 6, $k = 4$ for scenario 4 and $k = 3$ for scenario 5), middle ($\mathrm{Z_{ME}}$ with $k_1 = 1$ and $k_2 = 6$ for scenarios 1-2, $k_1 = 1$ and $k_2 = 7$ for scenario 3, $k_1 = 1$ and $k_2 = 4$ for scenarios 4 and 6, $k_1 = 0$ and $k_2 = 3$ for scenario 5) and delayed effects ($\mathrm{Z_{DE}}$ with $k = 2$ for scenarios 1-2, $k = 1$ for scenarios 3-4 and 6 and $k = 3$ for scenario 5), crossing hazards ($\mathrm{Z_{CH}}$), RMST-based test ($\tau=7$) and max-Combo test (Hochberg and multivariate normal integration) under uncertainties in the parameter $\lambda$ of the exponential distribution used to generate the external control group and with a true HR of 0.5. Black horizontal lines indicate the nominal 5\% type I error (scenario 1) and 80\% power (scenarios 2-6).
    \label{Power_sensitivity}}
\end{figure*}

\begin{figure*}[!h]
    \centerline{\includegraphics[width=\textwidth]{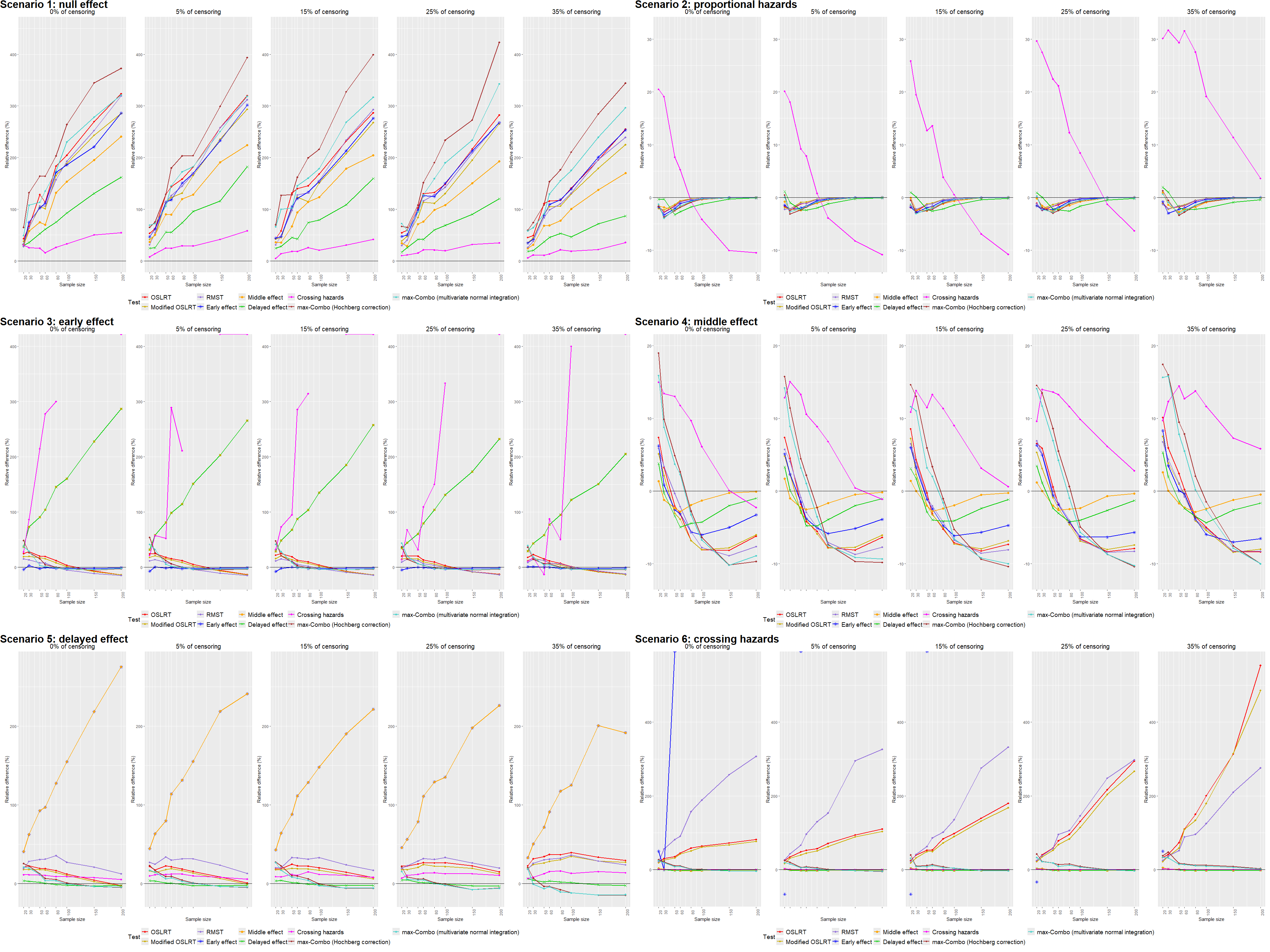}}
    \caption{Relative difference in type I error (scenario 1) and power (scenarios 2-6) between the case where uncertainty is introduced in the parameter of the exponential distribution of the external control group and the case where the true parameter is used, for the OSLRT, mOSLRT, developed score tests for an early ($\mathrm{Z_{EE}}$ with $k = 4$ for scenarios 1-2, $k = 1$ for scenarios 3 and 6, $k = 4$ for scenario 4 and $k = 3$ for scenario 5), middle ($\mathrm{Z_{ME}}$ with $k_1 = 1$ and $k_2 = 6$ for scenarios 1-2, $k_1 = 1$ and $k_2 = 7$ for scenario 3, $k_1 = 1$ and $k_2 = 4$ for scenarios 4 and 6, $k_1 = 0$ and $k_2 = 3$ for scenario 5) and delayed effects ($\mathrm{Z_{DE}}$ with $k = 2$ for scenarios 1-2, $k = 1$ for scenarios 3-4 and 6 and $k = 3$ for scenario 5), crossing hazards ($\mathrm{Z_{CH}}$), RMST-based test ($\tau$ = 7) and max-Combo test (Hochberg and multivariate normal integration) with a true HR of 0.5.}
    \label{Diff_sensibility_comp}
\end{figure*}

\begin{figure*}[!h]
    \centerline{\includegraphics[width=\textwidth]{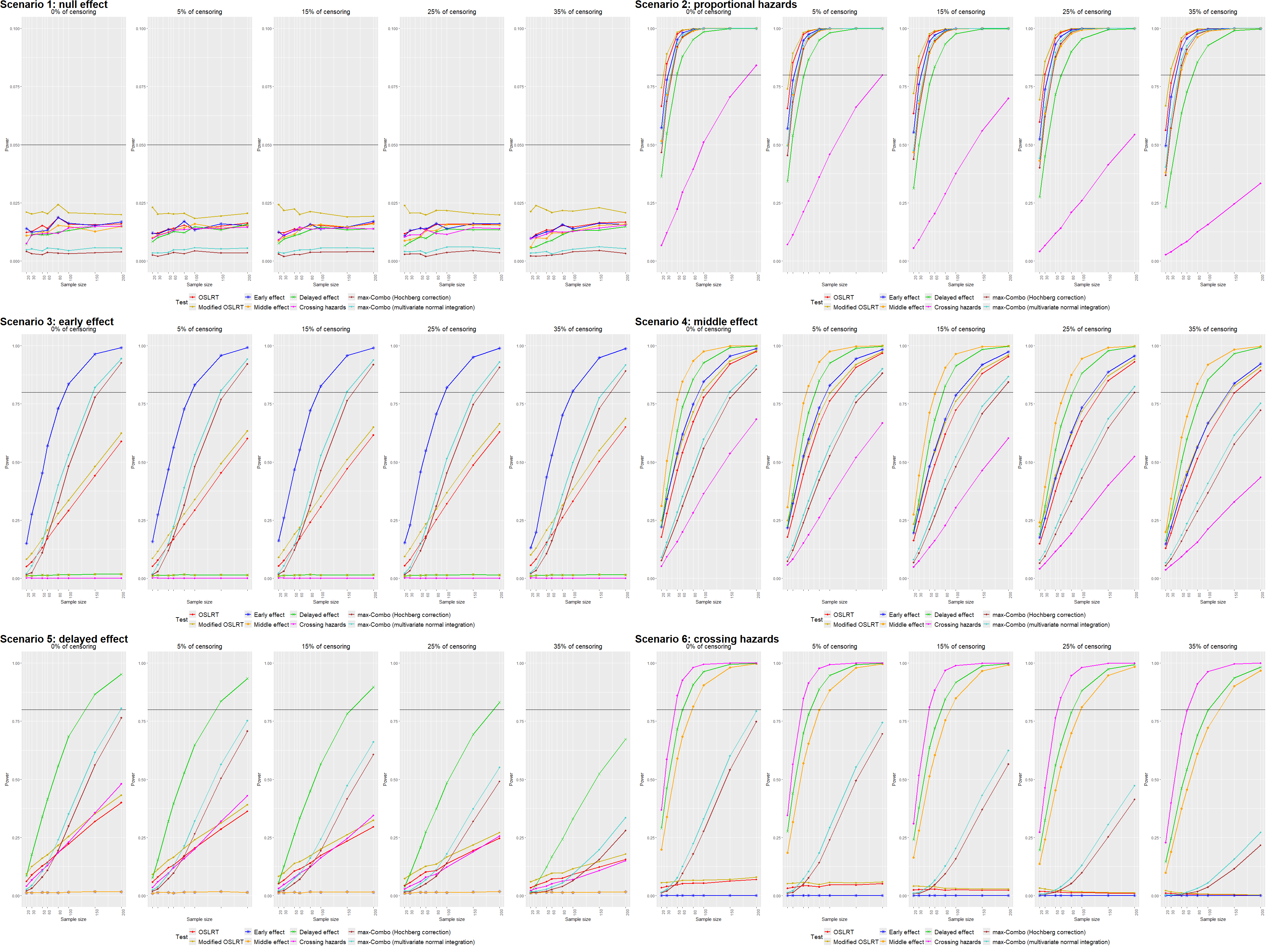}}
    \caption{Type I error (scenario 1) and power (scenario 2-6) for the OSLRT, mOSLRT, developed score tests for an early ($\mathrm{Z_{EE}}$ with $k = 4$ for scenarios 1-2, $k = 1$ for scenarios 3 and 6, $k = 4$ for scenario 4 and $k = 3$ for scenario 5), middle ($\mathrm{Z_{ME}}$ with $k_1 = 1$ and $k_2 = 6$ for scenarios 1-2, $k_1 = 1$ and $k_2 = 7$ for scenario 3, $k_1 = 1$ and $k_2 = 4$ for scenarios 4 and 6, $k_1 = 0$ and $k_2 = 3$ for scenario 5) and delayed effects ($\mathrm{Z_{DE}}$ with $k = 2$ for scenarios 1-2, $k = 1$ for scenarios 3-4 and 6 and $k = 3$ for scenario 5), crossing hazards ($\mathrm{Z_{CH}}$) and max-Combo test (Hochberg and multivariate normal integration) including the sampling variability of the external control group in the tests with a true HR of 0.5 and $\pi = 0.6$. Black horizontal lines indicate the nominal 5\% type I error (scenario 1) and 80\% power (scenarios 2-6).
    \label{Power_variability}}
\end{figure*}

\begin{figure*}[!h]
\centerline{\includegraphics[width=\textwidth]{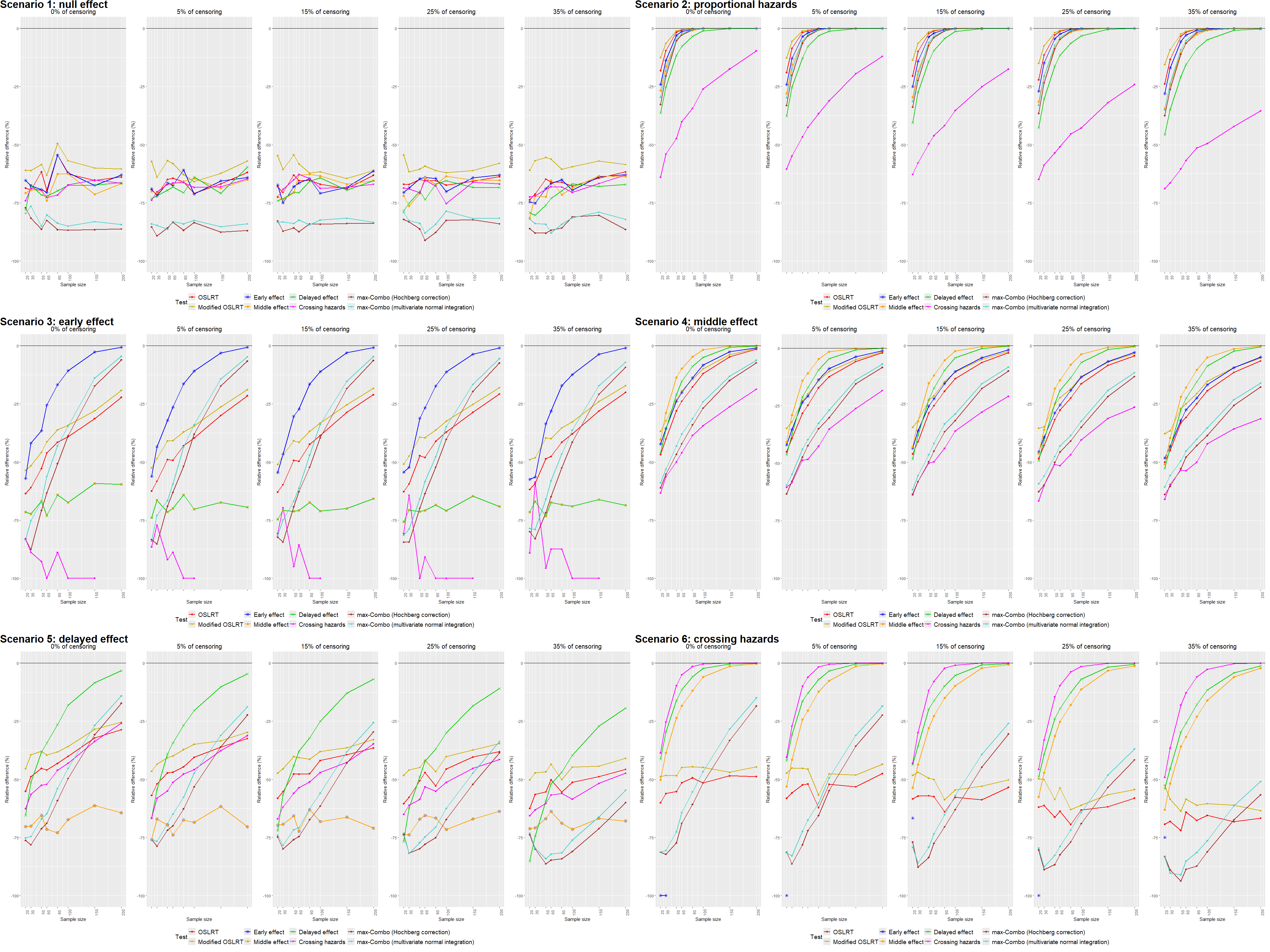}}
\caption{Relative difference in type I error (scenario 1) and power (scenarios 2-6) between the case where sampling variability of the external control group is included and the case where it is not included, for the OSLRT, mOSLRT, developed score tests for an early ($\mathrm{Z_{EE}}$ with $k = 4$ for scenarios 1-2, $k = 1$ for scenarios 3 and 6, $k = 4$ for scenario 4 and $k = 3$ for scenario 5), middle ($\mathrm{Z_{ME}}$ with $k_1 = 1$ and $k_2 = 6$ for scenarios 1-2, $k_1 = 1$ and $k_2 = 7$ for scenario 3, $k_1 = 1$ and $k_2 = 4$ for scenarios 4 and 6, $k_1 = 0$ and $k_2 = 3$ for scenario 5) and delayed effects ($\mathrm{Z_{DE}}$ with $k = 2$ for scenarios 1-2, $k = 1$ for scenarios 3-4 and 6 and $k = 3$ for scenario 5), crossing hazard ($\mathrm{Z_{CH}}$) and max-Combo test (Hochberg and multivariate normal integration) with a true HR of 0.5 and $\pi = 0.6$.
\label{Diff_variability_comp}}
\end{figure*}

\begin{figure*}[!h]
    \centerline{\includegraphics[width=\textwidth]{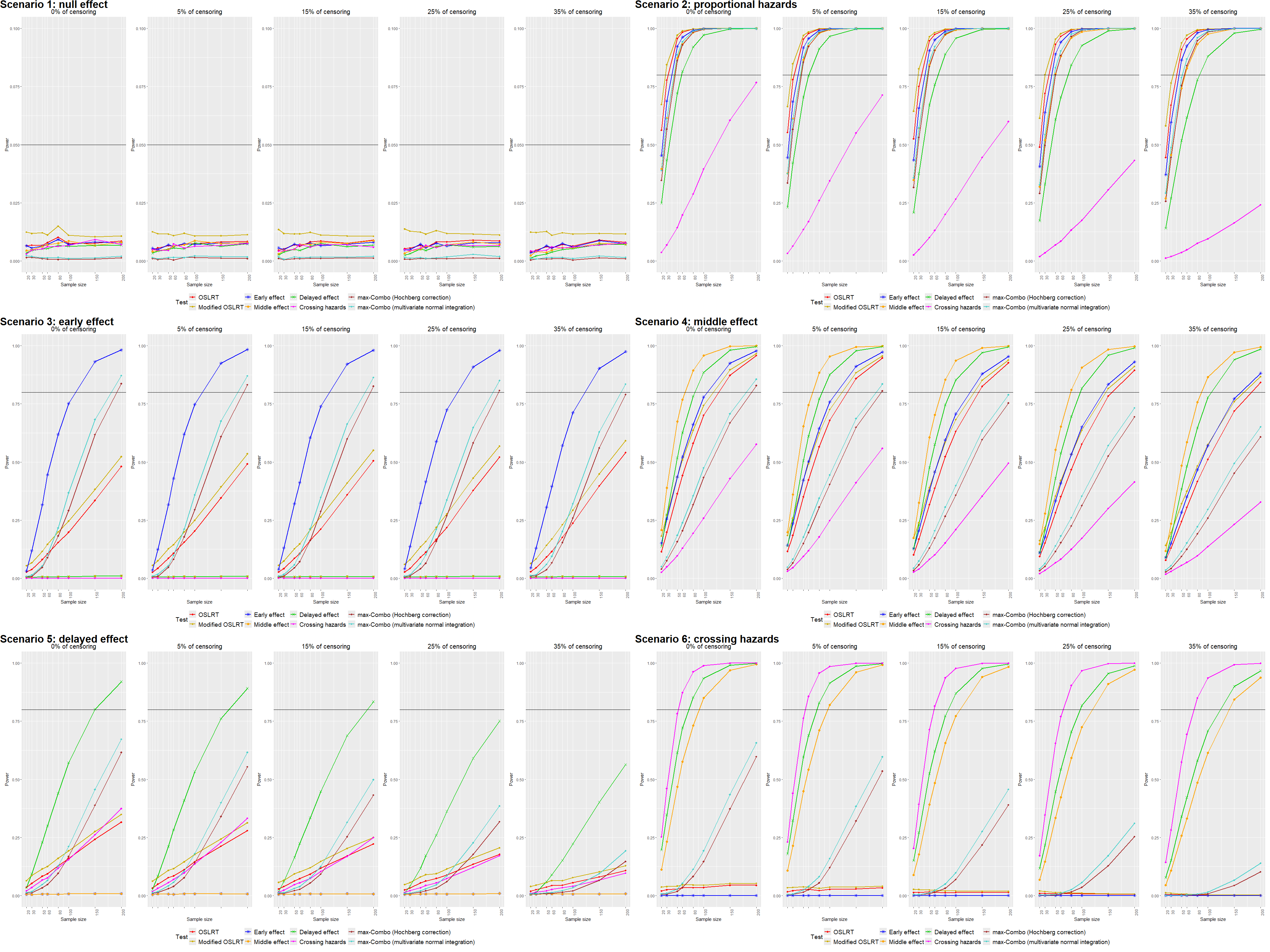}}
    \caption{Type I error (scenario 1) and power (scenario 2-6) for the OSLRT, mOSLRT, developed score tests for an early ($\mathrm{Z_{EE}}$ with $k = 4$ for scenarios 1-2, $k = 1$ for scenarios 3 and 6, $k = 4$ for scenario 4 and $k = 3$ for scenario 5), middle ($\mathrm{Z_{ME}}$ with $k_1 = 1$ and $k_2 = 6$ for scenarios 1-2, $k_1 = 1$ and $k_2 = 7$ for scenario 3, $k_1 = 1$ and $k_2 = 4$ for scenarios 4 and 6, $k_1 = 0$ and $k_2 = 3$ for scenario 5) and delayed effects ($\mathrm{Z_{DE}}$ with $k = 2$ for scenarios 1-2, $k = 1$ for scenarios 3-4 and 6 and $k = 3$ for scenario 5), crossing hazards ($\mathrm{Z_{CH}}$) and max-Combo test (Hochberg and multivariate normal integration) including the sampling variability of the external control group in the tests with a true HR of 0.5 and $\pi = 1$. Black horizontal lines indicate the nominal 5\% type I error (scenario 1) and 80\% power (scenarios 2-6).
    \label{Power_variability_pi1}}
\end{figure*}

\begin{figure*}[!h]
\centerline{\includegraphics[width=\textwidth]{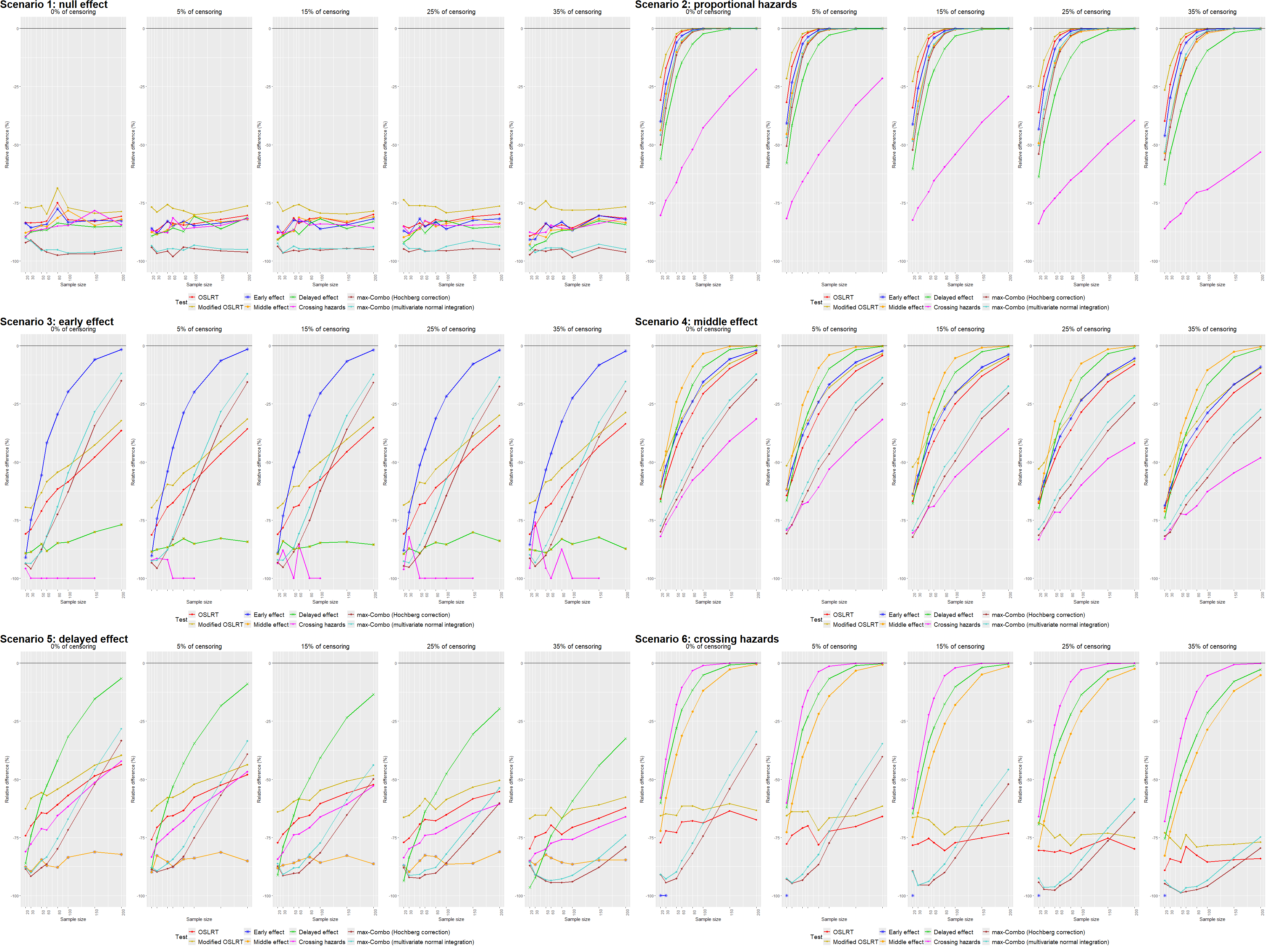}}
\caption{Relative difference in type I error (scenario 1) and power (scenarios 2-6) between the case where the sampling variability of the external control group is included and the case where it is not included for the OSLRT, mOSLRT, developed score tests for an early ($\mathrm{Z_{EE}}$ with $k = 4$ for scenarios 1-2, $k = 1$ for scenarios 3 and 6, $k = 4$ for scenario 4 and $k = 3$ for scenario 5), middle ($\mathrm{Z_{ME}}$ with $k_1 = 1$ and $k_2 = 6$ for scenarios 1-2, $k_1 = 1$ and $k_2 = 7$ for scenario 3, $k_1 = 1$ and $k_2 = 4$ for scenarios 4 and 6, $k_1 = 0$ and $k_2 = 3$ for scenario 5) and delayed effects ($\mathrm{Z_{DE}}$ with $k = 2$ for scenarios 1-2, $k = 1$ for scenarios 3-4 and 6 and $k = 3$ for scenario 5), crossing hazard ($\mathrm{Z_{CH}}$) and max-Combo test (Hochberg and multivariate normal integration) with a true HR of 0.5 and $\pi = 1$.
\label{Diff_variability_pi1}}
\end{figure*}

\begin{figure*}[!h]
    \centerline{\includegraphics[width=\textwidth]{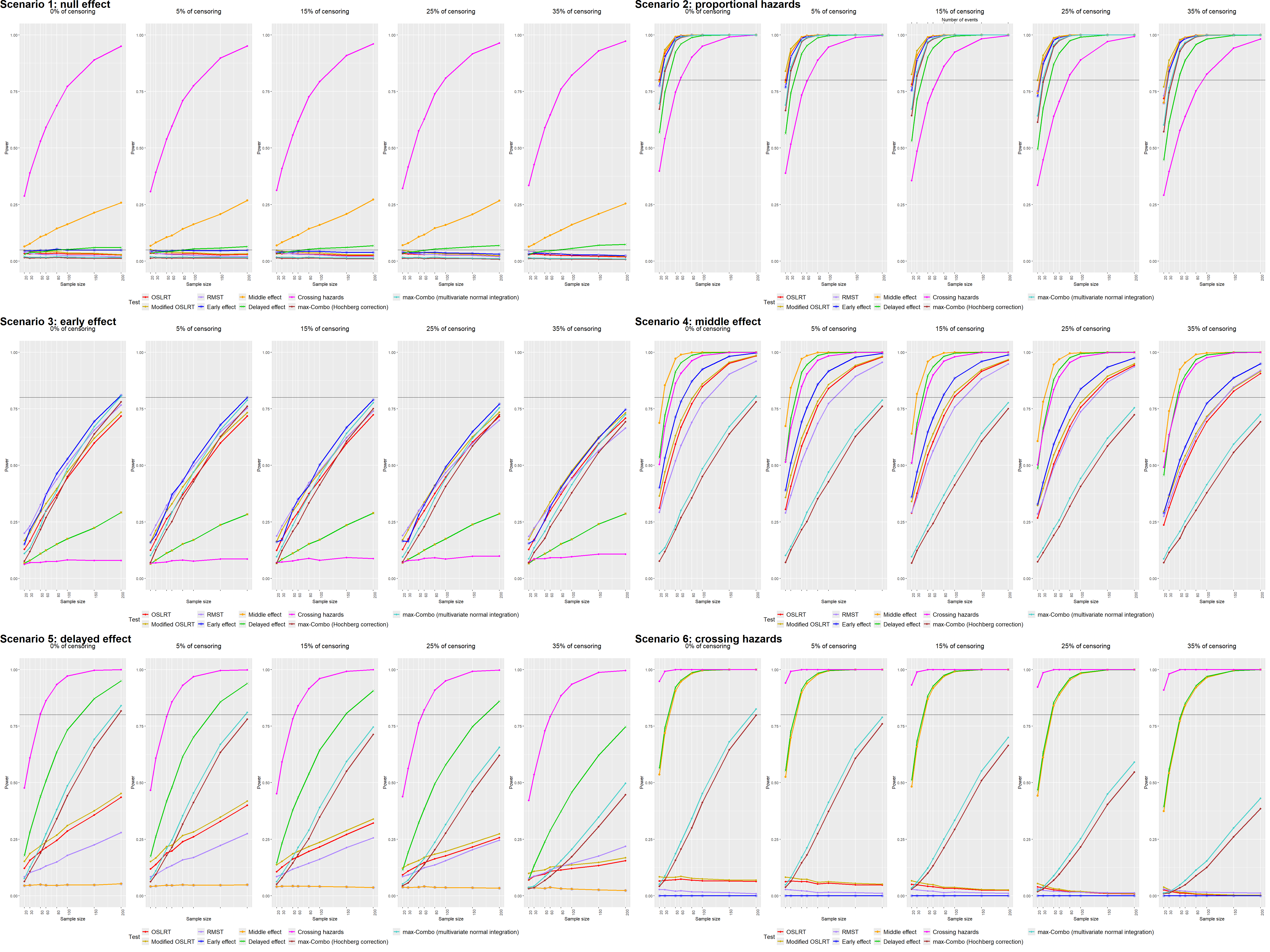}}
    \caption{Type I error (scenario 1) and power (scenario 2-6) for the OSLRT, mOSLRT, developed score tests for an early($\mathrm{Z_{EE}}$ with $k = 4$ for scenarios 1-2, $k = 1$ for scenarios 3 and 6, $k = 4$ for scenario 4 and $k = 3$ for scenario 5), middle ($\mathrm{Z_{ME}}$ with $k_1 = 1$ and $k_2 = 6$ for scenarios 1-2, $k_1 = 1$ and $k_2 = 7$ for scenario 3, $k_1 = 1$ and $k_2 = 4$ for scenarios 4 and 6, $k_1 = 0$ and $k_2 = 3$ for scenario 5) and delayed effects ($\mathrm{Z_{DE}}$ with $k = 2$ for scenarios 1-2, $k = 1$ for scenarios 3-4 and 6 and $k = 3$ for scenario 5), crossing hazards ($\mathrm{Z_{CH}}$), RMST-based test ($\tau=7$) and max-Combo test (Hochberg and multivariate normal integration) when the cumulative hazard function is modelled using a log-logistic distribution, not an exponential, and with a true HR of 0.5. Black horizontal lines indicate the nominal 5\% type I error (scenario 1) and 80\% power (scenarios 2-6).}
    \label{Power_misspe}
\end{figure*}

\begin{figure*}[!h]
    \centerline{\includegraphics[width=\textwidth]{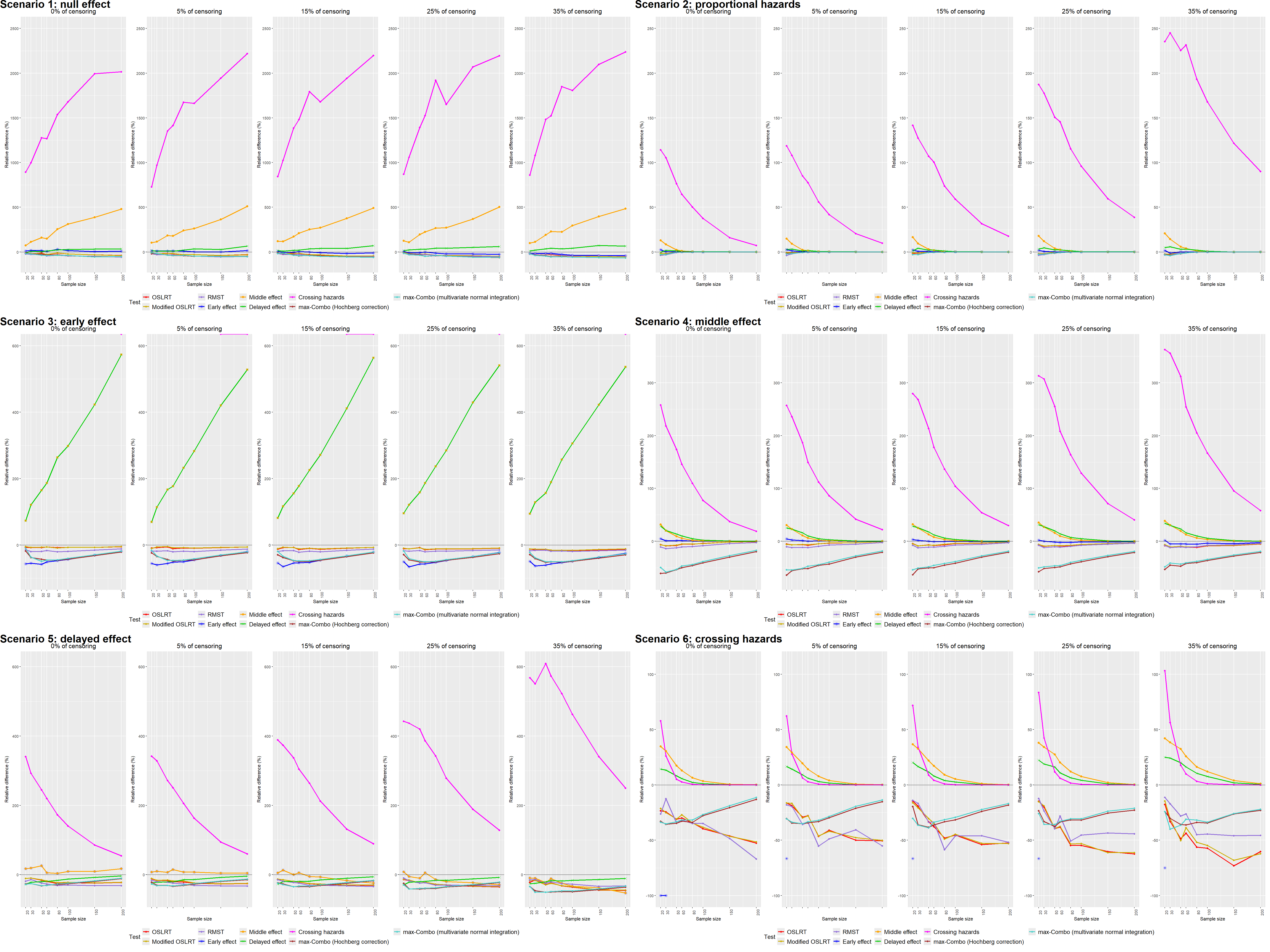}}
    \caption{Relative difference in power between the case where the survival distribution of the historical control group is misspecified (log-logistic) and the case where the survival distribution of the historical control group is correctly specified (exponential) for the OSLRT, mOSLRT, developed score tests for an early ($\mathrm{Z_{EE}}$ with $k = 4$ for scenarios 1-2, $k = 1$ for scenarios 3 and 6, $k = 4$ for scenario 4 and $k = 3$ for scenario 5), middle ($\mathrm{Z_{ME}}$ with $k_1 = 1$ and $k_2 = 6$ for scenarios 1-2, $k_1 = 1$ and $k_2 = 7$ for scenario 3, $k_1 = 1$ and $k_2 = 4$ for scenarios 4 and 6, $k_1 = 0$ and $k_2 = 3$ for scenario 5) and delayed effect ($\mathrm{Z_{DE}}$ with $k = 2$ for scenarios 1-2, $k = 1$ for scenarios 3-4 and 6 and $k = 3$ for scenario 5), crossing hazards ($\mathrm{Z_{CH}}$), RMST-based test ($\tau$ = 7) and max-Combo test (Hochberg and multivariate normal integration) with a true HR of 0.5.}
    \label{Diff_misspe_comp}
\end{figure*}

\end{document}